\documentclass[11pt]{article}

\usepackage[margin=1in]{geometry}
\usepackage[utf8]{inputenc}
\usepackage{cite}
\usepackage{amsmath,amssymb,amsfonts}
\usepackage{algorithmic}
\usepackage{graphicx}
\usepackage{textcomp}
\usepackage{url}
\usepackage{tabularx}
\usepackage{array}
\usepackage{booktabs}
\usepackage{bm}
\usepackage{xparse}
\usepackage{enumitem}
\usepackage{placeins}
\usepackage[hidelinks]{hyperref}

\newcommand{\code}[1]{\textsf{#1}}

\NewDocumentCommand{\Figure}{O{htbp} D(){} O{width=\linewidth} m m}{%
  \begin{figure}[#1]
    \centering
    \includegraphics[#3]{#4}
    \caption{#5}
  \end{figure}%
}

\newenvironment{keywords}
  {\par\medskip\noindent\textbf{Keywords: }}
  {\par\medskip}

\NewDocumentEnvironment{IEEEbiography}{O{} m}{%
  \par\bigskip\noindent
  \begin{minipage}[t]{1.1in}\centering #1\end{minipage}\hfill
  \begin{minipage}[t]{\dimexpr\textwidth-1.3in\relax}\textbf{#2}\ }{%
  \end{minipage}\par\medskip
}

\newcommand{\EOD}{}

\def\BibTeX{{\rm B\kern-.05em{\sc i\kern-.025em b}\kern-.08em
    T\kern-.1667em\lower.7ex\hbox{E}\kern-.125emX}}

\title{Agentic Graph Retrieval-Augmented Generation for Auditable Commercial Registry Analysis}
\author{%
Arthur Capozzi\\
\small Computational Social Science, ETH, Zurich, Switzerland\\
\small \texttt{arthur.capozzi@gess.ethz.ch}
\and
Dirk Helbing\\
\small Computational Social Science, ETH, Zurich, Switzerland\\
\small Complexity Science Hub, Vienna, Austria
}
\date{}

\begin{document}
\maketitle

\begin{abstract}
Public commercial registries are formally open, yet their practical analysis remains difficult because relevant facts are scattered across millions of records that combine structured metadata, multilingual legal notices, temporal events, and entity aliases. This paper presents a controlled, tool-mediated agentic GraphRAG architecture for auditable natural-language analysis of such registries. The proposed pipeline transforms publications from the Swiss Official Gazette of Commerce into a Neo4j knowledge graph comprising over five million nodes and 4.7 million relationships. It combines deterministic ingestion of structured registry fields, LLM-assisted extraction of latent actors from unstructured notices, and a deterministic identity-resolution layer. An analytical agent operates on this graph through intent routing, restricted graph tools, bounded reflection, and state-machine-guided response synthesis. We evaluate the system using a multi-tier protocol covering answer quality, retrieval behavior, entity resolution, and preliminary multi-turn conversational performance. The complete architecture is compared with dense, lexical, and hybrid flat-retrieval baselines and with controlled architectural ablations. On a manually curated 60-question benchmark, graph-mediated retrieval raises the mean Correctness score from 0.26 for the strongest flat-retrieval baseline to 0.83 for the complete system, with comparable improvements in relevance and completeness. Ablation results show that bounded reflection improves answer quality while intent routing and LLM-based graph enrichment improve reliability in difficult entity-resolution tasks. An exploratory dashboard displays the graph evidence and execution traces underlying each response, allowing users to inspect how answers were produced. Transfer to other national registries remains to be evaluated.
\end{abstract}


\section{Introduction}
\label{sec:introduction}

Recent advances in large language models (LLMs) have made agentic AI systems increasingly capable of performing complex retrieval and reasoning tasks~\cite{singh2025agentic, ferrag2026llmreasoningautonomousai}. At the same time, Retrieval-Augmented Generation (RAG) has emerged as a practical strategy for grounding these systems in external knowledge sources. However, conventional vector-based RAG architectures remain limited when the target domain contains highly structured relations, temporal dependencies, and entity-resolution challenges. In such settings, graph-based retrieval offers a more suitable foundation, as it enables multi-hop traversal over explicit relationships while still supporting natural-language interaction~\cite{Zhu_Guo_Cao_Li_Gong_2024, DongYingZouBiqing}.

% In this paper, we present a controlled agentic AI architecture built on a GraphRAG framework for expert analysis of commercial registry data.

In this paper, we present a system architecture for controlled, auditable agentic GraphRAG over large-scale commercial registry data.
We test this system with the Swiss Official Gazette of Commerce (SHAB).
The dataset spans seven years and comprises more than seven million registry publications. Within this extensive corpus, important facts are often scattered across multiple notices, expressed in German, French, or Italian, and distributed between structured metadata and free-text legal descriptions.
Consequently, although Switzerland is widely recognized for the transparency and accessibility of its public records~\cite{Cahlikova03052020}, free access in principle does not translate into effective analytical access in practice.
% An intelligent agent capable of navigating this material can therefore provide substantial value for exploratory analysis, corporate due diligence, and investigative discovery.

The proposed architecture addresses this problem through an end-to-end pipeline. First, it constructs a Neo4j knowledge graph from SHAB records by combining deterministic ingestion of structured registry fields with LLM-assisted extraction of latent actors from unstructured event text.
Second, an analytical agent operates on the graph through intent routing, restricted tool access, bounded reflection, and state-aware response synthesis. The resulting system supports natural-language questions, corporate-network exploration, event-history tracing, and higher-level analytical queries over the registry.

Although an agentic system can synthesize information quickly, commercial and financial analyses require transparency and traceability~\cite{MorgenthalPapenbrock, staley2025role}. To connect LLM-generated responses with deterministic graph evidence, the proposed system includes a human-in-the-loop exploratory dashboard. The interface enables users to inspect entity profiles, graph neighborhoods, event histories, retrieved evidence, and execution traces.
% The LLM therefore operates as one component of an auditable analytical environment rather than as an independent decision maker.

The modular architecture is designed to support adaptation beyond the Swiss setting. Its ingestion and retrieval components can be configured for other commercial gazettes and public registries that combine structured records with unstructured legal publications. Because the Swiss registry is multilingual and decentralized, the system is already exposed to substantial variation in terminology and document formatting. Nevertheless, deployment in another jurisdiction would still require adaptation of the schema, legal terminology, supported languages, and extraction prompts.
Because some regulatory and corporate environments impose stricter data-handling requirements, the proposed architecture also includes an optional privacy-abstraction layer that can reduce direct exposure of sensitive textual attributes.

A second contribution concerns evaluation. Although agentic RAG systems are increasingly common in research and practice, their assessment remains fragmented~\cite{NEURIPS2023_91f18a12, GU2026101253, ErsoyErsahin2025RAGArchitectures}. We introduce a multi-tier framework that measures entity-resolution consistency, tool behavior, answer grounding, mean Correctness, Answer Relevance, Information Recall, and preliminary multi-turn conversational behavior. The evaluation compares dense, lexical, and hybrid flat retrieval with controlled graph-architecture variants and the complete agentic system.
% LLM-based scores are treated as scalable comparative indicators rather than absolute truth labels.

To support scientific transparency and reproducibility, all code used for data collection, graph construction, agent implementation, evaluation, and dashboard exploration has been made open access~\cite{capozzi_shab_code}.

% The contribution is not a new foundation model or a new graph algorithm.
% The contribution of this paper is an evaluated system architecture for controlled, auditable agentic GraphRAG over large-scale commercial registry data.

The remainder of this paper is organized as follows. Section~\ref{sec:Related Work} reviews related work on retrieval-augmented generation, knowledge graphs, GraphRAG, and agentic AI.
Section~\ref{sec:objectives} defines the research objectives and summarizes the main architectural and empirical contributions of the study.
Section~\ref{sec:dataset} introduces the SHAB dataset, and Section~\ref{sec:architecture} presents the proposed architecture, including the data pipeline, the analytical agent, and the exploratory dashboard. Section~\ref{sec:Experimental Setup and Evaluation Methodology} introduces the evaluation methodology, and Section~\ref{sec:results} reports the empirical results across all evaluation tiers. Finally, Sections~\ref{sec:discussion}--\ref{sec:Conclusion} discuss implications, limitations, and future work.

\section{Related Work}
\label{sec:Related Work}

Analyzing commercial and corporate records is difficult because companies change names, ownership structures, and legal status over time. Relevant evidence is often dispersed across large volumes of structured and unstructured data.
Complex economic relationships may therefore remain implicit in legal text, registry metadata, and numerical records.
Traditional analysis relies heavily on manual processing and domain expertise, making multi-hop information extraction time-consuming and error-prone~\cite{10252608}.

Researchers have long used expert systems to navigate the complexity of commercial datasets. These systems have expanded beyond basic analysis and forecasting to applications such as fraud detection and bankruptcy prediction.
For example, Kirkos et al.~\cite{KIRKOS2007995} explored how decision trees and neural networks could effectively identify management fraud and anomalous reporting in corporate financial statements.
Similarly, Jo et al.~\cite{JO199797} demonstrated the effectiveness of algorithmic expert systems in evaluating financial health and predicting corporate insolvency.

Because financial decisions require accountability, explainability is an important aspect of bankruptcy prediction. Several studies therefore focus on models that provide interpretable explanations.
For example, Cho and Shin~\cite{CHO2023119390} proposed a genetic algorithm that generates clear, feature-weighted counterfactual explanations, making bankruptcy prediction models more interpretable and trustworthy.
In the same direction, Liu et al.~\cite{LiuJiamingChengzhang} showed that tree-based gradient boosting models can achieve accurate predictive performance, while TreeSHAP and Shapley regression effectively interpret the complex financial indicators driving those predictions.

\subsection{Retrieval-augmented generation}

Retrieval-augmented generation supplements an LLM with information retrieved from an external source. A typical RAG pipeline retrieves documents or passages relevant to a user query and supplies them to the model during response generation. This process can improve grounding and provide access to specialized or current information that is not reliably represented in the model parameters~\cite{LewisPerezPetroni}. However, the quality of the response still depends on the retrieval method, the retrieved evidence, and the generation procedure~\cite{SahaEtAl2024QuIMRAG}.

RAG has been applied to several financial and commercial technology tasks.
To assist private investors with financial report analysis, Iaroshev et al.~\cite{app14209318} showed that a tailored RAG system with high-quality components can substantially improve the accuracy and relevance of answers, especially for qualitative queries on well-structured documents.
Other specialized RAG frameworks combine multi-modal preprocessing, multi-path retrieval, and domain-specific re-ranking to overcome the challenges of extracting information from complex, multi-modal financial documents~\cite{WangChiTaiKwok}.
Hota~\cite{11380167} combined RAG with contextual financial knowledge graphs for portfolio optimization and investment analysis.
Another line of work unifies tabular and textual data processing by maintaining structural awareness instead of flattening tables into plain text~\cite{informatics13020030}.
Further RAG approaches retrieve financial information from third-party sources, such as news platforms or social media, to improve the accuracy of sentiment analysis of financial reports~\cite{ZhangBoyuYang}.
Similarly, recent work in legal tech emphasizes that, when RAG is applied to legal documents, retrieval precision, interpretability, and traceable evidence are central requirements~\cite{HindiEtAl2025LegalRAG}.

Standard vector-based RAG systems face important limitations when applied to financial databases. Retrieval primarily based on semantic similarity may return generic passages rather than the specific legal or structural evidence required by a query.
Naive document chunking can also break semantic connections and obscure the structure of financial documents~\cite{JIANG2025103909, YANG2025103904}. In particular, independently embedding metadata, text, and tables can separate numerical values from their textual and relational context.
Flat retrieval systems also do not directly support deterministic corpus-wide aggregation. For example, answering ``What are the top ten cryptocurrency companies sorted by share capital in city X?'' requires filtered ranking over structured attributes rather than only passage similarity. More generally, heterogeneous data spaces require retrieval designs that coordinate multiple data representations and lifecycle stages~\cite{AlQatfEtAl2025RAG4DS}.

\subsection{Knowledge graphs}

A knowledge graph represents entities as nodes and their relationships as edges. This structure preserves explicit context and hierarchies and supports multi-hop traversal across large datasets. Knowledge graphs are therefore well suited to the interconnected nature of financial markets and corporate structures.

Several studies have investigated the creation of knowledge graphs from unstructured financial data.
Fu et al.~\cite{8588771} proposed an integrated stochastic optimization technique based on a financial knowledge graph to build a market return prediction system.
Financial knowledge graphs have also been used in combination with event embeddings learned from financial news to support quantitative investments~\cite{ChengYangXiaoyangZhang}.

In addition to graph construction, graph-based methods have been used to detect fraud~\cite{weber2019anti,capozzi2025flowseries,ValeriaRonchiadin,vilella2025weirdnodes, LIU2025104198}, predict stock prices~\cite{FengFuliXiangnan} and new firms~\cite{capozzi2026beyond}.
Graph attention networks have also been used to model economic connections derived from financial documents and analyze market-momentum spillovers~\cite{ChungAndyKumiko}.

Other relevant applications include financial question answering~\cite{CHENG2026114670}, systemic-risk measurement~\cite{LIU2025102924}, and explainable fraud detection from commercial statements~\cite{CAI2024123126}. Further studies demonstrate that question answering over knowledge graphs can be achieved by decomposing questions into structured graph operations, eliminating the need for task-specific supervision~\cite{RonyEtAl2022TreeKGQA}.

\subsection{GraphRAG}

GraphRAG extends retrieval-augmented generation by incorporating graph-structured representations and graph-based retrieval.
GraphRAG systems generally follow a three-step process~\cite{ZhouYaodongYouranShuTaotao}.
First, during the construction phase, an AI model processes a collection of text documents, identifying key concepts and the links between them, to form a unified network.
Second, during retrieval, the system identifies relevant nodes, edges, paths, or graph communities. Third, during synthesis, an LLM combines the retrieved graph evidence with the user's question to produce a response.

Early GraphRAG work showed that flat RAG can struggle with query-focused summarization when evidence is distributed across a corpus~\cite{edge2025localglobalgraphrag}. Graph-based retrieval can address this limitation by traversing explicit relationships rather than relying only on vector similarity.
Other benchmarks report advantages for graph-aware generation in multi-hop reasoning tasks~\cite{huetal2025grag}. Domain-specific systems have also combined knowledge graphs with RAG to retrieve interconnected technical evidence, including cybersecurity vulnerabilities in power networks~\cite{GaoChang2025KGRAG}.

\subsection{Agentic AI}

Agentic AI refers to systems that use language models together with planning, tool use, memory, or feedback mechanisms to pursue multi-step objectives~\cite{AcharyaKuppanDivya2025AgenticAI}. Such agents can decompose complex tasks, select actions, inspect intermediate results, and revise their plans when an initial action fails.

Piccialli et al.~\cite{PICCIALLI2025128404} surveyed how autonomous agents enhance scalability and decision-making in complex industrial and analytical applications.
Agentic capabilities are also attracting growing interest in expert systems and finance~\cite{11167368, 11125703, ElgendyMohamedSharafi}.
For example, autonomous systems now process financial information, support decisions, and execute workflows~\cite{RIZINSKI20251}. They have moved beyond simple task automation to formulate and evaluate end-to-end trading and risk management strategies.
In this direction, Han et al.~\cite{XuewenCheYangHongyang} developed a multi-agent collaboration system designed to enhance decision-making in financial investment research.

The intersection of agentic AI and RAG systems is a rapidly developing frontier~\cite{singh2025agentic}. In 2024, Asai et al.~\cite{asai2024selfrag} introduced Self-RAG, a framework that trains a model to adaptively retrieve information on demand and use special reflection tokens to evaluate the retrieved text and its own generated responses. A similar example is RAG-Critic, which uses an automated critic workflow to detect errors and prompt the system to self-correct before producing a final response~\cite{dongetal2025rag}. Papageorgiou et al.~\cite{PapageorgiouEtAl2025Faithfulness} evaluated statement-level faithfulness in modular agentic RAG pipelines for e-governance, using multiple LLM judges, illustrating the importance of source-aware evaluation in tool-mediated retrieval systems.
Another application is dynamic network optimization, in which an interactive AI framework uses RAG to maintain a real-time knowledge base for autonomous network management~\cite{10531073}.

\section{Research Objectives and Contributions}
\label{sec:objectives}

Agentic AI, knowledge graphs, and GraphRAG have each been studied in different contexts, but few studies evaluate their controlled integration for large-scale registry exploration~\cite{PeiLiuLiuMengJun,app152312547}.
This gap is especially relevant for commercial and financial technology, where registries and documents contain explicit relationships, temporal events, and entity-resolution challenges.
The aim of this paper is therefore to evaluate whether controlled graph-mediated retrieval improves registry exploration compared with vector-based retrieval in tasks requiring identity resolution, temporal lookup, aggregation, and multi-hop traversal.

The contribution of this work is both architectural and empirical. We integrate knowledge-graph construction, agentic retrieval, and evidence inspection into a controlled framework for large-scale national registry analysis. We also conduct an ablation study to assess the contribution of the main architectural components.

This study defines four research objectives (ROs) that target the limitations of standard vector-based RAG architectures in commercial-registry analysis:

\begin{itemize}
    \item \textbf{RO1 (Data Architecture):} We construct a large-scale multilingual knowledge graph from Swiss commercial-registry publications by combining deterministic ingestion of structured metadata with LLM-assisted identification of latent actors from unstructured legal notices.
    \item \textbf{RO2 (System Design):} We introduce a controlled, tool-mediated agentic GraphRAG architecture that combines intent-gated tool access, bounded reflection, and state-machine-guided response synthesis for registry exploration.
    \item \textbf{RO3 (Audit Workflow):} We provide a dashboard-supported audit workflow that exposes entity dossiers, event histories, graph neighborhoods, retrieved evidence, and execution traces to support expert inspection.
    \item \textbf{RO4 (Evaluation):} We evaluate the system with a multi-tier protocol covering entity-resolution consistency, tool trajectories, answer quality, controlled architecture variants, strengthened flat baselines, human validation, weak-node extraction validation, and preliminary multi-turn conversational behavior.
\end{itemize}

\section{The Swiss Official Gazette of Commerce Dataset}\label{sec:dataset}

To validate the proposed GraphRAG architecture, we collect and process Open Government Data from the Swiss Official Gazette of Commerce\footnote{\url{https://www.shab.ch/}}. SHAB is Switzerland's central federal publication platform for legally binding notices concerning corporate life-cycle events, including company formation, bankruptcy, management changes, and liquidation. The code used to download, parse, and process the registry data is available in a public repository~\cite{capozzi_shab_code}.

The SHAB dataset spans seven years, from September 2018 to September 2025, and contains 7,014,137 raw records.
Each official publication record includes rich structured metadata, such as its primary rubric and sub-rubric, creation date, geographic location, and language. The rubric code identifies the broad legal nature of the event, such as the registration of a new company, a change of board members, or a corporate liquidation. However, the specific details of each event, together with the identities of the connected companies and secondary actors, are embedded in an unstructured free-text payload.
To reconstruct the corporate network, these hidden relationships can be extracted from the gazette text. We use an LLM-driven pipeline because the notices lack a uniform structure, vary substantially in length, and are published in German, French, or Italian. For example, a standard bankruptcy suspension notice (rubric KK06) can be as short as 163 characters, whereas a complex bankruptcy proceeding (rubric KK08) can be more than 100 times longer and may refer to several creditors and liquidators.
This LLM-driven approach to identifying entities and their relationships is supported by recent advances in document-level relation extraction, including confidence-based revision methods for refining relational facts in noisy, large-scale datasets~\cite{JIANG2025103909}.

% Section~\ref{sec:Data Pipeline} describes the pipeline used to compile the final dataset, which comprises over 2.1 million events, over one million company mentions, and more than 650,000 individual mentions.

\section{Proposed System Architecture}
\label{sec:architecture}
% exploratory
% This section outlines the architecture designed to transform raw registry files into an interactive analytical tool.
The proposed system architecture consists of three main components. First, the data pipeline (Section~\ref{sec:Data Pipeline}) ingests data and performs topological resolution to populate a Neo4j knowledge graph. Second, the analytical agent (Section~\ref{sec:Investigator Agent}) interprets human questions and queries the knowledge graph iteratively through controlled tools. Third, the exploratory dashboard (Section~\ref{sec:Forensic_Dashboard}) links the system to a user interface that allows users to visually analyze evidence and graph neighborhoods.

\subsection{Data ingestion pipeline}
\label{sec:Data Pipeline}

As shown in Fig.~\ref{fig:pipeline}, the data pipeline has three phases. It creates strong nodes from structured fields, extracts weak nodes from unstructured text, and resolves entity identities through tokenization and database cleanup.

\Figure[t!](topskip=0pt, botskip=0pt, midskip=0pt)[width=0.97\textwidth]{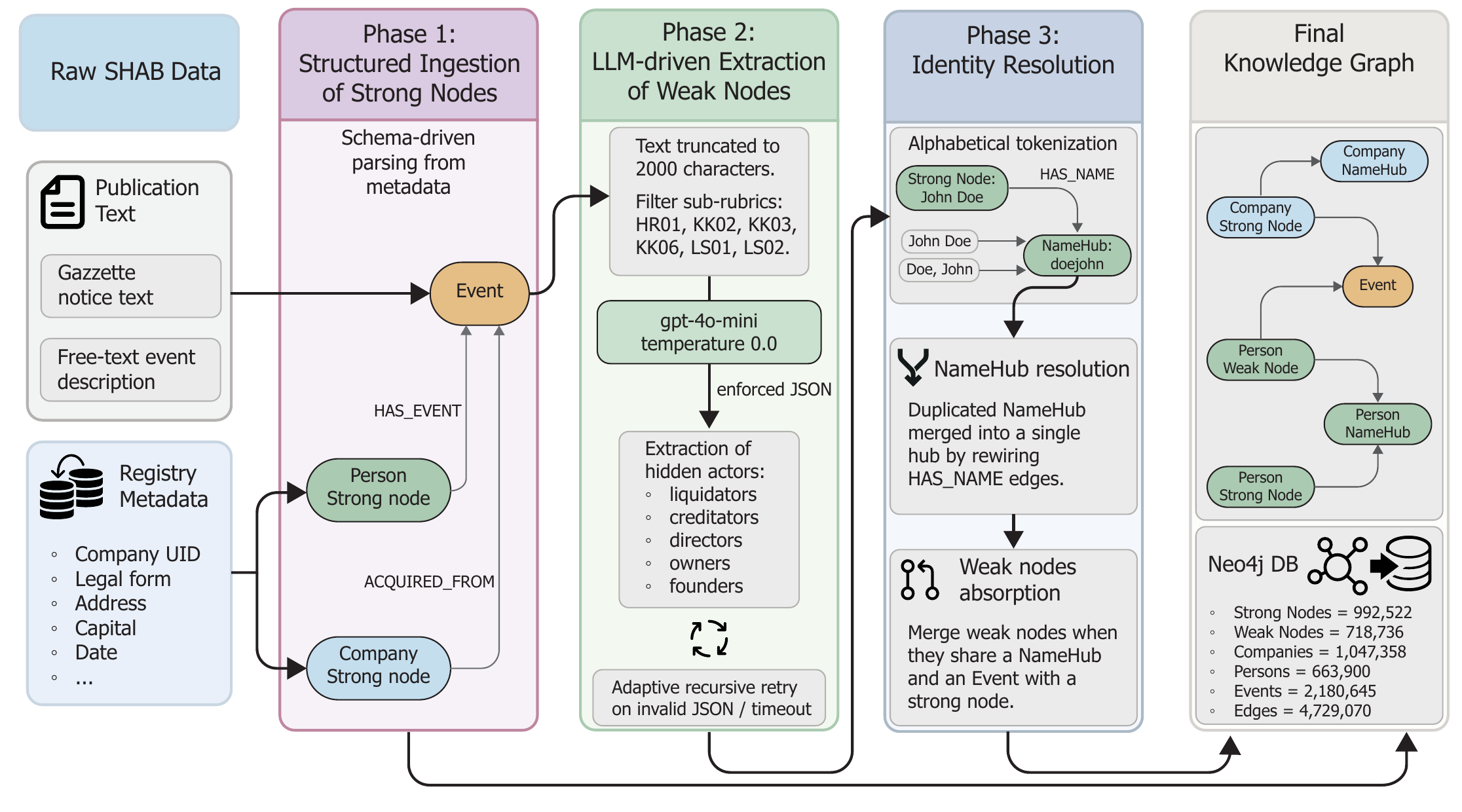}{Data ingestion pipeline used to construct the knowledge graph. Phase 1 performs schema-driven ingestion to create strong Company, Person, and Event nodes from verified fields. Phase 2 applies an LLM to commercial texts to extract hidden actors as weak nodes under constrained JSON output. Phase 3 resolves identity through alphabetical tokenization, NameHub merging, and weak-node absorption. The final output is a deduplicated Neo4j knowledge graph. \label{fig:pipeline}}

% \subsubsection{Graph schema and node taxonomy}

The proposed agentic GraphRAG system is built on a structured Neo4j knowledge graph.
All nodes in the knowledge graph inherit a \code{BaseNode} label, which supports a global unique-identifier index for efficient lookup across millions of entities. The schema organizes the corporate network into four primary node types: \code{Company}, \code{Person}, \code{Event}, and \code{NameHub}.

Company nodes represent legal entities. They store verified metadata extracted directly from the official registry, including the Swiss UID, legal form, nominal and paid-in capital, business purpose, and registered address. 

Person nodes represent individuals involved in corporate events, such as founders, board members, and liquidators.
NameHub nodes are abstract bridging entities for identity resolution. Because historical documents frequently contain spelling variations and transcription errors, these hubs link different string mentions to a shared normalized name anchor used during graph traversal.

Event nodes represent the official publication records. Rather than storing only the raw text, these nodes aggregate sparse metadata, such as publication titles, resolution dates, and affected land parcels, along with the main legal text in a single \code{full\_text} property.

Semantic edges interconnect the nodes to map the relational logic of the network.
Person nodes and Company nodes are never connected directly to each other, but always through an Event node.
For example, \code{HAS\_EVENT} links companies to their chronological publications, \code{ACTED\_IN} connects individuals to specific corporate resolutions, and \code{HAS\_NAME} binds entities to their corresponding NameHub.
The complete mapping of semantic roles to edge types is provided in the Supplementary Table S1.

\subsubsection{Structured Ingestion (Strong Nodes)}

The first phase of the data ingestion pipeline processes the original registry data row by row to generate ``strong nodes.'' These nodes represent entities derived directly from structured fields in the official register.
The metadata for Company nodes includes specific attributes, such as the standardized business identification number (UID), legal form, registered address, and capital. For Person nodes, the pipeline ensures database ingestion that is idempotent (repeatable without duplication).
Event nodes are constructed by aggregating sparse metadata columns into a \code{full\_text} string. This maximizes the contextual density of the event for the downstream LLM analysis (Phase 2). Finally, deterministic edges are generated based on explicit column mappings. For example, a buyer is linked to an acquired company via an \code{ACQUIRED\_FROM} edge, and a debtor is connected to a bankruptcy event via an \code{INVOLVED\_IN} edge.

\subsubsection{LLM-Driven Unstructured Extraction (Weak Nodes)}
\label{sec:phase_2}
A major challenge in processing historical commercial data is that secondary actors, such as liquidators, creditors, and temporarily appointed directors, are often only mentioned in the unstructured free text of the gazette notice. To identify these latent actors, Phase 2 employs a large language model, GPT-4o mini (version 2024-07-18), with enforced JSON object formatting and a temperature of 0.0. The model performs named entity recognition and relationship extraction on the \code{full\_text} field of the Event nodes.

In summary, the 7,014,137 source records are consolidated into 2,180,645 unique \code{Event} nodes because multiple input rows can refer to the same registry event. To minimize API costs and maximize the signal-to-noise ratio, the extraction is limited to 541,345 events from selected, high-value subrubrics. These include new foundations (rubric HR01 with 350,717 events), bankruptcies (rubrics KK03 with 47,608 events, KK02 with 36,050 events, and KK06 with 31,409 events), and liquidations (rubrics LS01 with 30,451 events and LS02 with 45,110 events).

A detailed system prompt instructs the LLM to extract Person and Company entities and infer their semantic roles (e.g., ``Creditor'' or ``Liquidator''). To ensure high-quality data, the prompt enforces three main extraction rules. First, the model must split compound names and extract individuals listed conjunctively as separate JSON objects. Second, the model must disambiguate generic legal authorities (e.g., a bankruptcy office, or ``Konkursamt'') by appending the relevant city identified in the text. Third, the system applies a deterministic stoplist (e.g., ``SHAB'') to filter out official gazette publishers and prevent metadata from polluting the graph.

To process the large number of Event nodes efficiently, the extraction pipeline uses batch processing with two main constraints. Complex alphanumeric UIDs are mapped to small integers in the prompt payload to reduce context length and avoid identifier corruption. Event texts are also truncated to 2,000 characters. The LLM returns a JSON object keyed by these temporary integers, which the pipeline maps back to the original database UIDs. An adaptive recursive-retry strategy is also employed. If a large batch times out or produces invalid JSON, the algorithm divides it into smaller batches and retries the extraction.

Entities extracted dynamically in this phase are identified as ``weak nodes.''
During their identification, every Company and Person node, both weak and strong, is linked via a \code{HAS\_NAME} edge to a central NameHub node. Each hub is keyed by a normalized version of the entity name, generated by an algorithm described in Phase 3. The NameHub layer therefore acts as an intermediate identity layer.
Combining structured records from Phase 1 with LLM-extracted entities from Phase 2 comes with substantial variation in entity names. The same person may appear as ``John Doe,'' ``Doe, John,'' or ``John A. Doe.'' Dynamically extracted entities may also duplicate structured nodes that already exist in the graph. If left unresolved, these variants would fragment graph traversal paths and reduce retrieval reliability.

\subsubsection{Identity Resolution}
\label{sec:phase_3}

Name inconsistencies are resolved by applying an alphabetical hub-key generation algorithm to all Company and Person nodes, whether they originate from strong structured data or weak LLM extractions. This algorithm normalizes the name string through \textit{alphabetical tokenization}: it lowercases the text, strips non-alphanumeric characters, splits the string into discrete tokens, sorts them alphabetically, and concatenates them.
After this, both ``Doe, John'' and ``John Doe'' deterministically yield the identical key \code{doejohn}. This mechanism is intentionally strict. Nodes are linked to the same central NameHub only when the resulting normalized key is byte-for-byte identical; at this step, no fuzzy tolerance is introduced. As a result, borderline cases such as names differing only by a middle initial are conservatively rejected.

% The NameHub resolution strategy is conservative by design: exact normalized-key matching prioritizes precision over recall. This reduces the risk of false entity merges, which are especially problematic in financial investigation, but it can miss true aliases that differ by initials, spelling variants, or transliteration.

Because ingestion occurs in parallel batches during the LLM extraction stage, duplicate NameHub nodes can occasionally be created for the same normalized key. To eliminate this redundancy, the pipeline includes a \textit{NameHub resolution} step. A deduplication query scans all existing hubs, recomputes their alphabetical tokenization keys, groups those that share the same canonical key, and selects a single surviving master hub for each duplicate set.
All incoming \code{HAS\_NAME} edges are then rewired from the redundant hubs to the master hub.

Following the merger of duplicate hubs, the pipeline performs a \textit{weak-node absorption} step to further clean up the graph topology. This procedure targets any weak node that shares both a NameHub and an event connection with a strong node. The system recognizes this pattern and infers that the weak node is simply an LLM-extracted duplicate of a structured registry entity that is already present in the event context. The redundant weak node is then deleted and its edges reconnected, leaving only the verified strong node.
These two cleanup stages transform the NameHub architecture into a conservative identity-resolution mechanism that unifies exact token-order aliases across structured and unstructured ingestion.

The final Neo4j knowledge graph comprises over 5 million nodes and 4.7 million relationships. Of these, 992,522 are strong nodes and 718,736 are weak nodes. Complete statistics can be found in Supplementary Table S2.

\subsubsection{Privacy-Preserving Abstraction Layer}
\label{sec:Privacy_Preserving_Abstraction_Layer}

Although public registries, such as SHAB, promote transparency, downstream analytical applications may require stricter handling of personally identifiable information. To support this deployment setting, we introduce an optional abstraction layer that obscures selected textual attributes before final deployment.
This layer runs as a post-processing operation immediately following the structural identity resolution phase.

The abstraction mechanism uses HMAC-SHA256~\cite{BellareCanetti}. When activated, the pipeline extracts selected textual properties, including entity names, registered locations, and event descriptions. It converts these values into fixed-length tokens and writes the tokens back to the database through bulk update operations.
This targeted approach aligns with recent work on privacy-aware and permission-aware RAG frameworks~\cite{HE2025104150, JeongLee2025PermissionAwareRAG}, which control access to sensitive information at the entity, document, or resource level.
Consequently, the knowledge graph can retain the topological structure required for graph analysis while reducing direct exposure of raw textual attributes to the reasoning agent.

Because SHAB is public Open Government Data and the quality evaluation requires human-readable evidence, the abstraction layer was not activated during the experiments reported in this paper. Its effects on retrieval quality and latency were therefore not evaluated. When activated, the framework can save a translation table that maps hashed tokens back to their original plain-text values for authorized users only.

\subsection{Analytical agent}
\label{sec:Investigator Agent}

The analytical agent interprets user questions and retrieves evidence from the Neo4j knowledge graph through a controlled, tool-based workflow. Instead of querying the database directly, the agent translates natural-language requests into executable back-end API calls.
These calls are mapped to highly optimized queries written in Cypher, a declarative graph database language~\cite{NadimeGuagliardoLibkin}.

LLM systems that rely on a single monolithic prompt often struggle with complex database interactions. They may select unsuitable tools, generate invalid parameters, traverse the graph inefficiently, or produce unsafe queries. To address these limitations, the analytical agent uses the multi-layer architecture shown in Fig.~\ref{fig:agent_arch}. A zero-shot \textit{intent router} first classifies the query and restricts the available tools. An \textit{agentic reflection loop} then selects tools, generates JSON arguments, and responds to failed actions. Finally, the \textit{response-synthesis} stage uses a strict state machine to control follow-up behavior and output formatting.
This architecture is designed to let the agent retrieve multi-step factual information from the graph while maintaining a flexible natural-language interface. At the same time, the separation between routing, tool execution, and synthesis controls the model's behavior, making the retrieval process easier to inspect.

Table~\ref{tab:llm_usage} summarizes the configuration, purpose, and invocation limits of each LLM call.
For all model invocations, the top-\textit{p} parameter is set to $1$.
The frequency and presence penalties are set to 0.0 to avoid penalizing repeated legal names and structural terminology.

\begin{table}[htbp]
\centering
\caption{Summary of LLM calls within the agent architecture.}
\label{tab:llm_usage}
\footnotesize
\begin{tabularx}{\columnwidth}{@{} 
    >{\raggedright\arraybackslash}p{2.2cm} 
    c 
    >{\raggedright\arraybackslash}X 
    >{\raggedright\arraybackslash}p{2.0cm} @{}}
\hline
\textbf{Architecture Phase} & \textbf{Calls} & \textbf{Purpose} & \textbf{Configuration} \\
\hline
1. Intent Router & 1 & Classifies user intent to dynamically restrict the available sandbox tools. & GPT-4o mini \newline (2024-07-18) \newline $T=0.0$ \\
\hline
2. Reflection Loop & 0--4 & Autonomously structures JSON tool requests and interprets raw graph responses. & GPT-4o mini \newline (2024-07-18) \newline $T=0.0$, tool calling \\
\hline
3. Response Synthesis & 1 & Formats the accumulated evidence according to UI constraints (e.g., Markdown tables or dossiers). & GPT-4o mini \newline (2024-07-18) \newline $T=0.0$, tool calling \\
\hline
\multicolumn{4}{@{}p{\columnwidth}@{}}{\scriptsize \textit{Note: Total calls per user query range from 2 to 6, bounded by the reflection-loop limit of four.}} \\
\end{tabularx}
\end{table}

The agentic reflection loop may execute zero tool calls if the requested information is already present in the conversational history or if the user submits a query entirely unrelated to the database. Zero-call executions also occur when the graphical interface injects a priority context rule containing the active entity identifier. In that case, the agent can answer directly from the active UI state. This mechanism is detailed in Section~\ref{sec:Forensic_Dashboard}.

\Figure[t!](topskip=0pt, botskip=0pt, midskip=0pt)[width=0.97\textwidth]{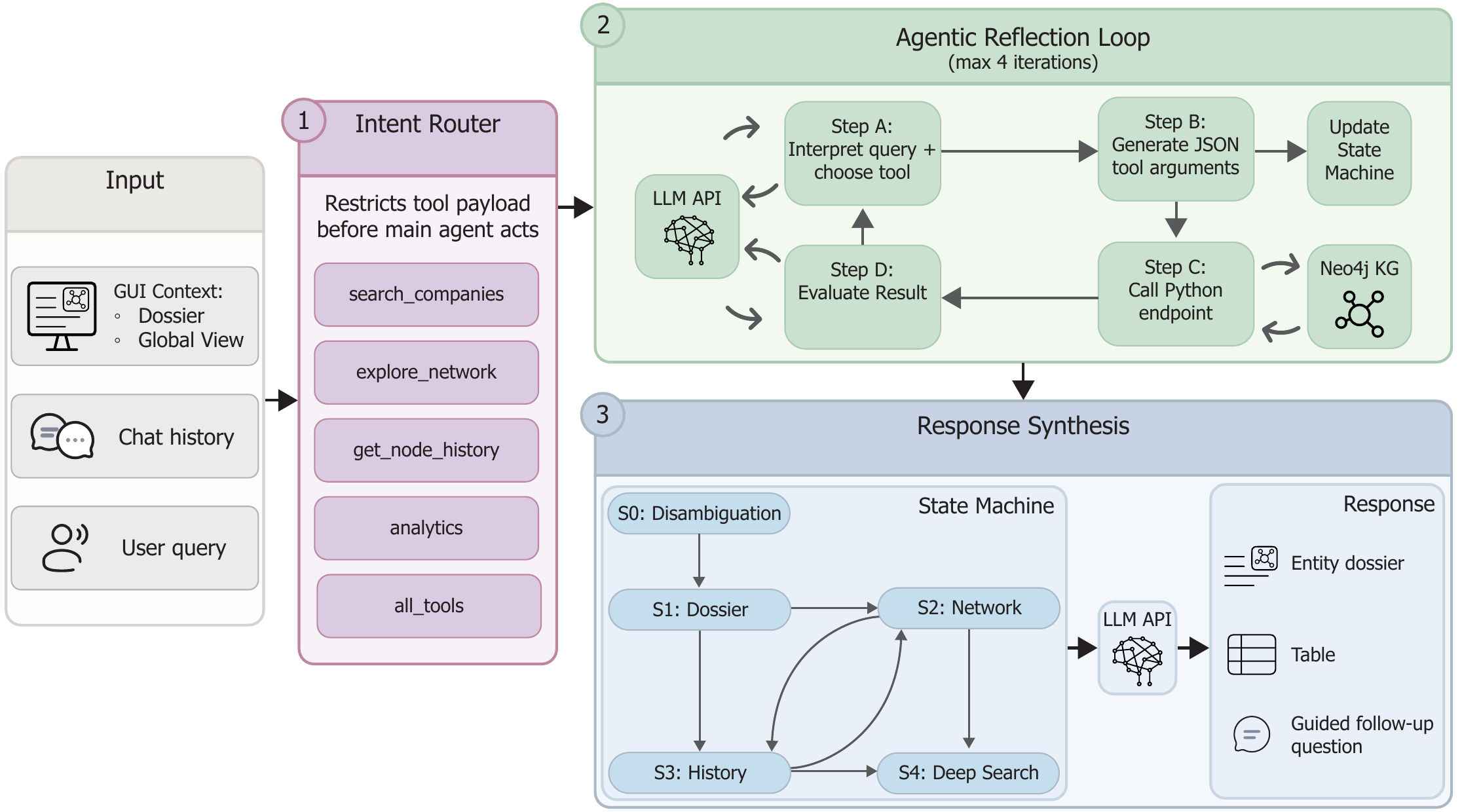}{Architecture of the analytical agent. In Phase 1, a zero-shot intent router restricts the available tools. In Phase 2, the bounded reflection loop selects secure Python endpoints, generates JSON arguments, queries Neo4j, and incorporates execution feedback. The state machine tracks the analytical trajectory and constrains the final response synthesis in Phase 3. \label{fig:agent_arch}}

\subsubsection{Intent router}
\label{sec:The Intent Router}

A large tool set can make tool selection more difficult for an LLM agent. The model may select an inappropriate endpoint or generate arguments merely to satisfy a tool schema. To address this issue, the proposed architecture uses a zero-shot intent router.
As illustrated in Fig.~\ref{fig:agent_arch}, before exploring the knowledge graph, an LLM evaluates the user query against a concise system prompt.
The router strictly classifies the request into one of five predefined semantic intent categories: entity disambiguation, network exploration, event-history retrieval, macro-level analytics, or unrestricted multi-hop exploration.
Supplementary Table S3 provides a comprehensive breakdown of the semantic intents and their associated tools.

Based on this classification, the system dynamically restricts the tool payload injected into the primary agent's context.
For standard explorations focused on specific entities, the router loads a suite of foundational search and traversal tools. This helps the agent to identify an entity and map its corporate connections before moving to synthesis.
Conversely, for macro-analytics involving high-level statistical questions, such as determining the top companies by capital, the system blocks access to exploratory tools and loads only statistical aggregation functions.
This enforced restriction prevents the LLM from performing inefficient operations, such as manually paginating through hundreds of company records to calculate a sum.

\subsubsection{Agentic reflection loop}
\label{sec:Agentic_Reflection_Loop}

Even when the intent router restricts the action space, one tool call may be insufficient. An entity name may be misspelled, the selected tool may require an unknown identifier, or the first retrieval attempt may return no results. The second phase therefore consists of a bounded agentic reflection loop.

As illustrated in Fig.~\ref{fig:agent_arch}, the loop receives the subset of tools selected by the intent router and runs for at most four iterations using GPT-4o mini. Each iteration has four steps, and each one builds on the results of the previous iteration. In step A, the LLM interprets the query and the accumulated conversational and retrieval context. In step B, it generates a JSON payload containing the selected tool and its arguments. In step C, the backend maps the payload to a predefined semantic endpoint, which queries Neo4j. In step D, the backend returns the execution result. Successful results are added to the context. Failed or empty results produce deterministic feedback that identifies the failure and suggests a valid recovery step. The model can then search for a missing identifier, revise the query terms, or switch to broader text retrieval.

The loop can revise its retrieval strategy by selecting among predefined semantic endpoints. The entity-disambiguation endpoint uses fuzzy matching to identify candidates from partial or slightly misspelled names. After identifying an entity, the agent can retrieve its network or event history. If structured lookup fails, the loop can query a Neo4j Lucene index over raw publication text. A read-only custom-query endpoint is reserved for requests that cannot be handled by the predefined pathways. Regex-based guards block mutation keywords such as \code{CREATE}, \code{DELETE}, and \code{MERGE}.

% Each iteration builds on the preceding results. When a tool returns no data, the backend adds the failed result and a recovery instruction to the context. The model can then search for a missing identifier, revise the query terms, or switch to broader text retrieval. The loop ends when sufficient evidence has been retrieved or the maximum number of iterations is reached.

\subsubsection{Response synthesis}
\label{sec:response_synthesis}

The final stage combines state control with response synthesis. After each tool execution, the backend inspects the returned JSON, the execution trace, and the conversational history. It then updates the current analytical state and adds a deterministic instruction to the model context. This instruction can restrict the next retrieval action, prevent repetition of an exhausted path, or define the follow-up option that may be presented to the user. Once the retrieval process is complete, a separate LLM call receives the accumulated evidence and final state instruction, generating the user-facing response.

This control step is necessary because multi-turn LLM agents may lose procedural context even when the dialogue history is available. They may repeat a tool call, revisit an exhausted retrieval path, or propose a next step that is inconsistent with the evidence already shown~\cite{yao2022react, zhu2025llm}. Therefore, the architecture employs a strict state machine (SSM) to govern the transition from tool execution to final response generation. The SSM does not write the answer. Rather, it determines the valid procedural state and constrains the context used by the synthesis model.

Formally, the SSM is defined by the tuple $(S, \Sigma, \delta, s_0)$, where $S = \{S_0, S_1, S_2, S_3, S_4\}$ is the finite set of operational states, $\Sigma$ is the input alphabet of tool-execution events and user intents inferred from the conversational history, $\delta: S \times \Sigma \rightarrow S$ is the deterministic transition function, and $s_0 = S_0$ is the initial state. The transition logic determines which instruction is added before the next retrieval or synthesis step:
\begin{itemize}
    \item \textbf{State $S_0$ (Disambiguation Mode):} This state is entered when a query returns multiple matching entities. The injected instruction forces the synthesis step to output a complete Markdown table of the entities and to conclude with the suggestion: \textit{``Which of these entities would you like to investigate further?''}
    \item \textbf{State $S_1$ (Dossier Mode):} $S_0 \rightarrow S_1$ is triggered when a single entity is identified or when the user selects one entity from the disambiguation table. The synthesis step is then constrained to output a dossier-style entity profile and to offer the two valid next directions: network exploration or event history.
    \item \textbf{State $S_2$ (Network Explored):} $S_1 \rightarrow S_2$ is triggered when the network has already been explored and the network-explored flag is true. If the history-explored flag is still false, the injected instruction directs the synthesis step toward event history; otherwise, the system transitions toward deep text search.
    \item \textbf{State $S_3$ (History Explored):} $S_1 \rightarrow S_3$ is triggered when node history has already been shown and the history-explored flag is true. If the network-explored flag is still false, the synthesis step is instructed to suggest network exploration; otherwise, it transitions toward deep text search.
    \item \textbf{State $S_4$ (Deep Text Search):} This state is reached when both the network-explored and history-explored flags are true, or when the user explicitly requests further corroboration after structured retrieval has been exhausted. The state instruction then permits or requires the lexical full-text fallback. After that retrieval step, the synthesis call presents the resulting unstructured evidence.
\end{itemize}

Once retrieval is complete and the final state instruction has been added, the synthesis LLM produces the response. Its input contains the retrieved JSON records, the current-turn execution trace, the prior conversational history, and, when available, the GUI-provided entity anchor described in Section~\ref{sec:Forensic_Dashboard}. The model converts this evidence into the required presentation format, such as a Markdown table, an entity dossier, or a concise explanatory summary. The state instruction also determines which procedural follow-up, if any, may be offered.

\subsection{Dashboard-Supported Audit Workflow}
\label{sec:Forensic_Dashboard}

We develop a dashboard-supported audit workflow for expert commercial analysis. The interface itself does not make the agent intrinsically trustworthy. Instead, it supports auditability by providing the evidence, graph neighborhoods, event histories, and execution traces used to produce an answer.
Recent work on human supervision of agentic AI supports this design direction, emphasizing the value of ``medium autonomy'' systems that keep users involved in consequential analytical workflows~\cite{GENINATTICOSSATIN2026104681}.

\begin{figure}[ht!]
		\centering
		\includegraphics[width=.99\columnwidth]{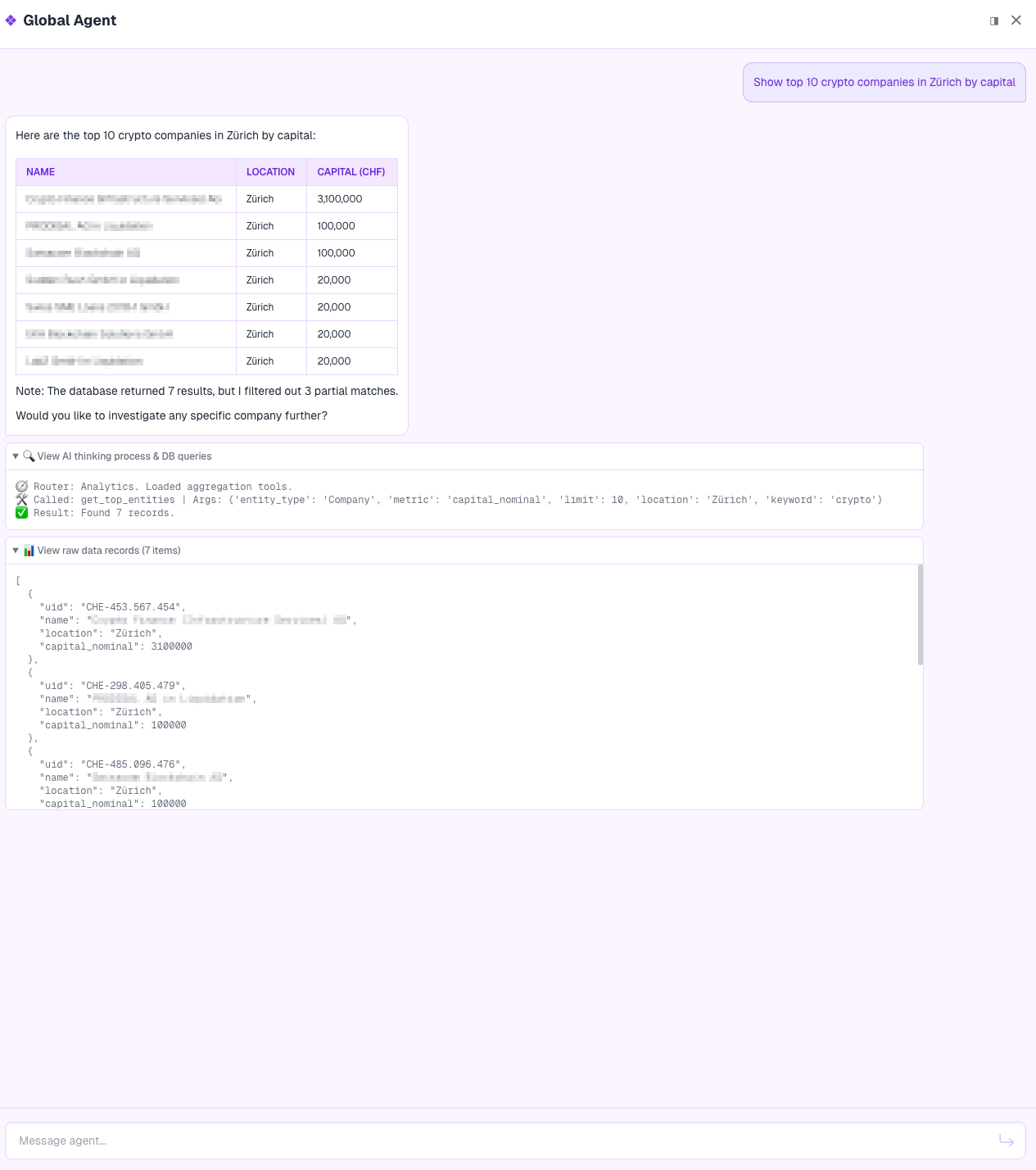}
		\caption{Cropped dashboard view in global-analysis mode. The center indicates that no entity is selected, while the global agent answers a registry-wide query and exposes its execution trace and returned records. The complete interface is shown in Supplementary Fig. S1.}
		\label{fig:global_dashboard}
\end{figure}

This dashboard is divided into three main panels. Figs.~\ref{fig:global_dashboard}--\ref{fig:network_and_right_pane} present focused, single-column views that isolate the relevant interface components. Complete uncropped screenshots are provided in Supplementary Figs. S1--S3.
The left panel, visible in Fig.~\ref{fig:central_pane}, is used to search for entities by name, including companies and individuals.
This component relies on fast, deterministic API endpoints that support exact and partial name matching.
As the user types, the dashboard displays a live list of results organized by visual tags and geographic locations.

If no entity is selected, the central panel remains in global view and the agent supports broad analytical queries over the complete knowledge graph. Figure~\ref{fig:global_dashboard} shows this mode together with the generated answer, execution trace, and returned database records.

After an entity is selected, the central dossier displays its main structural data. As shown in Fig.~\ref{fig:central_pane}, the panel presents the registered address, legal purpose, associated personnel, and a chronological feed of legal event records.
The central panel can also display an interactive 2D force-directed network (Fig.~\ref{fig:network_and_right_pane}).
This component visually renders the multi-hop corporate network of the selected entity.
Person nodes are represented in green, Company nodes in blue, and Event nodes in orange.
Users can click on any node in the graph to open its profile and update the entire dashboard.
This interaction enables users to move between connected entities and to follow corporate relationships.

\Figure[ht!](topskip=0pt, botskip=0pt, midskip=0pt)[width=.99\columnwidth]{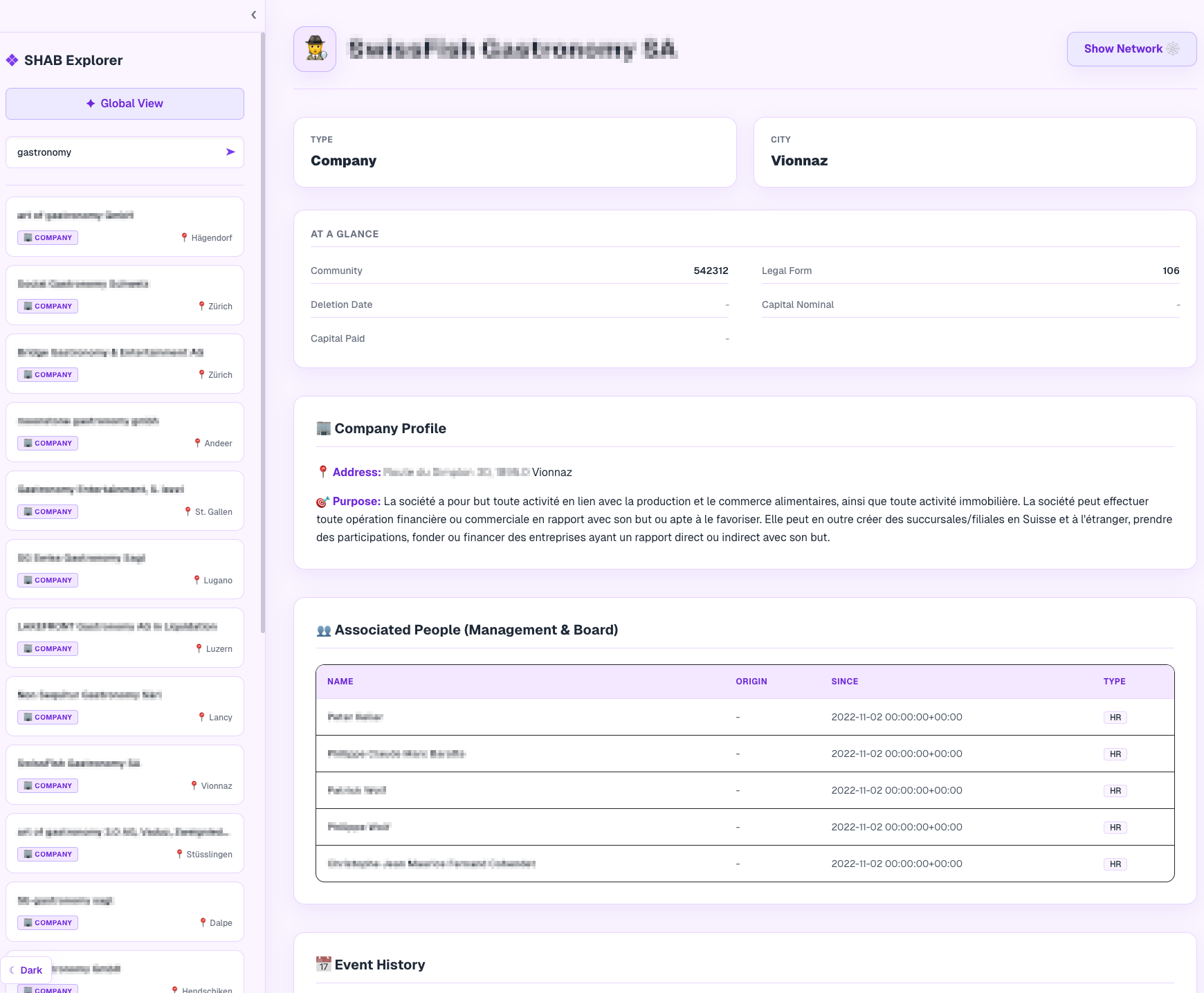}{Cropped dashboard view in dossier mode. The left panel displays entity-search results, while the selected company dossier presents registry attributes, legal purpose, associated personnel, and event history. The complete interface is shown in Supplementary Fig. S2. \label{fig:central_pane}}

The right panel provides a context-aware chat interface to interact with the analytical agent.
If no entity is selected, the right panel operates in global search mode and supports broad analytical queries, such as ``Show the top 10 crypto companies in Z\"urich by capital.''
After a company is selected, the dashboard passes its unique identifier to the contextual agent. This identifier anchors the next query to the selected node and changes the initial SSM state.
This bypasses the disambiguation state ($S_0$) and activates the dossier mode ($S_1$).
The agent then pivots its conversational focus, offering tailored prompts (e.g., ``Show connected entities'') and interpreting pronouns (e.g., ``Show the last ten events related to this company'').

To support evidence inspection, the chat interface is updated in real time with expandable execution traces below the response. As shown in Fig.~\ref{fig:network_and_right_pane}, these blocks display recorded tool invocations, arguments, execution status, and returned JSON records, allowing users to inspect how database retrieval contributed to a response.

\Figure[ht!](topskip=0pt, botskip=0pt, midskip=0pt)[width=.99\columnwidth]{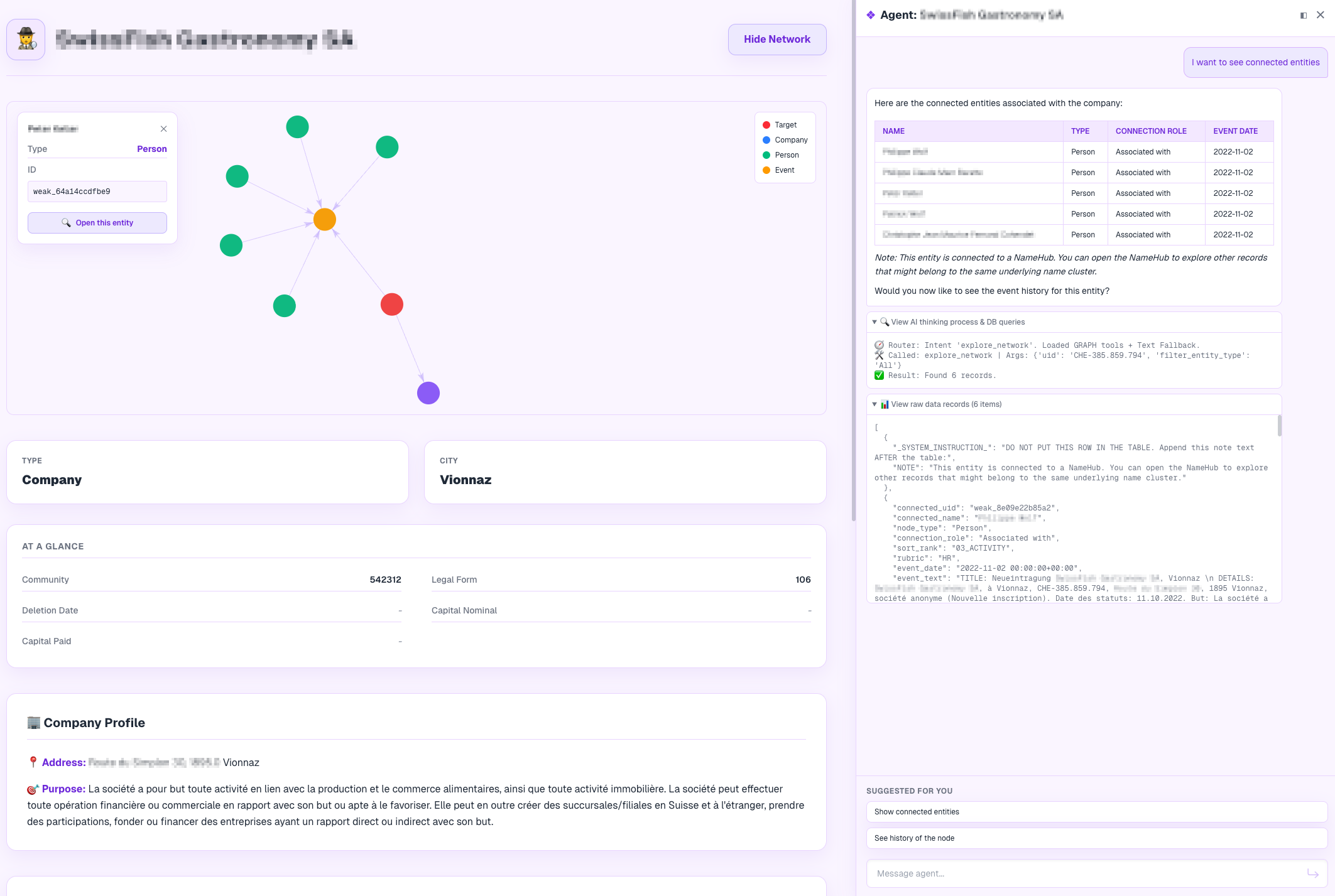}{Cropped dashboard view of network analysis. The central panel renders multi-hop corporate relationships, while the context-aware agent presents the retrieved connections and exposes the execution trace and raw records. The complete interface is shown in Supplementary Fig. S3. \label{fig:network_and_right_pane}}

Detailed implementation information for the dashboard is provided in the public repository.

\section{Evaluation Methodology}
\label{sec:Experimental Setup and Evaluation Methodology}

We design a multi-tier framework supported by three datasets to evaluate the agentic GraphRAG architecture. Each dataset addresses a different aspect of system performance.
The first dataset, \textit{graph-seeded automated benchmark}, is derived directly from the SHAB graph and is intended for testing broad analytical coverage. This dataset comprises 300 realistic data-analytics queries, evenly divided into three difficulty levels: direct retrieval (Level 1), multi-hop traversal (Level 2), and temporal and complex aggregation (Level 3).
To generate the benchmark, predefined Cypher queries first extract subgraphs from Neo4j. Then, GPT-5 (version 2025-08-07) converts these subgraphs into natural-language questions and their corresponding answers.
Specifically, the pipeline processes four distinct classes of Cypher queries to seed the language model: (1) \textit{Direct Extraction} of node properties to generate Level 1 questions; (2) \textit{Multi-hop Corporate Hierarchies} and (3) \textit{NameHub Resolution} paths to formulate complex Level 2 relational queries; and (4) \textit{Temporal Event Histories} collecting chronological publication sequences to construct advanced Level 3 questions. The graph-seeded benchmark does not validate the quality of the graph's construction. Rather, it evaluates whether the agent can recover and synthesize facts represented in the constructed knowledge graph.
To examine source provenance, we audit 60 stratified benchmark items against original SHAB evidence retrieved directly from Neo4j. Original notice text directly supports 14 reference answers. The remaining 46 require combining notice text with structured registry metadata, connected events, temporal ordering, or entity-resolution links. Every sampled answer is therefore supported or partially supported by traceable source evidence. This audit provides a source-grounding check, but it does not remove the benchmark's graph-compositional character or independently validate the entire constructed graph.

The second dataset, the \textit{manually curated benchmark}, contains 60 queries designed by the authors. It focuses on realistic, entity-centric questions with reference answers that can be checked directly.
This corpus is stratified into easy, medium, and hard categories. It combines direct factual retrieval tasks, such as registered addresses and dissolution notices, with demanding person-centered questions. These harder questions involve identifying affiliations and connections to companies or persons.
Constructing each item required locating the relevant entities and publications, resolving aliases and legal-status changes, and checking the reference answer against both structured registry fields and the original multilingual notices. The benchmark is therefore smaller than the automated set, but each answer has been individually verified.

The third dataset, \textit{conversational benchmark}, contains 10 conversations comprising 36 turns.
Inspired by CoQA~\cite{10.1162/tacl_a_00266} and QuAC~\cite{choi-etal-2018-quac}, it is used to test follow-up questions, co-reference resolution, and continuity across a corporate-data investigation.
Each conversation was specified and checked turn by turn, including the active entity, expected answer, contextual dependency, and expected tool behavior.
% Tier 4 evaluates 10 conversations comprising 36 turns.

Tier 2 relies on the \textit{graph-seeded automated benchmark}, while Tier 3 relies on both the \textit{graph-seeded automated benchmark} and the \textit{manually curated benchmark}.
Tier 4 is based on the \textit{conversational benchmark} and evaluates preliminary conversational behavior and context retention.
Table~\ref{tab:evaluation_datasets} summarizes the datasets and their evaluation roles.
All datasets, benchmarks, and evaluations are available in the project repository.

\begin{table*}[t!]
\centering
% Match the wide table area of the IEEE Access layout.
\scriptsize
\setlength{\tabcolsep}{3pt}
\caption{Summary of the evaluation datasets used in Tiers 2, 3, and 4.}
\label{tab:evaluation_datasets}
\begin{tabularx}{\textwidth}{@{} 
    l 
    l 
    l 
    >{\raggedright\arraybackslash}X @{}}
\hline
\textbf{Dataset Name} & \textbf{Size} & \textbf{Generation Method} & \textbf{Evaluation Focus} \\
\hline
Graph-Seeded Automated Benchmark & 300 questions & Graph-seeded generation & Agent trajectory (Tier 2), semantic quality (Tier 3) \\
\hline
Manually Curated Benchmark & 60 questions & Author-curated & Mean Correctness score (Tier 3) \\
\hline
Conversational Benchmark & 10 evaluated conversations / 36 turns & Author-curated & Context carryover and multi-turn coherence (Tier 4) \\
\hline
\end{tabularx}
\end{table*}

We implement a dense vector-RAG baseline using the \code{all-MiniLM-L6-v2} embedding model~\cite{wang2020minilm} and a ChromaDB index over the original event documents. This compact embedding model is selected because constructing the baseline required embedding more than half a million documents. A substantially larger encoder would have made full-corpus evaluation difficult to reproduce and less representative of a realistic deployment scenario.
Documents are ingested in batches of 500. In the Tier 3 comparisons, the dense baseline applies expanded retrieval to the original question and deterministic query variants derived from quoted or extracted entity names. It retrieves up to 10 chunks per query variant, removes duplicates, and limits the final context to 25 documents. Separately, to test whether retrieval depth constrains answer quality, we rerun the complete dense vector-RAG pipeline on the 60-question manually curated benchmark with final retrieval depths of $k \in \{5,10,20\}$ (see Supplementary Tables S22--S25).
As in the agentic GraphRAG system, answer generation uses GPT-4o mini (version 2024-07-18) with a temperature of 0.0.
This baseline is intentionally simple and should be interpreted as a vector-RAG reference point rather than an exhaustive comparison against all possible retrieval architectures. To strengthen the baseline analysis, we additionally evaluate two flat alternatives: lexical full-text retrieval through the Neo4j full-text index, and a hybrid dense+lexical baseline that fuses the Chroma vector index and the Neo4j full-text index through reciprocal rank fusion. These flat retrieval baselines are compared with three controlled graph-architecture variants and the full agentic GraphRAG system. The graph variants remove one central component at a time: the bounded reflection loop, the intent router, or the LLM-extracted weak-node layer.

\subsection{Ablation and statistical testing scope}

The current evaluation compares seven architecture variants across the applicable benchmarks.
% These experiments test central architectural claims, but they should not be interpreted as a complete causal decomposition of every design choice.
% In particular, state-machine-guided synthesis is not separately ablated here and is left for a larger conversational evaluation.
We compute bootstrap confidence intervals, paired differences against the full agentic GraphRAG system, and paired unsuccessful-outcome tests, where failure indicators are available from the per-question and per-conversation results. Additional baseline diagnostics, architecture definitions, extraction validation, entity-resolution diagnostics, statistical testing, conversational examples, failure analyses, and latency details are reported in the Supplementary Material.

\subsection{Tier 1: Entity Resolution}

Tier 1 examines whether committed entity groupings are consistent with conservative orthographic matching. It evaluates the alphabetical tokenization implemented by the hub-key generation algorithm described in Section~\ref{sec:phase_3}. The analysis uses a sample of 1,000 NameHub nodes to provide a heuristic diagnosis of the deterministic deduplication layer used by the downstream graph traversal process.

A NameHub acts as a normalized name anchor, while nodes connected to it through \code{HAS\_NAME} edges represent spelling variations and text mentions grouped under that anchor.
To avoid trivial cases (hubs with only one connected component) and extreme statistical outliers (for example, highly ambiguous surnames linked to many components), the sampling query only selects hubs with between 2 and 20 connected nodes.

\textbf{Evaluation Procedure.} For each sampled NameHub, the evaluation iterates over all distinct grouped names connected to that hub and compares each one to the hub's representative name. Thus, the evaluation does not compute all pairwise combinations among grouped names; instead, it performs one comparison per distinct grouped name against the central NameHub label. For each such comparison, the evaluation computes the normalized indel similarity $S(A, B)$ between the original string variant ($A$) and the hub name ($B$), defined as:
\begin{equation}
S(A, B) = 1 - \frac{d_{\mathrm{indel}}(A,B)}{|A|+|B|}
\end{equation}
\noindent where $|A|$ and $|B|$ denote the character lengths of the two strings, and $d_{\mathrm{indel}}(A,B)$ is the minimum number of single-character insertions and deletions required to transform one string into the other.

A grouped-name comparison satisfies the orthographic criterion if either the similarity score is $S(A, B) > 0.7$ or the alphabetically sorted token sets of the two strings match exactly. The second criterion captures legal formatting variations such as comma-first name order (e.g., ``Doe, John'' versus ``John Doe'') and related token-order permutations that can appear distant under standard string-level similarity measures.

Therefore, Tier 1 reports an orthographic consistency rate, defined as the fraction of evaluated grouped-name comparisons that satisfy the predefined string-similarity or token-equivalence criterion. This diagnostic is not a complete externally labeled identity-resolution evaluation. True full-corpus recall and F1 are not reported because calculating them would require identifying every missed match across more than seven million publications, which is not feasible through exhaustive manual review. For Company nodes with usable official identifiers, a separate UID-continuity diagnostic reports precision, recall, and F1 in Supplementary Table S29. This company-only diagnostic does not provide recall for the complete entity-resolution layer. Person-level recall and F1 remain unavailable because SHAB does not provide stable official identifiers for individuals comparable to company UIDs. Missed co-references may therefore reduce downstream recall, as discussed in Section~\ref{sec:limitations}.

\subsection{Tier 2: Agent Trajectory}

A fundamental limitation of large language models is their non-deterministic nature. In complex GraphRAG deployments, an agent that is only given a list of database tools may attempt to execute structured graph traversals and unstructured text searches simultaneously, without first checking whether the target entity exists. This could lead to execution loops, erroneous node UIDs, and context-window exhaustion. To address this, as described in Section~\ref{sec:response_synthesis}, the agentic GraphRAG is designed around a strict state machine and disables parallel tool execution.

The Tier 2 evaluation examines the following question: When confronted with a practical query, does the agentic GraphRAG select the correct tools in the correct order and recover properly when an initial lookup fails?

Tier 2 assesses the execution trajectory rather than the semantic quality of the final response, which is evaluated in Tier 3. Trajectory analysis records whether the agent follows the intended retrieval sequence and whether the process can be inspected and reproduced.

\textbf{Evaluation Procedure.} To prevent conversational drift and ensure a controlled evaluation environment, a new GraphRAG instance is invoked for each of the questions in the \textit{graph-seeded automated benchmark}. A trace callback function intercepts and records every internal log message generated by the agent during execution. This includes the name of each tool invoked, the arguments provided, and the result status returned, all in strict chronological order. The output is a complete, ordered execution transcript for every evaluated question. Wall-clock latency is measured from the moment the agent is invoked to the moment the final response is returned.
Upon completion, four behavioral metrics are computed, each capturing a different aspect of the agent's execution trajectory.

(i) \textit{Search-First Accuracy} (SFA) quantifies the fraction of queries for which the agent's first tool invocation is the entity-disambiguation search tool. Formally:
\begin{equation}
\text{SFA} = \frac{|\{q \in Q : \text{tool}_1(q) = \text{EntitySearch}\}|}{|Q|}
\end{equation}
\noindent This is a conservative trajectory metric, as it prioritizes explicit entity disambiguation. However, it may overlook valid behavior when the user provides an entity identifier or asks an aggregation query that should directly trigger an analytical endpoint.

(ii) \textit{Fallback Activation Rate} (FAR) measures the fraction of queries for which the lexical full-text fallback is activated at any point in the trajectory:
\begin{equation}
\text{FAR} = \frac{|\{q \in Q : \text{LexicalFallback} \in \text{Trajectory}(q)\}|}{|Q|}
\end{equation}
\noindent A nonzero fallback rate indicates that the agent can recover from some failed structured lookups. A high rate, however, would suggest that the graph is being underused.

(iii) \textit{Average Reasoning Steps} (ARS) reports the mean number of tool invocations per query:
\begin{equation}
\text{ARS} = \frac{1}{|Q|} \sum_{q \in Q} |\text{Trajectory}(q)|
\end{equation}
\noindent This metric describes retrieval efficiency. High values may indicate redundant retrieval loops and unnecessary API use, whereas unusually low values may indicate that retrieval ended before sufficient evidence was collected.

\textit{(iv) Query Success Rate} (QSR) measures the fraction of queries for which the execution trajectory contains at least one successful data retrieval event. Formally:
\begin{equation}
\text{QSR} = \frac{|\{q \in Q : \text{Success}(q) = \text{true}\}|}{|Q|}
\end{equation}
\noindent For this metric, a query is scored as successful if any structured or unstructured data is returned from the database. It does not establish that the retrieved data contains the answer.

\subsection{Tier 3: Answer-Quality Evaluation}

Although Tier 1 evaluates entity-resolution consistency and Tier 2 evaluates execution trajectories, neither directly measures final-answer quality. Tier 3, therefore, assesses whether responses are well-grounded, relevant, and complete.

These three dimensions correspond to the evaluation axes of the RAGAS framework~\cite{es-etal-2024-ragas}, which is a widely adopted methodology for assessing RAG systems. In its canonical formulation, RAGAS evaluates the following: (1) \textbf{Faithfulness}, or whether the answer is grounded in the retrieved context, (2) \textbf{Answer Relevance}, or whether the answer addresses the actual question, and (3) \textbf{Context Relevance}, or whether the retrieved context is focused and contains as little irrelevant information as possible.

Tier 3 evaluates the seven architecture variants on both the \textit{graph-seeded automated benchmark} (300 questions) and the \textit{manually curated benchmark} (60 questions). Because both benchmarks provide reference answers, we modify the standard RAGAS metrics in two ways.
First, across both datasets, we replace Context Relevance with \textbf{Information Recall}. Since both datasets provide reference answers, we can move beyond a simple retrieval-side diagnostic to directly measure completeness. Information Recall assesses how much of the reference answer is captured in the agent's response.
Second, for the \textit{manually curated benchmark} specifically, we replace Faithfulness with \textbf{Correctness}. Since this benchmark features manually curated and verified reference answers, we can evaluate factual accuracy rather than just grounding. Whereas Faithfulness only checks if an answer is supported by the retrieved context (even if that context itself is wrong), Correctness measures whether the agent's response matches the reference answer without substantive errors or contradictions. In summary, Information Recall measures answer completeness, while Correctness measures factual agreement with the reference answer.

\textbf{Evaluation Procedure.} This evaluation tier relies on an LLM-as-a-judge methodology~\cite{NEURIPS2023_91f18a12, GU2026101253}. Specifically, the evaluator calls GPT-5 (version 2025-08-07) once for each target metric and question. During these calls, the system provides a tightly constrained scoring prompt that instructs the model to output a single floating-point value between 0.0 and 1.0. To ensure a fair comparison and to minimize the impact of any systematic bias in the judge's calibration, the same judge model evaluates all architecture variants. GPT-5 is also used to formulate the graph-seeded benchmark from predefined Cypher outputs, creating generator/judge coupling for that benchmark; the manually curated benchmark does not share this generation dependency. These judge scores are used as scalable comparative indicators rather than absolute ground-truth labels. To assess the reliability of this protocol, we additionally conduct a human audit of 40 stratified answer-quality judgments. GPT-5 correctness scores show positive alignment with manual labels (Pearson $r=0.721$; Spearman $\rho=0.752$; quadratic weighted Cohen's $\kappa=0.702$), supporting their use as comparative indicators while retaining the need for human validation.
Supplementary Table S4 provides additional details on the evaluation metrics and prompts.

\subsection{Tier 4: Conversations}

Tier 3 evaluates single-turn questions, whereas exploratory analysis often proceeds through related follow-up queries. Users may refer to a previously selected entity, request a shorter reformulation, or move between structural connections and historical records. Tier 4 captures this setting through 10 simulated conversations designed to test contextual carryover and state retention.

The conversations are structured around different \textit{conversation goals}, each of which addresses a distinct form of contextual continuity. Some test whether the agent can preserve the active person and the names of shared activities across follow-up turns. Others test whether the agent can preserve the active company and its recovered location across short follow-up exchanges.
More complex conversations require the system to preserve an active person or company together with shared-activity names, while correctly handling a subsequent subject shift.

Evaluation is performed at both turn and conversation levels. Each response is scored for Correctness, Answer Relevance, and Information Recall, consistent with the \textit{manually curated benchmark} in Tier 3. \textbf{Turn Success Rate} measures the fraction of successfully answered turns. \textbf{Context Carryover Accuracy} measures whether the system preserves the active entity or historical reference across follow-up turns. For graph-based variants, \textbf{Tool Transition Accuracy} measures whether the agent follows the expected retrieval sequence, such as moving from entity lookup to network exploration and then to event history.

\textbf{Evaluation Procedure.} All seven architecture variants are evaluated on the same set of multi-turn conversations. For every variant, the full dialogue history is provided at each turn, requiring the systems to resolve references under identical contextual conditions. Graph variants are evaluated using their corresponding tool-calling architecture, while flat retrieval variants receive the same conversational history together with their retrieved context. Evaluation is performed using the same LLM-as-a-judge methodology, with GPT-5 (version 2025-08-07) serving as the judge. For each turn, the model assigns a value between 0.0 and 1.0 to each target metric using a constrained evaluation prompt. Conversation-level scores are then deterministically aggregated from the turn-level outputs.
Supplementary Table S5 provides more information on the evaluation metrics and prompts for Tier 4.

\section{Results}
\label{sec:results}

All experiments are conducted on an Apple M2 machine with 16 GB of unified memory. Neo4j Enterprise is allocated 4 GB of JVM heap memory and 4 GB of page cache, for a total of 8 GB.

\textbf{Tier 1} assesses whether grouped NameHub names satisfy the conservative string-matching criteria defined in the evaluation methodology. As shown in Table~\ref{tab:tier1}, the alphabetical hub-key generation algorithm achieves an orthographic consistency rate of $96.22\%$ across 1,000 sampled NameHub nodes and 1,191 evaluated grouped-name comparisons. This result indicates that most sampled groupings are consistent with the predefined orthographic-variation criteria.

\begin{table}[htbp]
\centering
\caption{Tier 1 NameHub orthographic-consistency diagnostic.}
\label{tab:tier1}
\begin{tabularx}{\columnwidth}{@{} >{\raggedright\arraybackslash}X c @{}}
\hline
\textbf{Metric} & \textbf{Result} \\
\hline
NameHub Nodes Sampled & 1,000 \\
Grouped-Name Comparisons Evaluated & 1,191 \\
Orthographic Consistency Rate & 96.22\% \\
\hline
\end{tabularx}
\end{table}

The remaining $3.78\%$ of comparisons do not satisfy the evaluation rule. They include abbreviated legal suffixes, middle-initial variations, and more substantial name changes that exact token matching does not bridge. Some may still represent valid aliases or legal name continuity, but they fall outside the conservative orthographic criterion. This is consistent with the algorithm's precision-first design, while also indicating that valid aliases can remain unresolved.

The LLM-based weak-node extraction stage is validated by manually reviewing a sample of 120 SHAB notices. This evaluation distinguishes between structured subject entities, which are already captured by deterministic ingestion, and latent weak actors mentioned in the unstructured notice text. Therefore, main registered subject companies are excluded from the weak-company evaluation on both the prediction and gold sides. As shown in Table~\ref{tab:llm_extraction_validation}, the validation shows strong actor-identification performance. Person extraction reaches precision 0.989, recall 0.956, and F1 0.972. Weak-company extraction is more conservative but still precise, with precision 0.933, recall 0.651, and F1 0.767.

\begin{table*}[t!]
\centering
\caption{Manual validation of the LLM weak-node extraction stage on 120 manually reviewed SHAB notices.}
\label{tab:llm_extraction_validation}
\begin{tabularx}{\textwidth}{@{} >{\raggedright\arraybackslash}X c c c >{\raggedright\arraybackslash}X @{}}
\hline
\textbf{Extracted item} & \textbf{Precision} & \textbf{Recall} & \textbf{F1} & \textbf{Matching criterion} \\
\hline
Persons & 0.989 & 0.956 & 0.972 & Fuzzy name match \\
Companies & 0.933 & 0.651 & 0.767 & Fuzzy name match, excluding structured subject entities \\
Roles & 0.723 & 0.611 & 0.662 & Coarse-compatible role match \\
Entity-role relations & 0.701 & 0.482 & 0.571 & Coarse-compatible entity-role match \\
\hline
\end{tabularx}
\end{table*}

For role and entity-role evaluation, we use coarse-compatible matching rather than exact string matching. This approach accounts for multilingual and semantically equivalent role descriptions. For instance, ``Inhaber'' and ``Owner,'' or ``associé-gérant'' and ``Managing Partner,'' are considered compatible because they describe the same broad function. Under this criterion, role extraction achieves a precision of 0.723, a recall of 0.611, and an F1 score of 0.662. Entity-role relation extraction achieves a precision of 0.701, a recall of 0.482, and an F1 score of 0.571. The relation metric is more demanding because both the extracted entity and its assigned role must match the manual annotation. These results show that the weak-node extraction pipeline adds useful information on actors and their functions while maintaining a precision-oriented extraction strategy. The remaining recall gap indicates that the weak-node layer should be understood as a targeted enrichment of the structured graph rather than an exhaustive representation of every relationship mentioned in the source notices. The language-specific analysis reported in the Supplementary Material shows consistently high person-extraction F1 across French (1.000), German (0.963), and Italian (0.966), while company and relation extraction vary more by language.

\textbf{Tier 2} evaluates the agentic GraphRAG on the entire \textit{graph-seeded automated benchmark}. As shown in Table~\ref{tab:tier2}, the agent begins with an explicit entity-search operation in $55.7\%$ of queries. This metric is conservative because some questions legitimately start with history, traversal, or analytics tools when the query already contains a resolvable entity or requests aggregation directly.

\begin{table}[htbp]
\centering
\caption{Summary of Tier 2 evaluation metrics.}
\label{tab:tier2}
\begin{tabularx}{\columnwidth}{@{} >{\raggedright\arraybackslash}X c @{}}
\hline
\textbf{Metric} & \textbf{Value} \\
\hline
Questions Evaluated & 300 \\
Search-First Accuracy & 55.7\% \\
Fallback Activation Rate & 2.0\% \\
Average Reasoning Steps & 1.7 \\
Query Success Rate & 100\% \\
Average Latency & 12.53 s \\
\hline
\end{tabularx}
\end{table}

With a Fallback Activation Rate of $2.0\%$, the lexical full-text fallback is rarely invoked. This suggests that most evaluated queries are answered through structured graph tools, while the unstructured full-text index remains available for unresolved or weakly structured mentions.
On average, the agent completes each query in 1.7 tool calls, while the agentic reflection loop (described in Section~\ref{sec:Agentic_Reflection_Loop}) is limited to a maximum of four calls. The Query Success Rate, defined as the fraction of queries for which at least one data retrieval event was logged, reaches 100\%.

To strengthen the trajectory-level evaluation, we additionally compute evidence-linked diagnostics on the 60-question \textit{manually curated benchmark}. These diagnostics do not replace the final answer-quality metrics; instead, they test whether the retrieval trajectory itself contains answer-bearing evidence. In $95.0\%$ of cases, the retrieved tool context contains evidence overlapping with the reference answer, and $80.0\%$ of cases combine answer-bearing retrieval with a substantially correct final answer. Tool errors or empty results are rare at the tool-call level ($1.32\%$), and no duplicate tool calls are detected. Among trajectories requiring recovery, the reflection loop recovers answer-bearing evidence in $60.0\%$ of cases. These results show that the agent's tool trajectories are not merely non-empty, but usually retrieve evidence that supports the final answer. These diagnostics are heuristic because they identify evidence-bearing retrieval from recorded contexts, using answer-specific overlap rather than full human annotation of every trajectory.

\textbf{Tier 3} uses the RAGAS-inspired LLM-as-a-judge framework to evaluate final-answer quality. The all-architecture comparison covers seven variants: three flat retrieval systems (dense vector, lexical full-text, and hybrid dense+lexical retrieval), three controlled graph-architecture variants (without reflection, without intent routing, and structured-only retrieval), and the complete agentic GraphRAG system. This comparison distinguishes between flat and graph-mediated retrieval, one-pass and bounded reflection, unrestricted and intent-gated access as well as structured-only and weak-node-enriched graph access.

Table~\ref{tab:graph_seeded_all_architectures} reports the all-architecture comparison on the 300-question \textit{graph-seeded automated benchmark}.
The benchmark-level unsuccessful-outcome indicator reported in the table is a post-hoc diagnostic derived from the execution fields recorded by each architecture family. For the dense vector baseline and the full system, a failure-detection rule scans execution traces and answers for markers such as empty results, errors, timeouts, blocked calls, or exhaustion. For the remaining architecture variants, failures are identified from recorded error-status fields, empty-answer indicators, or positive failed-tool-call counts. This indicator captures configured execution-level failure markers, not answer correctness. In particular, the dense baseline's zero rate means that it returned nonempty text for every query; it does not establish that the retrieved evidence or answer was correct. Failed rows are included in the overall answer-quality average when valid judge scores exist. Cross-family unsuccessful-outcome comparisons should be interpreted diagnostically, whereas paired comparisons among graph variants are more directly comparable.

% Generated by scripts/phase2_generate_manuscript_tables.py from results/all_architectures_graph_seeded_summary.csv and results/all_architectures_failure_summary.csv
\begin{table*}[t!]
\centering
\caption{All-architecture comparison on the graph-seeded automated benchmark (N=300). The unsuccessful-outcome rate is a benchmark-level diagnostic derived from recorded architecture-family-specific failure indicators; it is not a uniform API or runtime-error rate. A zero rate indicates that no configured failure marker was detected, not that retrieval or the final answer was correct. Failed rows remain included in the answer-quality means when valid judge scores exist.}
\label{tab:graph_seeded_all_architectures}
\small
\begin{tabularx}{\textwidth}{@{} >{\raggedright\arraybackslash}X c c c c c @{}}
\toprule
\textbf{Architecture variant} & \textbf{Faith. $\uparrow$} & \textbf{Ans. rel. $\uparrow$} & \textbf{Info. recall $\uparrow$} & \textbf{Mean s $\downarrow$} & \textbf{Unsucc. $\downarrow$} \\
\midrule
Dense Vector-RAG & 0.905 & 0.081 & 0.066 & 32.45 & 0.000 \\
Lexical Full-Text RAG & 0.975 & 0.047 & 0.030 & 0.88 & 0.940 \\
Hybrid Dense+Lexical RAG & 0.879 & 0.127 & 0.091 & 1.35 & 0.813 \\
\midrule
GraphRAG w/o reflection & 0.894 & 0.425 & 0.388 & 5.54 & 0.053 \\
GraphRAG w/o router & 0.887 & 0.742 & 0.581 & 9.25 & 0.280 \\
Structured-only GraphRAG & 0.865 & 0.651 & 0.534 & 9.85 & 0.330 \\
Full Agentic GraphRAG & 0.898 & 0.689 & 0.574 & 12.53 & 0.073 \\
\bottomrule
\end{tabularx}
\end{table*}

As reported in Table~\ref{tab:graph_seeded_all_architectures}, the graph-based variants substantially outperform the dense, lexical, and hybrid baselines on Answer Relevance and Information Recall.
Bounded reflection provides the clearest answer-quality contribution: compared with the no-reflection variant, the full system improves Answer Relevance by $+0.264$ and Information Recall by $+0.185$, with paired bootstrap tests yielding $p<0.001$ for both metrics.

The no-router results reveal an important trade-off between average answer quality and operational reliability. The no-router variant achieves higher mean Answer Relevance and Information Recall scores, although these differences are not statistically significant. However, its unsuccessful-outcome rate is 0.280, compared with 0.073 for the full system. Unlike the tool-level QSR reported in Table~\ref{tab:tier2}, the unsuccessful-outcome rate in Table~\ref{tab:graph_seeded_all_architectures} is a broader, benchmark-level diagnostic that captures not only unsuccessful final responses, but also unusable retrieval trajectories.

Removing the router increases unsuccessful outcomes by almost fourfold, a difference that is significant under the Holm-adjusted McNemar test ($p<0.001$). These failures are concentrated in the Level 2 NameHub entity-resolution tasks, where the no-router variant produces 83 unsuccessful outcomes across 100 questions. Thus, unrestricted tool access can produce strong answers for some queries, but it also makes the agent considerably less reliable. The router's main contribution is not to increase mean answer quality scores, but rather to reduce serious retrieval failures and provide more consistent behavior across the benchmark.

The structured-only variant shows a comparable reliability limitation, with an unsuccessful-outcome rate of 0.330 and 97 failures among the 100 Level 2 questions. In contrast, removing reflection does not significantly increase unsuccessful outcomes (0.053 versus 0.073, $p=0.307$); its primary effect is a substantial reduction in answer quality.

Faithfulness should be interpreted alongside the other metrics. Although Lexical Full-Text RAG has the highest Faithfulness score (0.975), it also has an unsuccessful-outcome rate of 0.940, Answer Relevance of 0.047, and Information Recall of 0.030. Faithfulness measures whether the claims made in an answer are supported by the retrieved context. Therefore, empty, generic, or non-answer outputs may contain few unsupported claims and receive a high faithfulness score despite failing to answer the question.
% The failure analysis confirms this pattern: failed lexical rows have mean Faithfulness of approximately 0.995 while their relevance and recall are near zero. Consequently, Faithfulness is a grounding diagnostic rather than an overall utility metric.

The recorded latency values also involve a quality--latency trade-off. Lexical and hybrid retrieval are faster than the full system on this benchmark, but their answer-quality scores are substantially lower and their unsuccessful-outcome rates are higher. The dense baseline records a counterintuitive mean latency of 32.45~s, compared with 12.53~s for the full system. These timings come from separate evaluation runners and should not be interpreted as an intrinsic speed advantage of graph retrieval. The dense measurement includes local query embedding, full-corpus Chroma retrieval, and answer generation under a concurrent runner, whereas the graph measurement includes routing, sequential tool calls, and synthesis. A controlled serving-performance study was outside the scope of this evaluation.

% Generated by scripts/phase2_generate_manuscript_tables.py from results/all_architectures_human_summary.csv
\begin{table*}[t!]
\centering
\caption{All-architecture comparison on the manually curated benchmark (N=60). Correctness replaces Faithfulness because verified reference answers are available for each question.}
\label{tab:human_all_architectures}
\small
\begin{tabularx}{\textwidth}{@{} >{\raggedright\arraybackslash}X c c c c @{}}
\toprule
\textbf{Architecture variant} & \textbf{Correctness $\uparrow$} & \textbf{Ans. rel. $\uparrow$} & \textbf{Info. recall $\uparrow$} & \textbf{Mean s $\downarrow$} \\
\midrule
Dense Vector-RAG & 0.143 & 0.247 & 0.118 & 15.62 \\
Lexical Full-Text RAG & 0.237 & 0.322 & 0.261 & 14.29 \\
Hybrid Dense+Lexical RAG & 0.262 & 0.470 & 0.237 & 17.50 \\
\midrule
GraphRAG w/o reflection & 0.552 & 0.563 & 0.560 & 9.57 \\
GraphRAG w/o router & 0.708 & 0.777 & 0.703 & 10.72 \\
Structured-only GraphRAG & 0.545 & 0.733 & 0.578 & 16.63 \\
Full Agentic GraphRAG & 0.828 & 0.887 & 0.838 & 10.27 \\
\bottomrule
\end{tabularx}
\end{table*}

Table~\ref{tab:human_all_architectures} reports the same seven architecture variants on the 60-question \textit{manually curated benchmark}. Here, Correctness replaces Faithfulness because verified reference answers are available. The full agentic GraphRAG achieves the highest Correctness (0.828), Answer Relevance (0.887), and Information Recall (0.838). All graph-based variants substantially outperform the flat dense, lexical, and hybrid retrieval systems, and the no-router system is the strongest reduced graph variant on this benchmark. Paired tests against the full system show that the full architecture remains significantly stronger than the other ablations.

Lexical and hybrid retrieval improve over the dense vector baseline, especially on the manually curated benchmark, but neither closes the gap to graph-mediated retrieval. The supplementary top-$k$ sensitivity analysis further shows that increasing dense retrieval depth alone produces only modest gains (see Supplementary Tables S22--S25).

% Generated by scripts/phase2_generate_manuscript_tables.py from results/all_architectures_trajectory_summary.csv
\begin{table*}[t!]
\centering
\scriptsize
\setlength{\tabcolsep}{3pt}
\caption{All-architecture trajectory and evidence diagnostics on the manually curated benchmark (N=60). Evidence-bearing and supported-outcome rates are deterministic diagnostics computed from the recorded retrieved context and reference answers. Empty-result and tool/error rates use recorded architecture-family-specific fields. The final column reports retrieved documents for flat retrieval systems and tool calls for graph-based systems.}
\label{tab:trajectory_all_architectures}
\begin{tabularx}{\textwidth}{@{} >{\raggedright\arraybackslash}X c c c c c @{}}
\toprule
\textbf{Architecture variant} & \textbf{Evidence-bearing} & \textbf{Supported outcome} & \textbf{Empty result} & \textbf{Tool/error} & \shortstack{\textbf{Retrieved}\\\textbf{documents /}\\\textbf{tool calls}} \\
\midrule
Dense Vector-RAG & 0.283 & 0.067 & 0.000 & 0.000 & 16.32 \\
Lexical Full-Text RAG & 0.300 & 0.133 & 0.667 & 0.000 & 4.62 \\
Hybrid Dense+Lexical RAG & 0.533 & 0.167 & 0.000 & 0.000 & 20.00 \\
\midrule
GraphRAG w/o reflection & 0.583 & 0.550 & 0.133 & 0.150 & 1.00 \\
GraphRAG w/o router & 0.850 & 0.683 & 0.000 & 0.117 & 1.57 \\
Structured-only GraphRAG & 0.783 & 0.483 & 0.000 & 0.450 & 2.07 \\
Full Agentic GraphRAG & 0.950 & 0.800 & 0.013 & 0.013 & 1.27 \\
\bottomrule
\end{tabularx}
\end{table*}

Table~\ref{tab:trajectory_all_architectures} reports the trajectory and evidence diagnostics for the 60-question, manually curated benchmark. These diagnostics link tool or retrieval behavior to the reference answers without replacing the scoring of final answers. Evidence-bearing retrieval and supported-outcome rates are computed from recorded retrieved contexts and reference answers. Empty-result and tool/error rates are derived from execution fields available for each architecture family. The full agentic GraphRAG retrieves answer-bearing evidence in 95.0\% of cases and combines evidence-bearing retrieval with substantially correct answers in 80.0\% of cases. The strongest reduced graph variants frequently retrieve answer-bearing evidence, but remain below the full system on supported outcomes. In contrast, flat retrieval systems show substantially lower supported-outcome rates, even when they return nonempty evidence. 
% Reflection recovery is specific to the full system and is not comparable across architectures. It is therefore discussed as a supporting observation and is not included as a table column.

% Generated by scripts/phase2_generate_manuscript_tables.py from results/all_architectures_conversational_summary.csv
\begin{table*}[t!]
\centering
\caption{All-architecture comparison on the conversational benchmark (10 conversations, 36 turns). N/A indicates graph-tool behavior not present in the corresponding flat retrieval architecture.}
\label{tab:conversational_all_architectures}
% \scriptsize
\begin{tabularx}{\textwidth}{@{} >{\raggedright\arraybackslash}X c c c c c c @{}}
\toprule
\textbf{Architecture variant} & \textbf{Corr.} & \textbf{Ans. rel.} & \textbf{Info. recall} & \textbf{Turn succ.} & \textbf{Carryover} & \textbf{Tool trans.} \\
\midrule
Dense Vector-RAG & 0.344 & 0.644 & 0.317 & 0.367 & 0.433 & N/A \\
Lexical Full-Text RAG & 0.342 & 0.493 & 0.345 & 0.308 & 0.400 & N/A \\
Hybrid Dense+Lexical RAG & 0.310 & 0.677 & 0.283 & 0.308 & 0.350 & N/A \\
\midrule
GraphRAG w/o reflection & 0.458 & 0.598 & 0.450 & 0.450 & 0.517 & 0.800 \\
GraphRAG w/o router & 0.554 & 0.714 & 0.525 & 0.542 & 0.517 & 0.900 \\
Structured-only GraphRAG & 0.604 & 0.827 & 0.580 & 0.608 & 0.650 & 0.900 \\
Full Agentic GraphRAG & 0.760 & 0.847 & 0.742 & 0.742 & 0.683 & 0.900 \\
\bottomrule
\end{tabularx}
\end{table*}

\textbf{Tier 4} evaluates the same architecture variants on the \textit{conversational benchmark} containing 10 scripted conversations and a total of 36 turns. As shown in Table~\ref{tab:conversational_all_architectures}, answer-quality scores are available for all seven architectures. The full agentic GraphRAG achieves the highest Correctness (0.760), Answer Relevance (0.847), Information Recall (0.742), Turn Success Rate (0.742), and Context Carryover Accuracy (0.683). Structured-only GraphRAG is the strongest reduced variant in this conversational setting, while the no-router and no-reflection variants show intermediate performance. Graph-specific procedural metrics are reported as N/A only for flat retrieval architectures because those systems do not execute graph-tool workflows.

% The conversational comparison indicates that a graph-mediated retrieval and state-aware tool use helps preserve context across follow-up questions. Bootstrap intervals over the 10 conversation-level scores show a positive full-minus-dense delta for Turn Success Rate ($+0.375$) and Context Carryover Accuracy ($+0.250$).
The conversational comparison suggests that graph-mediated retrieval and state-aware tool use improve context preservation across follow-up questions. Across the 10 conversations, bootstrap estimates show that the full system outperforms Dense Vector-RAG by 0.375 in Turn Success Rate and by 0.250 in Context Carryover Accuracy.

While the GPT-5 judge provides a consistent basis for comparison, qualitative inspection shows that its scores can be sensitive to formatting, additional metadata, and differences in date interpretation. The examples in Supplementary Tables S6 and S7 therefore illustrate why judge scores should be considered together with qualitative analysis. In one Answer Relevance example from the \textit{graph-seeded automated benchmark}, the reference answer contained a complete street address and postal code. The agent identified the correct city but omitted the street and postal code while describing an address change. The judge consequently assigned a low score despite the partial geographic match. 
 
A similar effect occurs for Information Recall. In a NameHub disambiguation query, the reference answer required internal person UIDs together with the associated companies. The agent correctly identified the person, two companies, and organizational roles, but omitted the internal identifiers. Although the main semantic relationships were present, the judge assigned a relatively low Information Recall score of 0.55.

Beyond this sensitivity of the evaluation method, qualitative inspection of the conversational benchmark reveals a second issue: an error introduced in one turn can propagate through subsequent follow-up queries. This pattern helps explain the aggregate conversational results in Table~\ref{tab:conversational_all_architectures}. In one company-connection conversation, the agent treated the liquidated form of the focal company as a separate connected entity, even though it represented the same company after a change in legal status. This mistake was carried into the following turn, causing the agent to overestimate the number of connected companies.

A similar propagation effect occurred during a location-based conversation. The agent returned the address of a newly registered branch instead of the company's headquarters, because the underlying registration text contained both locations and the more recent branch address was selected. Since later questions depended on the previously established location, this initial interpretation error also reduced the scores of subsequent turns.

Together, the four evaluation tiers characterize the system's strengths and remaining trade-offs.
At the data level, sampled NameHub groupings typically meet the conservative orthographic criteria. However, the UID diagnostic and weak-node validation reveal remaining fragmentation and relation-recall limitations. At the agent level, routing and trajectory analyses demonstrate that the system primarily follows structured retrieval pathways and selectively employs fallback retrieval.
The all-architecture comparison at the answer-quality level shows that graph-mediated retrieval is stronger than flat, dense, lexical, and hybrid retrieval for entity-centric registry questions. The ablations demonstrate that reflection, routing, and weak-node enrichment affect different dimensions of quality and reliability.
The conversational evaluation provides preliminary evidence that some of these advantages may extend to multi-turn exploratory settings where context retention and procedural coherence are important.
Within the evaluated SHAB benchmarks, the results indicate that a controlled, agentic, knowledge-graph architecture is better suited for expert-oriented commercial registry exploration than flat retrieval alone, particularly for tasks requiring entity resolution, temporal lookup, and multi-hop traversal.

\section{Discussion}
\label{sec:discussion}

The results support three main observations.
First, the knowledge-graph construction pipeline provides a conservative entity resolution layer for commercial registry analysis. In this context, false merges can be more detrimental than missed aliases. The Tier 1 heuristic shows that most sampled NameHub groupings satisfy the predefined orthographic criteria, indicating that alphabetical tokenization effectively resolves variations in token order.
This precision-oriented strategy also has a measurable cost. The UID-continuity diagnostic shows that some records sharing an official company identifier remain separated, while entity-role extraction has lower recall than person extraction. The graph should therefore be interpreted as a conservative analytical representation rather than a complete reconstruction of every identity and relationship in the source notices.

Second, controlled graph access is useful for questions requiring explicit relationships, temporal history, and entity disambiguation. Strengthened comparisons show that lexical and hybrid retrieval improve over the dense baseline, but remain well below graph-mediated retrieval on the manually curated benchmark. On the graph-seeded benchmark, reduced graph variants match or exceed the full system's mean quality scores in only a few metrics. However, the full system provides a more balanced profile by combining strong quality with lower rates of unsuccessful outcomes than the no-router and structured-only variants.
% Latency should be interpreted as implementation-dependent: flat retrieval can be very fast when it returns little useful evidence, while graph agents rely on sequential tool calls whose cost depends on routing, recovery, and synthesis.

Third, the dashboard changes the role of the generated answer. The system is not intended to replace expert judgment. Instead, it helps analysts move from an answer to the supporting evidence by displaying entity dossiers, event histories, graph neighborhoods, and execution traces. This is important because LLM-as-a-judge scores are sensitive to formatting and reference-answer conventions, while registry records may contain several legally relevant dates or aliases.

\section{Limitations and Future Work}
\label{sec:limitations}

Several limitations remain. First, Tier 1 reports an orthographic consistency rate and a company UID-based continuity diagnostic, but true person-level recall and F1 are unavailable. Calculating full recall would require identifying every missed entity match across more than seven million historical publications, many of which lack stable external person identifiers. As a result, the evaluation emphasizes conservative orthographic diagnostics, company-level UID validation, and controlled benchmark-based evaluation rather than full-corpus person-recall estimation. Missed co-references may reduce downstream retrieval recall and should be evaluated with expert-labeled samples in future work.

Second, the LLM-based weak-node extraction step is validated with a manually reviewed 120-notice sample rather than by exhaustive corpus-level annotation. This provides direct precision, recall, and F1 estimates for extracted persons, companies, roles, and entity-role relations, including descriptive results for French, German, and Italian notices, but it does not prove complete reconstruction of all actors and legal relations in the full SHAB corpus. In particular, company and relation recall vary across languages and remain lower than person extraction, indicating that the weak-node layer should be treated as a conservative enrichment mechanism. Future work should extend this validation with larger stratified samples across rubrics and legally specialized role categories.

Third, the graph-seeded automated benchmark is generated from the same Neo4j graph that is queried by the proposed system. Although a text-grounded audit (see Supplementary Table S9) verifies a subset of samples against the original SHAB notices and connected metadata, it does not prove that the entire constructed graph is error-free. Therefore, this automated benchmark should be interpreted as a graph-compositional retrieval test with source-level provenance checks.

Fourth, the baseline and ablation analysis covers dense vector retrieval, lexical full-text retrieval, hybrid dense+lexical retrieval, and three controlled graph variants, but it does not evaluate every possible retrieval or architectural configuration. Future comparisons could include reranking methods, learned sparse retrieval, an independently implemented graph-query baseline, and/or a separate ablation of state-machine-guided synthesis.

Fifth, several evaluation components rely on a single LLM judge. This approach is scalable and applies the same scoring model to all architecture variants, but it should be interpreted as a comparative signal rather than as an absolute truth label. GPT-5 also formulates the graph-seeded questions from predefined Cypher outputs, creating model-family coupling between benchmark generation and judging for that dataset. The examples in the Supplementary Material show that formatting, verbosity, internal identifiers, and date interpretation can affect scores. The evaluation includes bootstrap confidence intervals, paired deltas, and a 40-example human validation sample, but future work should still add expert adjudication, repeated judging, and multi-judge agreement analysis. Exact replication also depends on continued access to the proprietary GPT model versions used in the experiments.

Sixth, the conversational benchmark is preliminary. The reported Tier 4 evaluation covers 10 conversations and 36 turns. Although the results suggest improved context carryover relative to flat retrieval systems, the small number of conversations requires caution. Larger benchmarks with more entity switches, contradictory user instructions, and longer investigative sessions are needed. Similarly, the 60-question manually curated benchmark was designed by the authors rather than independently annotated. It provides a controlled factual evaluation with verified reference answers, but future work should include larger datasets annotated by independent domain experts.

Seventh, the dashboard provides an audit workflow by exposing evidence and execution traces, but its usability and effect on expert error detection have not been evaluated in a user study. The optional HMAC-SHA256 abstraction layer was also inactive during the experiments, so its effects on retrieval quality and latency remain untested.

Finally, the system was evaluated on a single registry, the SHAB dataset. Nevertheless, SHAB provides a large and diverse case study, comprising more than seven million publications in German, French, and Italian, as well as structured registry metadata and unstructured legal notices. Validation on registries from other countries is needed before claiming cross-registry generalization. Such deployments may also require adapting schema mappings, legal rubrics, extraction prompts, supported languages, and entity types.

\section{Conclusion}
\label{sec:Conclusion}

This paper presents a controlled, tool-mediated agentic GraphRAG system for large-scale commercial registry analysis. The system combines deterministic graph construction, LLM-assisted weak-node extraction, conservative identity resolution, intent-gated tool use, bounded reflection, state-machine-guided response synthesis, and a dashboard-supported workflow designed to support traceability.

Within the evaluated SHAB benchmarks, our findings suggest that graph-grounded retrieval is a promising foundation for commercial registry tasks that require multi-hop traversal, temporal tracking, and entity disambiguation.
The system builds a large Neo4j knowledge graph from more than seven million registry publications and outperforms flat dense, lexical, and hybrid retrieval baselines across the manually curated benchmark and the preliminary conversational evaluation. The all-architecture comparison also shows that surrounding control mechanisms matter. Bounded reflection improves answer quality, intent routing improves reliability in difficult graph-seeded tasks, and weak-node enrichment helps identify actors and relations embedded in unstructured legal text.

% Keep all main-paper figures and tables ahead of the references.
\FloatBarrier
\bibliographystyle{IEEEtran}
\bibliography{refs}

\begin{IEEEbiography}[{\includegraphics[width=1in,height=1.25in,clip,keepaspectratio]{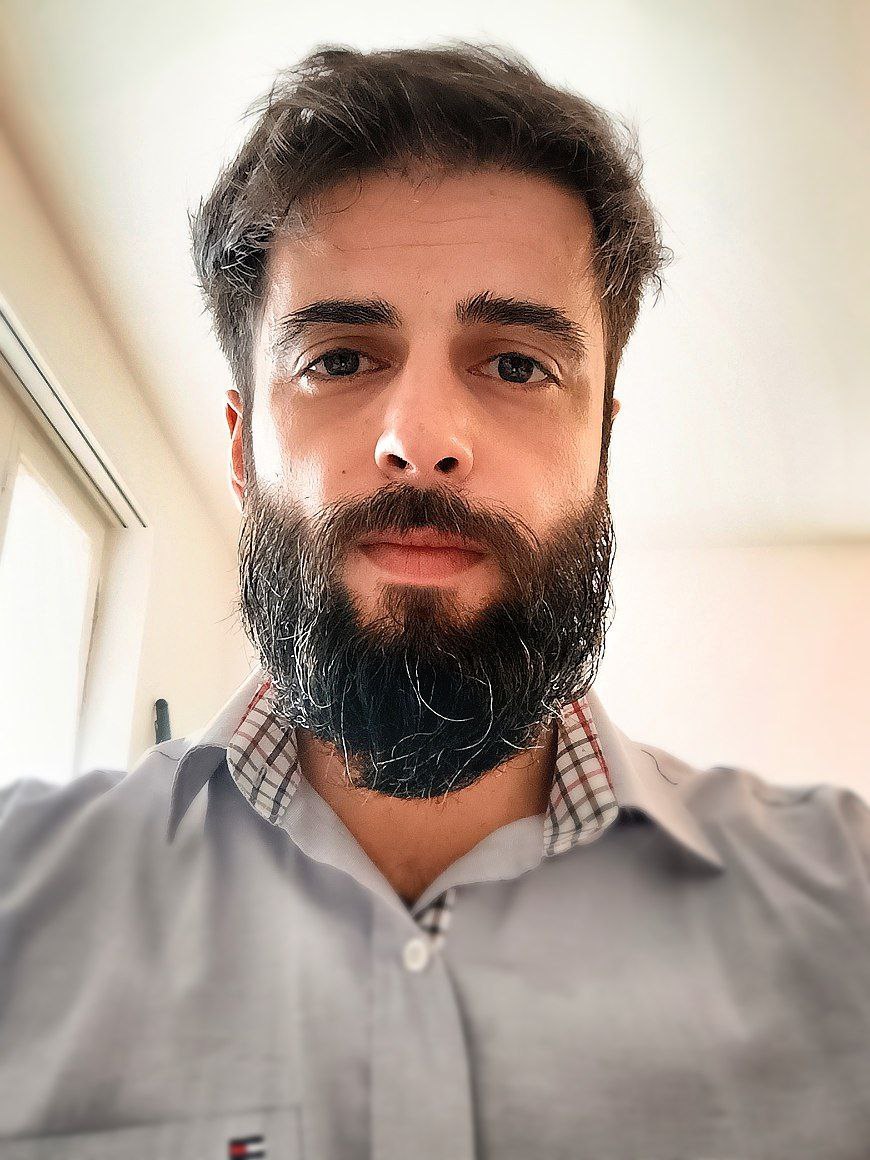}}]{Arthur Capozzi} is a Postdoctoral Researcher and Lecturer in the Computational Social Science group at the Department of Humanities, Social and Political Sciences at ETH Zurich. His research combines complex-network analysis, machine learning, and natural language processing to study financial networks, social media, political polarization, and other socio-economic systems. He received his Ph.D. in Computer Science from the University of Turin. His research has been recognized with Best Paper Awards at the ACM CHI Conference in 2021 and the International Conference on Social Informatics in 2020.
\end{IEEEbiography}

\begin{IEEEbiography}[{\includegraphics[width=1in,height=1.25in,clip,keepaspectratio]{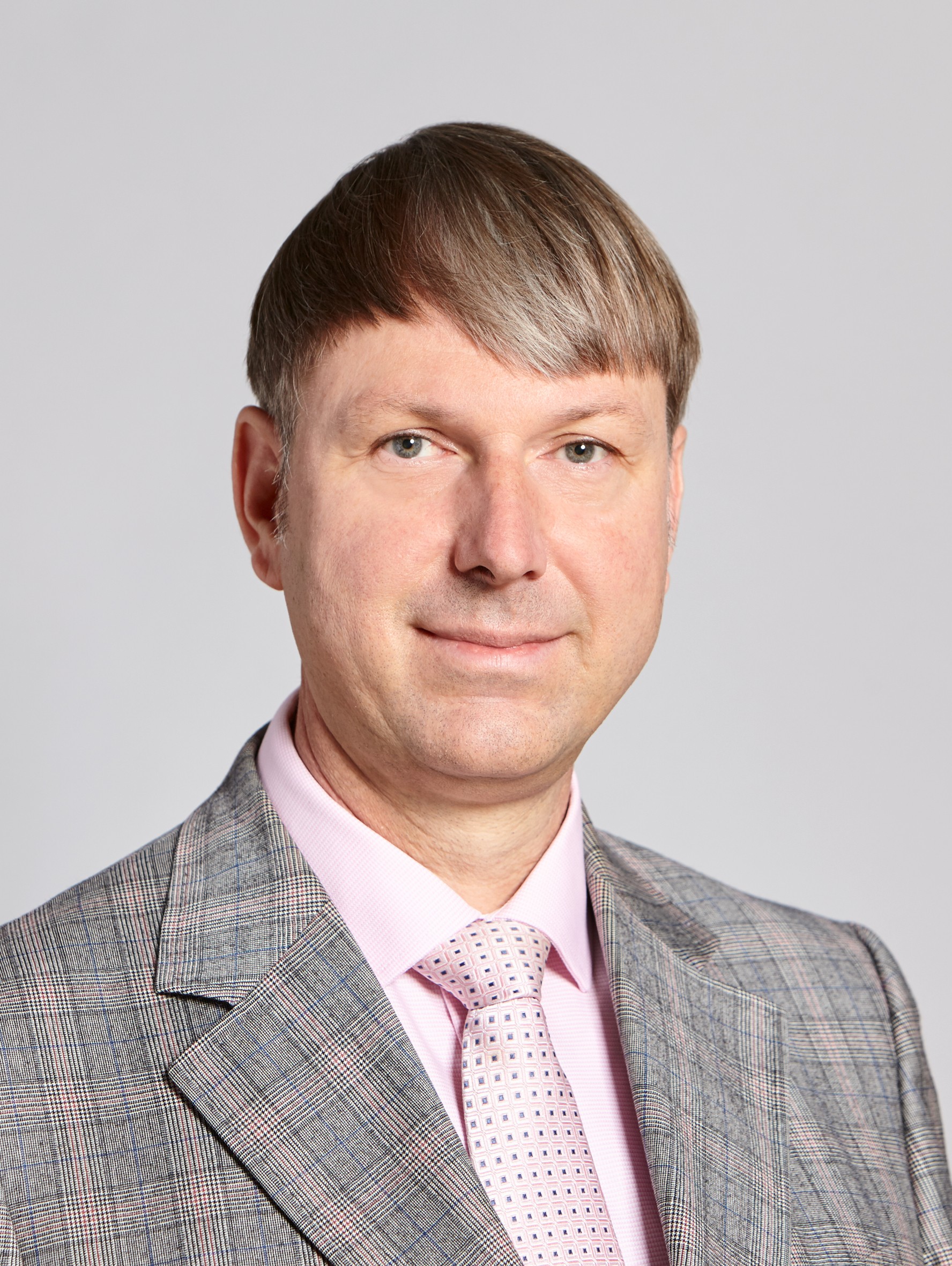}}]{Dirk Helbing} is Professor of Computational Social Science at the Department of Humanities, Social and Political Sciences and associated at ETH Zurich's Department of Computer Science. His current studies discuss the digital society, smart cities, as well as challenges and opportunities of digital democracies. Among others, he coordinated the doctoral program "Engineering Social Technologies for a Responsible Digital Future“ at TU Delft. Furthermore, he is an elected member of the German Academy of Sciences “Leopoldina” and a member of the External Faculty of the Complexity Science Hub Vienna.
\end{IEEEbiography}

\EOD

\clearpage
\FloatBarrier
\setcounter{section}{0}
\setcounter{table}{0}
\setcounter{figure}{0}
\renewcommand{\thetable}{S\arabic{table}}
\renewcommand{\thefigure}{S\arabic{figure}}
\renewcommand{\thesection}{S\arabic{section}}
\renewcommand{\theHtable}{supp.table.\arabic{table}}
\renewcommand{\theHfigure}{supp.figure.\arabic{figure}}
\renewcommand{\theHsection}{supp.section.\arabic{section}}
\renewcommand{\theHsubsection}{supp.subsection.\arabic{section}.\arabic{subsection}}

\newcolumntype{L}[1]{>{\raggedright\arraybackslash}p{#1}}
\newcolumntype{Y}{>{\raggedright\arraybackslash}X}
\setlist[itemize]{leftmargin=*, nosep}

\begin{center}
{\LARGE\bfseries Supplementary Material for\\[0.5em]
\emph{Agentic Graph Retrieval-Augmented Generation for Auditable Commercial Registry Analysis}\par}
\vspace{1em}
{\large Arthur Capozzi and Dirk Helbing\par}
\end{center}
\vspace{1em}

\section*{Overview}

This supplementary material provides extended implementation details,
schema mappings, evaluation prompts, benchmark examples, and reproducibility
information supporting the main manuscript.

\section{Knowledge Graph Schema and Statistics}
\label{sec:supp_knowledge_graph}

Table~\ref{tab:supp_edge_mapping} details the mapping rules used during the ingestion pipeline to convert semantic roles into deterministic graph edges. Table~\ref{tab:supp_node_stats} reports detailed statistics of the final Neo4j knowledge graph.

\begin{table}[htbp]
    \caption{Mapping of extracted semantic roles to deterministic graph edge types during the ingestion pipeline.}
    \label{tab:supp_edge_mapping}
    \centering
    \begin{tabular}{ll}
        \toprule
        \textbf{Extracted Role} & \textbf{Graph Edge Type} \\
        \midrule
        \texttt{SUBJECT} & \texttt{HAS\_EVENT} \\
        \texttt{PARENT} & \texttt{HEAD\_OFFICE\_OF} \\
        \texttt{SELLER} & \texttt{TRANSFERRED\_TO} \\
        \texttt{BUYER} & \texttt{ACQUIRED\_FROM} \\
        \texttt{DEBTOR\_COMPANY} & \texttt{HAS\_EVENT} \\
        \texttt{DEBTOR\_PERSON} & \texttt{INVOLVED\_IN} \\
        \texttt{DISSOLVED} & \texttt{DISSOLVED\_IN} \\
        \texttt{ASSURANCE} & \texttt{PROVIDES\_ASSURANCE\_TO} \\
        \texttt{NAME\_ALIAS} & \texttt{HAS\_NAME} \\
        \texttt{DEFAULT} & \texttt{RELATED\_TO} \\
        \bottomrule
    \end{tabular}
\end{table}

\begin{table}[htbp]
    \caption{Total number of edges and distribution of nodes by entity type and identification strength.}
    \label{tab:supp_node_stats}
    \centering
    \begin{tabular}{lrrr}
        \toprule
        \textbf{Node Type} & \textbf{Strong} & \textbf{Weak} & \textbf{Total} \\
        \midrule
        Company  & 935,014 & 112,344 & 1,047,358 \\
        Person   & 57,508  & 606,392 & 663,900 \\
        Event    & N/A     & N/A     & 2,180,645 \\
        NameHub  & N/A     & N/A     & 1,354,406 \\
        \midrule
        \textbf{Total Nodes}&         &         & \textbf{5,246,309} \\
        \textbf{Total Edges}&         &         & \textbf{4,729,070} \\
        \bottomrule
    \end{tabular}
\end{table}

\section{Tool Routing and Semantic Intents}
\label{sec:supp_tool_routing}

Table~\ref{tab:supp_tool_routing} details the mapping of semantic intents to activated database tools, underlying Cypher mechanisms, and architectural rationales. This dynamic routing reduces tool overload and improves query determinism. The intent names match the five categories introduced in the main manuscript.

\begin{table*}[htbp]
    \centering
    \small
    \caption{Mapping of semantic intents to activated database tools, underlying Cypher mechanisms, and architectural rationales.}
    \label{tab:supp_tool_routing}
    \begin{tabularx}{\textwidth}{@{} 
        >{\raggedright\arraybackslash}p{3.5cm} 
        >{\raggedright\arraybackslash}p{3.5cm} 
        >{\raggedright\arraybackslash}X 
        >{\raggedright\arraybackslash}X @{}}
        \toprule
        \textbf{Semantic Intent \newline (Router Output)} & \textbf{Activated Tools in Sandbox} & \textbf{Mechanism \& Description} & \textbf{Routing Rationale (Problem Solved)} \\
        \midrule
        
        \textbf{Entity Disambiguation} &
        Entity-disambiguation, network-exploration, event-history, and lexical full-text endpoints &
        Searches for Company and Person entities through fuzzy matching over names, locations, and purposes. It excludes orphan weak nodes through the NameHub hierarchy. &
        Supports initial entity identification while retaining the network and history endpoints required for the next analytical step. \\
        \addlinespace

        \textbf{Network Exploration} &
        Entity-disambiguation, network-exploration, event-history, and lexical full-text endpoints &
        Executes a three-branch \texttt{UNION ALL} Cypher query to identify corporate structures, event subjects, and lateral connections across NameHub aliases. &
        Keeps entity search available because network traversal requires a resolved target entity. \\
        \addlinespace

        \textbf{Event-History Retrieval} &
        Entity-disambiguation, network-exploration, event-history, and lexical full-text endpoints &
        Traverses an entity through its NameHub and collects chronologically ordered Event nodes associated with its legal aliases. &
        Retains entity search so that the agent can resolve the target UID before retrieving its timeline. \\
        \addlinespace

        \textbf{Macro-Level Analytics} &
        Top-entity aggregation, event-conditioned counting, and read-only custom-query endpoints &
        Provides statistical summaries by ranking entities by selected metrics (e.g., risk rank or nominal capital) or counting nodes associated with specific sub-rubrics. &
        Removes entity-search and traversal endpoints so that aggregation is performed by predefined server-side queries rather than manual pagination. \\
        \addlinespace

        \textbf{Deep Text Fallback} \newline (Nested in non-analytic intents) &
        Lexical full-text search endpoint &
        Runs a broad lexical search utilizing a Neo4j Lucene full-text inverted index directly against raw legal text snippets. &
        Provides a fallback when structured entity matching fails for an uncommon string or an entity that is not represented as a resolved graph node. \\
        \addlinespace

        \textbf{Unrestricted Multi-Hop Exploration} \newline (\texttt{all}) &
        All 7 tools loaded simultaneously &
        Grants the agent access to the standard retrieval suite, analytics endpoints, and the read-only Cypher sandbox. &
        Supports ambiguous or multi-part questions that cannot be assigned reliably to a narrower intent category. \\
        \bottomrule
    \end{tabularx}
\end{table*}

\FloatBarrier
\section{Complete Dashboard Views}
\label{sec:supp_dashboard_views}

\begin{figure}[p]
    \centering
    \includegraphics[width=\textwidth]{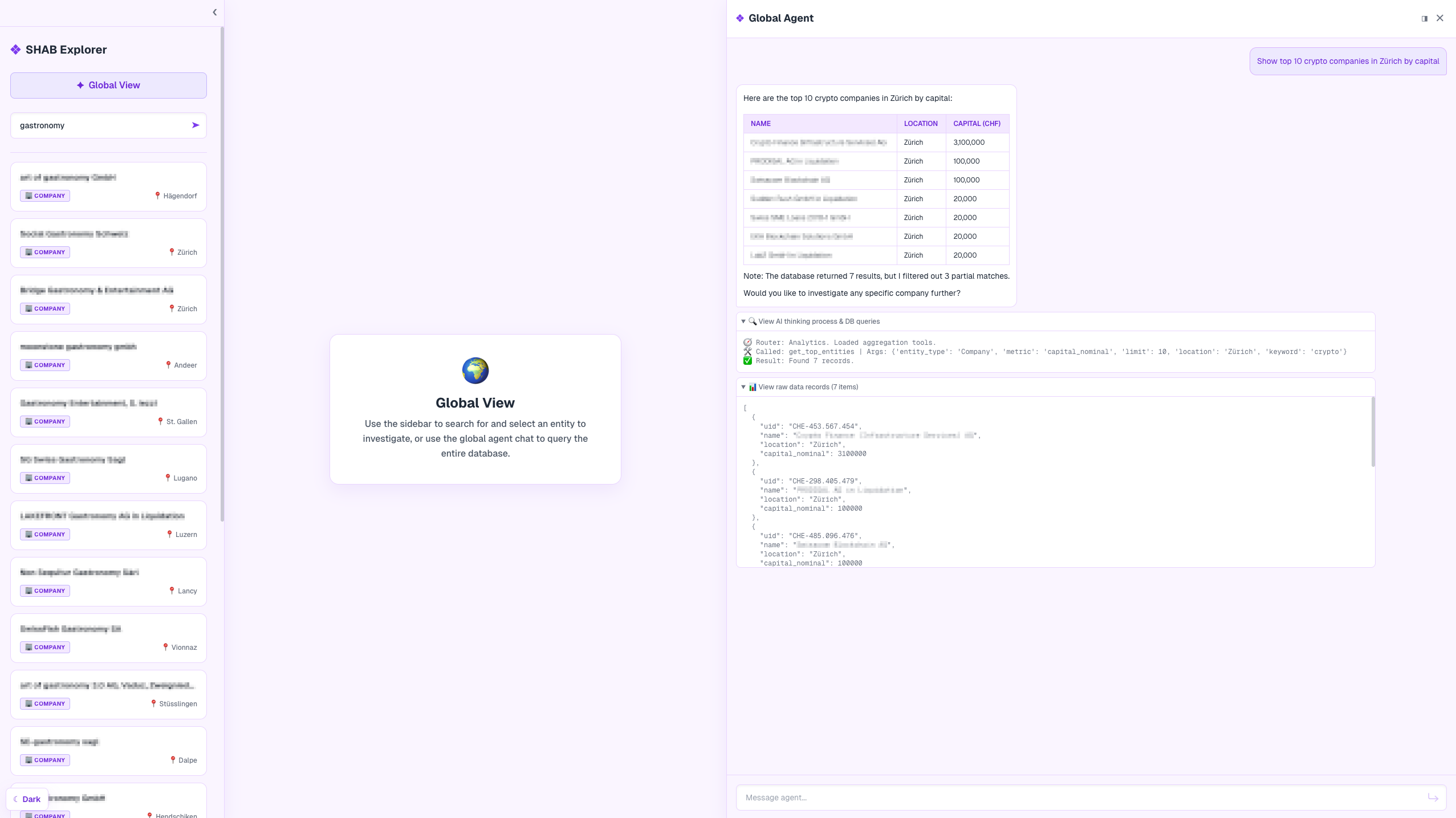}
    \caption{Complete dashboard in global-analysis mode, corresponding to Fig. 3 of the main paper. The left panel supports entity search, the center remains in global view because no entity is selected, and the right panel shows a registry-wide analytical result together with its execution trace and returned records.}
    \label{fig:supp_dashboard_global_full}
\end{figure}

\begin{figure}[p]
    \centering
    \includegraphics[width=\textwidth]{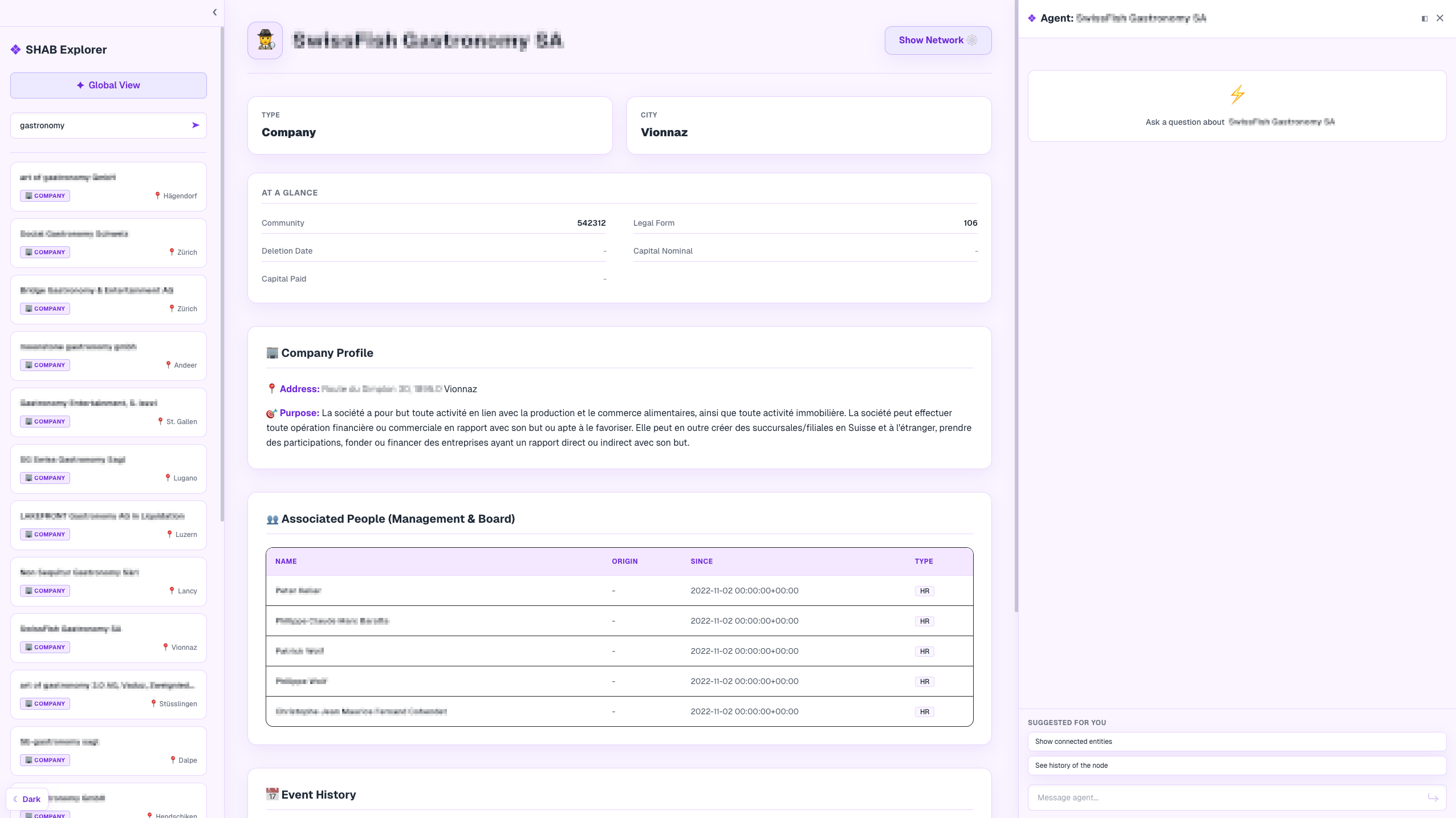}
    \caption{Complete dashboard in dossier mode, corresponding to Fig. 4 of the main paper. Entity-search results remain visible on the left, the selected company dossier occupies the central workspace, and the context-aware agent appears on the right.}
    \label{fig:supp_dashboard_dossier_full}
\end{figure}

\begin{figure}[p]
    \centering
    \includegraphics[width=\textwidth]{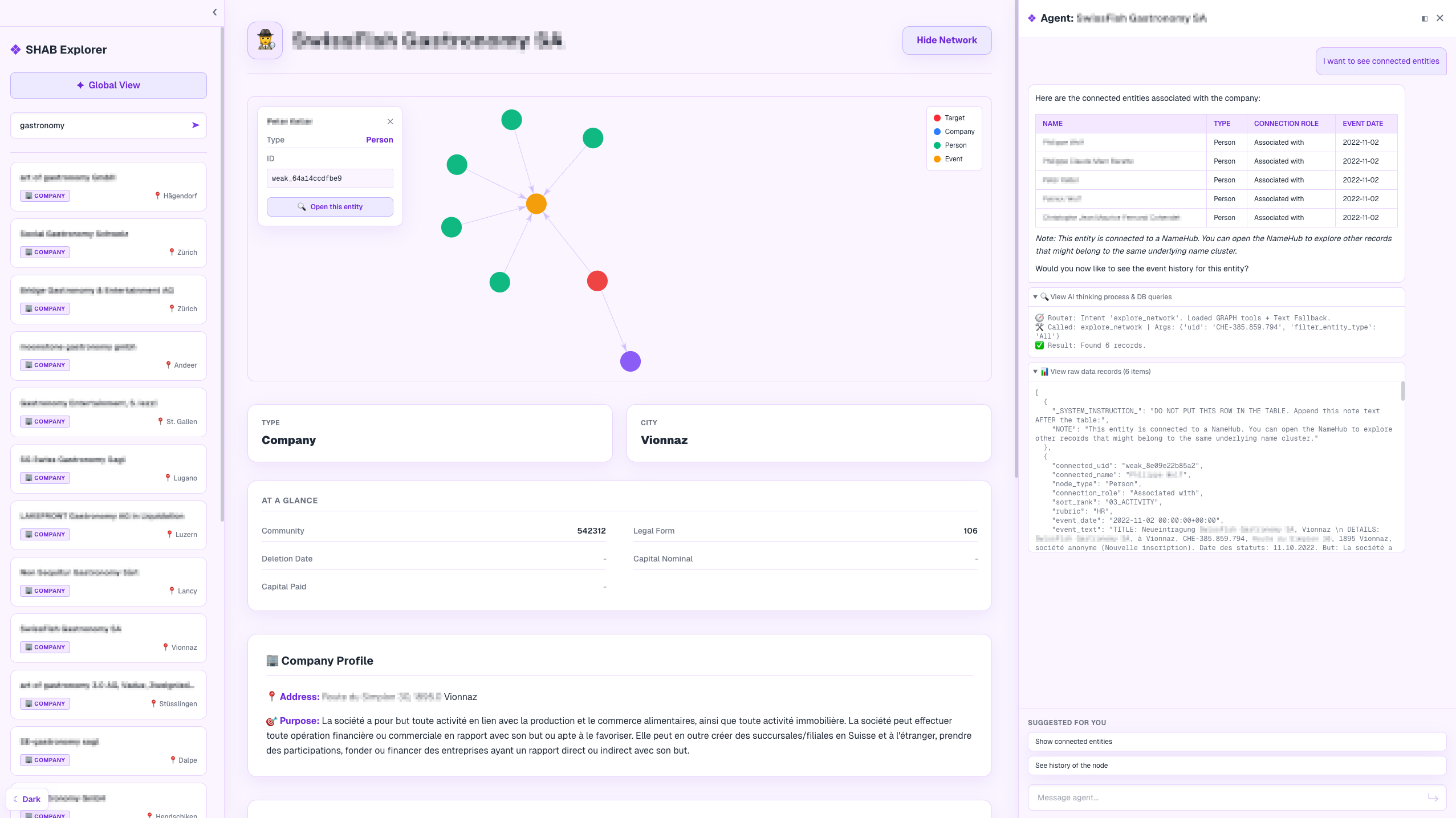}
    \caption{Complete dashboard during network exploration, corresponding to Fig. 5 of the main paper. The interface combines entity search, the interactive graph and dossier, and the context-aware agent with expandable execution evidence.}
    \label{fig:supp_dashboard_network_full}
\end{figure}

Figs.~\ref{fig:supp_dashboard_global_full}--\ref{fig:supp_dashboard_network_full} provide the complete, uncropped dashboard states corresponding to the focused views in Figs. 3--5 of the main paper. The main-paper figures use focused crops at the full column width, while the views below preserve the spatial relationship among entity search, the central workspace, and the analytical agent.

\FloatBarrier
\section{Answer-Quality Evaluation Metrics}
\label{sec:supp_answer_quality_metrics}

Table~\ref{tab:supp_rag_metrics_detailed} details the answer-quality evaluation metrics, their mechanisms, operational importance, and the LLM-as-a-judge prompts used to compute the final scores.

\begin{table*}[htbp]
    \centering
    \small
    \caption{Answer-quality evaluation metrics, their mechanisms, operational importance, and corresponding LLM-as-a-judge prompts.}
    \label{tab:supp_rag_metrics_detailed}
    \begin{tabularx}{\textwidth}{@{} 
        >{\raggedright\arraybackslash}p{2.5cm} 
        >{\raggedright\arraybackslash}p{3.5cm} 
        >{\footnotesize\raggedright\arraybackslash}X 
        >{\raggedright\arraybackslash}X @{}}
        \toprule
        \textbf{Metric} & \textbf{What it measures} & \textbf{Why it matters} & \textbf{Scoring prompt} \\
        \midrule
        
        \textbf{Faithfulness} & 
        Whether the agent's answer is fully grounded in the retrieved context, meaning that every factual claim in the answer can be traced back to the data the system actually retrieved from the database. & 
        A fluent response may still contain claims that are absent from the retrieved evidence. A score of 1.0 indicates that all factual claims are supported by the context, whereas 0.0 indicates that the response is unsupported. This metric is used for the graph-seeded automated benchmark, where grounding to retrieved evidence is central to the evaluation. &
        ``Given the context and the answer, compute a faithfulness score between 0.0 and 1.0. 1.0 means the answer is 100\% derived from the context. 0.0 means the answer states completely fabricated facts. Output ONLY a float number.'' \\
        \addlinespace
        
        \textbf{Correctness} & 
        Whether the agent's answer is factually consistent with the verified reference answer, regardless of how the supporting context was retrieved. & 
        Correctness is used for the manually curated benchmark, where each question has a verified reference answer. It measures factual agreement with that answer rather than only support from the retrieved context. &
        ``Given the question, the expected answer, and the actual answer, compute a correctness score between 0.0 and 1.0. 1.0 means the actual answer is factually correct and captures the expected answer well. 0.0 means the actual answer is factually wrong or fails to answer the question correctly. Be tolerant to differences in wording, formatting, ordering, or level of detail, as long as the substance is correct. Output ONLY a float number.'' \\
        \addlinespace
        
        \textbf{Answer Relevance} & 
        Whether the answer directly and completely addresses the user's question, regardless of the retrieved context. & 
        A factually supported response can still address the wrong question or obscure the requested information with irrelevant detail. Answer Relevance isolates whether the response directly addresses the question. &
        ``Given the question and the answer, compute an answer relevance score between 0.0 and 1.0. 1.0 means it perfectly answers the question directly. 0.0 means it dodges the question entirely. Output ONLY a float number.'' \\
        \addlinespace
        
        \textbf{Information Recall} & 
        The fraction of the expected answer's core information that the agent's response successfully captures. & 
        Grounding, correctness, and relevance do not establish that an answer is complete. Information Recall compares the response with the reference answer and measures how much of its core information is present. It is therefore sensitive to incomplete retrieval and incomplete synthesis. &
        ``Compare the Expected Answer to the Actual Answer. Score from 0.0 to 1.0 how much of the Expected Answer's core information is present in the Actual Answer. Output ONLY a float number.'' \\
        \bottomrule
    \end{tabularx}
\end{table*}

\section{Conversational Evaluation Metrics}
\label{sec:supp_conversational_evaluation_metrics}

Table~\ref{tab:supp_conversational_metrics_detailed} details the conversational evaluation metrics that extend the turn-level RAGAS-style analysis with dialogue-aware measures of continuity, procedural consistency, and multi-turn task completion.

\begin{table*}[htbp]
    \centering
    \small
    \caption{Conversational evaluation metrics, their mechanisms, operational importance, and corresponding scoring logic.}
    \label{tab:supp_conversational_metrics_detailed}
    \begin{tabularx}{\textwidth}{@{} 
        >{\raggedright\arraybackslash}p{3.0cm} 
        >{\raggedright\arraybackslash}p{3.8cm} 
        >{\raggedright\arraybackslash}X 
        >{\raggedright\arraybackslash}X @{}}
        \toprule
        \textbf{Metric} & \textbf{What it measures} & \textbf{Why it matters} & \textbf{Scoring prompt / logic} \\
        \midrule
        
        \textbf{Turn Success Rate} & 
        The fraction of turns in a conversation that are successfully answered according to the conversational scoring rule. & 
        A system may answer the final question correctly after making errors in earlier turns. Turn Success Rate captures consistency across the full interaction and reflects the propagation of local errors through later turns. &
        For each turn, success is assigned deterministically after the judge scores the answer. A turn is marked successful if the normalized exact match equals 1.0, or if \textbf{Correctness} $\geq 0.80$, or if \textbf{Correctness} $\geq 0.70$ and \textbf{Answer Relevance} $\geq 0.70$. The final Turn Success Rate is the average of these binary turn outcomes over all turns in the conversation. \\
        \addlinespace
        
        \textbf{Context Carryover Accuracy} & 
        Whether the system correctly preserves the active entity, prior references, and relevant facts across follow-up turns that require memory of earlier dialogue state. & 
        Conversational systems may answer isolated questions correctly but lose track of the active person, company, event, or previously retrieved fact. This metric evaluates continuity across follow-up questions involving counting, summarization, or co-reference resolution. &
        This metric is computed deterministically over the subset of turns tagged for context carryover, entity memory, or temporal memory. For these turns, the same binary success rule used in Turn Success Rate is applied. Context Carryover Accuracy is then defined as the average success over this memory-sensitive subset only. \\
        \addlinespace
        
        \textbf{Tool Transition Accuracy} & 
        Whether the GraphRAG Agent follows the expected structured tool sequence for a given conversational turn. & 
        The graph-based architecture assigns different tools to different semantic intents. Tool Transition Accuracy evaluates whether the agent follows the expected structured retrieval path, thereby testing the state machine and sequential tool logic. &
        This metric is only computed for the agentic GraphRAG. For each turn that specifies an expected tool behavior, the execution trace is inspected to determine whether at least one of the expected tools was actually called. A hit receives a score of 1.0 and a miss receives 0.0. Tool Transition Accuracy is the average of these binary hits across all turns for which an expected tool behavior is defined. \\
        \bottomrule
    \end{tabularx}
\end{table*}

\section{Examples of LLM-Judge Sensitivity}
\label{sec:supp_llm_judge_sensitivity}

Tables~\ref{tab:supp_edge_cases_automated} and~\ref{tab:supp_edge_cases_human} present questions, reference answers, GraphRAG responses, and the corresponding LLM-judge scores for the graph-seeded and manually curated benchmarks. The examples illustrate sensitivity to formatting, additional metadata, internal identifiers, and date interpretation. Company names, personal names, and registry identifiers are anonymized.

\begin{table*}[htbp]
    \centering
    \small
    \caption{Examples of LLM-judge sensitivity on the graph-seeded automated benchmark.}
    \label{tab:supp_edge_cases_automated}
    \begin{tabularx}{\textwidth}{@{} 
        >{\raggedright\arraybackslash}p{2.5cm} 
        >{\raggedright\arraybackslash}p{3.5cm} 
        >{\raggedright\arraybackslash}X 
        >{\raggedright\arraybackslash}X @{}}
        \toprule
        \textbf{Metric \& Score} & \textbf{Question} & \textbf{Reference Answer} & \textbf{Agent Answer} \\
        \midrule
        
        \textbf{Answer Relevance} \newline 
        Score: \textbf{0.10} & 
        What is the registered address of the company with anonymized UID CHE-000.000.000? &
        Street Name 19, 5300 City A & 
        The registered address of the company \textbf{Company A GmbH} (UID: CHE-000.000.000) has changed to \textbf{City A} as of January 12, 2024. Previously, it was located in \textbf{City B}. \newline\newline Would you like to investigate anything else related to this company? \\
        \addlinespace
        \midrule
        \addlinespace
        
        \textbf{Information Recall} \newline 
        Score: \textbf{0.55} & 
        Within the NameHub 'Person X', do the company links resolve to one or multiple distinct person profiles, and for each profile (by person\_uid), which company is it associated with? & 
        Multiple distinct profiles: two in total. weak\_1 is associated with Company B Sàrl; weak\_2 is associated with Foundation C. & 
        The search for "Person X" returned a weak node, which has been explored to reveal connections to other individuals. Here are the distinct person profiles associated with Person X, along with the companies they are linked to:
        \begin{itemize}[leftmargin=*, nosep]
            \item Person X: Company B Sàrl (Role: Associate)
            \item Person X: Foundation C (Role: Board Member)
        \end{itemize}
        \textbf{Note:} This entity is connected to a NameHub. You can open the NameHub to explore other records that might belong to the same underlying name cluster. \newline\newline Would you like to investigate further into any specific company or person? \\
        
        \bottomrule
    \end{tabularx}
\end{table*}
\clearpage
\begin{table*}[htbp]
    \centering
    \small
    \caption{Examples of LLM-judge sensitivity on the manually curated benchmark. These cases illustrate how verbosity or technically accurate details, such as a legal effective date versus a publication date, can affect scores.}
    \label{tab:supp_edge_cases_human}
    \begin{tabularx}{\textwidth}{@{} 
        >{\raggedright\arraybackslash}p{2.5cm} 
        >{\raggedright\arraybackslash}p{3.5cm} 
        >{\raggedright\arraybackslash}X 
        >{\raggedright\arraybackslash}X @{}}
        \toprule
        \textbf{Metric \& Score} & \textbf{Question} & \textbf{Reference Answer} & \textbf{Agent Answer} \\
        \midrule
        
        \textbf{Answer Relevance} \newline 
        Score: \textbf{0.40} & 
        Show all the companies with which Person Y is connected. & 
        Person Y is connected with the following companies: Person Y, Company D Sàrl. & 
        The entities shown below are connected to various records sharing this exact name, but the database cannot determine with 100\% certainty that they belong to the exact same physical person.
        \begin{itemize}[leftmargin=*, nosep]
            \item Company D Sàrl (Location: City C, Connection: HAS\_EVENT, Date: 2023-12-20)
            \item Person Y (Location: City C, Connection: HAS\_EVENT, Date: 2023-12-12)
        \end{itemize}
        Would you like to investigate one of these companies further? \\
        \addlinespace
        \midrule
        \addlinespace
        
        \textbf{Information Recall} \newline 
        Score: \textbf{0.00} & 
        Have any notices of dissolution or bankruptcy been published for Company E? & 
        Yes, a notice of dissolution for Company E was published on December 17, 2020. & 
        The company "Company E" located in City D has a history of bankruptcy. Here are the relevant details: \newline\newline
        \textbf{1. Bankruptcy Notice:} \newline
        The bankruptcy was officially opened by the Regional Court F, effective from December 10, 2020, at 14:02. \newline\newline
        \textbf{2. Previous Changes:} \newline
        On June 5, 2020, the company changed its name from "Company G" to "Company E" and relocated to Street Name 10, City D. \newline\newline
        \textbf{3. Initial Registration:} \newline
        The company was initially registered as "Company G" on September 19, 2019, in City E. \newline\newline
        If you need more information or further investigation into related entities or events, please let me know! \\
        
        \bottomrule
    \end{tabularx}
\end{table*}

\subsection{Human validation of LLM-as-a-judge scores}
\label{sec:supp_human_judge_validation}

To assess whether the LLM-as-a-judge scores were aligned with human judgment, we manually audited 40 stratified answer-quality judgments. The sample included high-, medium-, and low-scoring examples and compared GPT-5 correctness scores against manual labels on a three-level scale: correct, partially correct, and incorrect. Rows marked as uncertain were excluded from quantitative agreement statistics. Table~\ref{tab:human_judge_validation} reports the resulting agreement statistics.

\begin{table}[t]
\centering
\caption{Human validation of GPT-5 judge scores on a stratified sample of answer-quality evaluations.}
\label{tab:human_judge_validation}
\begin{tabular}{lr}
\hline
Metric & Value \\
\hline
Labeled examples & 40 \\
Pearson correlation & 0.721 \\
Spearman correlation & 0.752 \\
Mean absolute error & 0.168 \\
Within-half-point agreement & 0.875 \\
Quadratic weighted Cohen's $\kappa$ & 0.702 \\
\hline
\end{tabular}
\end{table}

The audit shows positive alignment between GPT-5 correctness scores and the manual labels. Pearson correlation is $r=0.721$, Spearman correlation is $\rho=0.752$, and the mean absolute error is 0.168. Within-half-point agreement is 0.875. For the categorical agreement analysis, continuous GPT-5 scores were mapped to the same three-level scale using thresholds of 0.25 and 0.75, resulting in a quadratic weighted Cohen's $\kappa$ of 0.702. These results support the use of GPT-5 scores as comparative indicators, but not as absolute ground truth.

\subsection{Text-grounded audit of graph-seeded benchmark items}
\label{sec:supp_text_grounded_validation}

To assess whether the graph-seeded benchmark was grounded in original registry evidence, we audited a stratified sample of 60 benchmark items, with 20 items from each difficulty level. For each item, we retrieved the associated original SHAB event text from the Neo4j graph, together with connected structured metadata and adjacent evidence when the reference answer required graph composition. Table~\ref{tab:text_grounded_validation} summarizes the support labels by benchmark level.

\begin{table}[t!]
\centering
\caption{Text-grounded validation sample for the graph-seeded automated benchmark. Labels reflect conservative provenance checks against original SHAB notice text and connected structured metadata; they are not independent human annotations.}
\label{tab:text_grounded_validation}
\small
\begin{tabular}{lrr}
\toprule
Support label & Count & Share \\
\midrule
Supported & 14 & 0.233 \\
Partially supported & 46 & 0.767 \\
Not supported & 0 & 0.000 \\
Uncertain & 0 & 0.000 \\
\midrule
Supported or partially supported & 60 & 1.000 \\
\bottomrule
\end{tabular}
\end{table}

The audit found that all sampled benchmark items were at least partially supported by original SHAB evidence or directly connected registry metadata. Fourteen items were directly supported by retrieved notice text. The remaining 46 items were partially supported: their key facts appeared in the original evidence, but the full reference answer required graph-level composition, such as combining notice text with structured metadata, connected events, temporal ordering, or entity-resolution mappings. No sampled item remained unsupported after auditing the initially negative cases. Specifically, among the six initially unsupported cases, one was supported by an additional connected event, three were supported by structured metadata rather than \texttt{Event.text}, and two required graph metadata for internal weak-node UID mappings.

The audit establishes source-level provenance for the sampled benchmark items. Its scope remains limited: it evaluates graph-compositional retrieval from the constructed registry graph rather than independently validating the complete graph-construction pipeline.

\section{All-Architecture Benchmark Details}
\label{sec:supp_all_architecture_benchmarks}

The all-architecture evaluation compares seven variants across the graph-seeded automated benchmark, the manually curated benchmark, the conversational benchmark, and trajectory-level diagnostics. Table~\ref{tab:supp_architecture_variants} defines the evaluated systems. The first three rows are flat retrieval baselines, while the remaining rows are controlled graph variants and the complete agentic GraphRAG system.

% Generated by scripts/phase2_generate_manuscript_tables.py from architecture summary CSVs
\begin{table*}[t!]
\centering
\caption{Complete architecture variant definitions used in the all-architecture benchmark comparison.}
\label{tab:supp_architecture_variants}
\small
\begin{tabularx}{\textwidth}{@{} L{3.6cm} L{3.2cm} X @{}}
\toprule
\textbf{Architecture variant} & \textbf{Family} & \textbf{Controlled configuration} \\
\midrule
Dense Vector-RAG & Flat retrieval & Dense retrieval over event documents using the Chroma index. \\
Lexical Full-Text RAG & Flat retrieval & Neo4j full-text index retrieval at the selected $k=20$ setting. \\
Hybrid Dense+Lexical RAG & Flat retrieval & Reciprocal rank fusion of Chroma dense and Neo4j lexical candidates at $k=20$. \\
\midrule
GraphRAG w/o reflection & Graph ablation & Full graph tools and router retained, but bounded recovery is limited to one retrieval/planning attempt. \\
GraphRAG w/o router & Graph ablation & Reflection and graph tools retained, but the agent receives the full tool set for every query. \\
Structured-only GraphRAG & Graph ablation & Router and reflection retained, but LLM-extracted weak nodes are ignored during retrieval. \\
Full Agentic GraphRAG & Proposed system & Intent routing, bounded reflection, weak-node-enriched graph access, and state-machine-guided synthesis. \\
\bottomrule
\end{tabularx}
\end{table*}

The all-architecture tables are computed from the recorded per-question outputs of the benchmark runs rather than from new answer-generation runs. For the conversational benchmark, the Dense Vector-RAG and Full Agentic GraphRAG scores come from the initial Tier 4 evaluation, while the five additional architecture variants were scored in a subsequent run with the same benchmark and judging configuration. Thus, N/A is used only for structurally inapplicable graph-tool metrics in flat retrieval systems, not for missing answer-quality scores.

The trajectory diagnostics are computed from the recorded retrieved contexts, tool traces, answer strings, and reference answers on the manually curated benchmark. Evidence-bearing retrieval and supported-outcome rates are generic diagnostics that can be computed for both flat and graph-based architectures. Empty-result and tool/error rates are derived from architecture-family-specific execution fields and therefore should be interpreted as reliability diagnostics rather than as identical API-level error counters. Reflection recovery is not included in the all-architecture trajectory table because this diagnostic was designed for the full reflective graph agent and is not directly comparable across non-reflective or flat architectures.

Tables~\ref{tab:supp_all_arch_graph_seeded_detail} and~\ref{tab:supp_all_arch_human_detail} report the detailed single-turn results for the graph-seeded and manually curated benchmarks, respectively. Table~\ref{tab:supp_all_arch_conversational_detail} reports the conversational results, and Table~\ref{tab:supp_all_arch_trajectory_detail} reports the trajectory and evidence diagnostics.

% Generated by scripts/phase2_generate_manuscript_tables.py from results/all_architectures_all_arch_graph_seeded_detail_summary.csv
\begin{table*}[t!]
\centering
\caption{Detailed graph-seeded automated benchmark results with bootstrap confidence intervals.}
\label{tab:supp_all_arch_graph_seeded_detail}
\scriptsize
\begin{tabularx}{\textwidth}{@{} >{\raggedright\arraybackslash}X c c c c c c c @{}}
\toprule
\textbf{Architecture variant} & \textbf{N} & \textbf{Faith.} & \textbf{95\% CI} & \textbf{Ans. rel.} & \textbf{95\% CI} & \textbf{Info. recall} & \textbf{95\% CI} \\
\midrule
Dense Vector-RAG & 300 & 0.905 & [0.877, 0.931] & 0.081 & [0.055, 0.109] & 0.066 & [0.043, 0.090] \\
Lexical Full-Text RAG & 300 & 0.975 & [0.959, 0.988] & 0.047 & [0.026, 0.072] & 0.030 & [0.016, 0.046] \\
Hybrid Dense+Lexical RAG & 300 & 0.879 & [0.848, 0.908] & 0.127 & [0.094, 0.164] & 0.091 & [0.065, 0.119] \\
\midrule
GraphRAG w/o reflection & 300 & 0.894 & [0.866, 0.923] & 0.425 & [0.373, 0.477] & 0.388 & [0.339, 0.438] \\
GraphRAG w/o router & 300 & 0.887 & [0.865, 0.908] & 0.742 & [0.704, 0.782] & 0.581 & [0.540, 0.625] \\
Structured-only GraphRAG & 300 & 0.865 & [0.838, 0.892] & 0.651 & [0.607, 0.696] & 0.534 & [0.486, 0.582] \\
Full Agentic GraphRAG & 300 & 0.898 & [0.876, 0.920] & 0.689 & [0.647, 0.730] & 0.574 & [0.530, 0.616] \\
\bottomrule
\end{tabularx}
\end{table*}

% Generated by scripts/phase2_generate_manuscript_tables.py from results/all_architectures_all_arch_human_detail_summary.csv
\begin{table*}[t!]
\centering
\caption{Detailed manually curated benchmark results with bootstrap confidence intervals.}
\label{tab:supp_all_arch_human_detail}
\scriptsize
\begin{tabularx}{\textwidth}{@{} >{\raggedright\arraybackslash}X c c c c c c c @{}}
\toprule
\textbf{Architecture variant} & \textbf{N} & \textbf{Corr.} & \textbf{95\% CI} & \textbf{Ans. rel.} & \textbf{95\% CI} & \textbf{Info. recall} & \textbf{95\% CI} \\
\midrule
Dense Vector-RAG & 60 & 0.143 & [0.075, 0.213] & 0.247 & [0.152, 0.346] & 0.118 & [0.064, 0.181] \\
Lexical Full-Text RAG & 60 & 0.237 & [0.147, 0.336] & 0.322 & [0.212, 0.437] & 0.261 & [0.158, 0.370] \\
Hybrid Dense+Lexical RAG & 60 & 0.262 & [0.176, 0.357] & 0.470 & [0.358, 0.581] & 0.237 & [0.157, 0.324] \\
\midrule
GraphRAG w/o reflection & 60 & 0.552 & [0.428, 0.676] & 0.563 & [0.444, 0.680] & 0.560 & [0.435, 0.685] \\
GraphRAG w/o router & 60 & 0.708 & [0.598, 0.813] & 0.777 & [0.677, 0.868] & 0.703 & [0.592, 0.812] \\
Structured-only GraphRAG & 60 & 0.545 & [0.429, 0.659] & 0.733 & [0.629, 0.825] & 0.578 & [0.461, 0.695] \\
Full Agentic GraphRAG & 60 & 0.828 & [0.742, 0.913] & 0.887 & [0.823, 0.940] & 0.838 & [0.751, 0.923] \\
\bottomrule
\end{tabularx}
\end{table*}

% Generated by scripts/phase2_generate_manuscript_tables.py from results/all_architectures_all_arch_conversational_detail_summary.csv
\begin{table*}[t!]
\centering
\caption{Detailed conversational benchmark results with bootstrap confidence intervals where available.}
\label{tab:supp_all_arch_conversational_detail}
\scriptsize
\begin{tabularx}{\textwidth}{@{} >{\raggedright\arraybackslash}X c c c c c c c c c @{}}
\toprule
\textbf{Architecture variant} & \textbf{N} & \textbf{Corr.} & \textbf{95\% CI} & \textbf{Ans. rel.} & \textbf{95\% CI} & \textbf{Turn succ.} & \textbf{95\% CI} & \textbf{Carryover} & \textbf{95\% CI} \\
\midrule
Dense Vector-RAG & 10 & 0.344 & N/A & 0.644 & N/A & 0.367 & [0.192, 0.558] & 0.433 & [0.250, 0.633] \\
Lexical Full-Text RAG & 10 & 0.342 & N/A & 0.493 & N/A & 0.308 & [0.175, 0.483] & 0.400 & [0.233, 0.567] \\
Hybrid Dense+Lexical RAG & 10 & 0.310 & N/A & 0.677 & N/A & 0.308 & [0.142, 0.508] & 0.350 & [0.183, 0.533] \\
\midrule
GraphRAG w/o reflection & 10 & 0.458 & N/A & 0.598 & N/A & 0.450 & [0.292, 0.617] & 0.517 & [0.350, 0.683] \\
GraphRAG w/o router & 10 & 0.554 & N/A & 0.714 & N/A & 0.542 & [0.367, 0.708] & 0.517 & [0.350, 0.683] \\
Structured-only GraphRAG & 10 & 0.604 & N/A & 0.827 & N/A & 0.608 & [0.400, 0.808] & 0.650 & [0.467, 0.833] \\
Full Agentic GraphRAG & 10 & 0.760 & N/A & 0.847 & N/A & 0.742 & [0.642, 0.833] & 0.683 & [0.600, 0.783] \\
\bottomrule
\end{tabularx}
\end{table*}

% Generated by scripts/phase2_generate_manuscript_tables.py from results/all_architectures_all_arch_trajectory_detail_summary.csv
\begin{table*}[t!]
\centering
\caption{Detailed trajectory and evidence diagnostics with bootstrap confidence intervals where available. Reflection recovery is omitted here because it is a full-system-only diagnostic rather than a comparable all-architecture metric.}
\label{tab:supp_all_arch_trajectory_detail}
\tiny
\setlength{\tabcolsep}{2.5pt}
\renewcommand{\arraystretch}{0.95}
\begin{tabularx}{\textwidth}{@{} L{2.6cm} c c c c c c c c c @{}}
\toprule
\textbf{Architecture variant} & \textbf{N} & \textbf{Evidence} & \textbf{95\% CI} & \textbf{Supported} & \textbf{95\% CI} & \textbf{Empty} & \textbf{95\% CI} & \textbf{Tool/error} & \textbf{95\% CI} \\
\midrule
Dense Vector-RAG & 60 & 0.283 & [0.167, 0.400] & 0.067 & [0.017, 0.133] & 0.000 & [0.000, 0.000] & 0.000 & [0.000, 0.000] \\
Lexical Full-Text RAG & 60 & 0.300 & [0.183, 0.417] & 0.133 & [0.050, 0.217] & 0.667 & [0.550, 0.783] & 0.000 & [0.000, 0.000] \\
Hybrid Dense+Lexical RAG & 60 & 0.533 & [0.417, 0.650] & 0.167 & [0.083, 0.267] & 0.000 & [0.000, 0.000] & 0.000 & [0.000, 0.000] \\
\midrule
GraphRAG w/o reflection & 60 & 0.583 & [0.450, 0.717] & 0.550 & [0.433, 0.683] & 0.133 & [0.050, 0.217] & 0.150 & [0.067, 0.250] \\
GraphRAG w/o router & 60 & 0.850 & [0.750, 0.933] & 0.683 & [0.567, 0.800] & 0.000 & [0.000, 0.000] & 0.117 & [0.050, 0.200] \\
Structured-only GraphRAG & 60 & 0.783 & [0.667, 0.883] & 0.483 & [0.350, 0.617] & 0.000 & [0.000, 0.000] & 0.450 & [0.333, 0.567] \\
Full Agentic GraphRAG & 60 & 0.950 & [0.883, 1.000] & 0.800 & [0.700, 0.900] & 0.013 & N/A & 0.013 & [0.000, 0.050] \\
\bottomrule
\end{tabularx}
\end{table*}

The unsuccessful-outcome diagnostic is derived from recorded execution fields and should be interpreted as a benchmark-level reliability signal rather than as a uniform API or runtime-error rate. Table~\ref{tab:supp_failure_definitions} summarizes the source fields used for each result family. Tables~\ref{tab:supp_failure_summary}, \ref{tab:supp_failure_pairwise}, \ref{tab:supp_failure_categories}, \ref{tab:supp_failure_faithfulness}, and~\ref{tab:supp_failure_breakdown} report the aggregate rates, paired comparisons, failure categories, faithfulness diagnostics, and task-level breakdowns. Table~\ref{tab:supp_pairwise_quality_all_architectures} reports paired answer-quality comparisons between the full system and every other architecture. This distinction is important because flat retrievers expose retrieved-document counts and answer strings, whereas graph agents expose tool traces, empty-result markers, and recovery flags.

% Generated by scripts/phase2_generate_manuscript_tables.py from results/all_architectures_failure_summary.csv schema notes
\begin{table*}[t!]
\centering
\caption{Source-field definitions used for unsuccessful-outcome diagnostics. These diagnostics are comparable as benchmark-level reliability signals, but their raw source fields differ across architecture families because flat retrievers, graph agents, and conversational traces expose different execution metadata.}
\label{tab:supp_failure_definitions}
\scriptsize
\begin{tabularx}{\textwidth}{@{} L{3.2cm} L{4.0cm} X @{}}
\toprule
\textbf{Result family} & \textbf{Primary recorded fields} & \textbf{Interpretation} \\
\midrule
Flat dense/lexical/hybrid runs & Retrieved-document counts, generated-answer text, explicit no-answer strings, API/judge errors & Captures empty retrieval, empty or explicit no-answer outputs, and execution errors; it does not represent graph-tool failures. \\
Graph-seeded graph-agent runs & Recorded trace logs, failed or empty tool-call indicators, recovery flags, generated answers, judge scores & Captures graph-tool retrieval failures or empty intermediate results; these may occur even when the final answer still receives a valid judge score. \\
Manually curated benchmark graph-agent runs & Recorded tool traces, answer text, judge scores, latency, and ablation-specific execution fields & Captures the same graph-tool reliability signals on the manually curated benchmark and keeps failed rows in aggregate quality means when scored. \\
Conversational runs & Turn-level answer text, trace fields, explicit no-answer strings, and graph-tool transition records & Captures turn-level unsuccessful outcomes; aggregate conversational metrics are computed separately at the conversation/turn level. \\
\bottomrule
\end{tabularx}
\end{table*}

% Generated by scripts/phase2_generate_manuscript_tables.py from results/all_architectures_failure_summary.csv
\begin{table*}[t!]
\centering
\caption{Graph-seeded benchmark-level unsuccessful-outcome rates with Wilson confidence intervals and conditional non-failure quality.}
\label{tab:supp_failure_summary}
\scriptsize
\begin{tabularx}{\textwidth}{@{} >{\raggedright\arraybackslash}X c c c c c @{}}
\toprule
\textbf{Architecture variant} & \textbf{Failed} & \textbf{Rate} & \textbf{Wilson 95\% CI} & \textbf{Non-fail rel.} & \textbf{Non-fail recall} \\
\midrule
Dense Vector-RAG & 0/300 & 0.000 & [0.000, 0.013] & 0.081 & 0.066 \\
Lexical Full-Text RAG & 282/300 & 0.940 & [0.907, 0.962] & 0.791 & 0.494 \\
Hybrid Dense+Lexical RAG & 244/300 & 0.813 & [0.765, 0.853] & 0.673 & 0.484 \\
\midrule
GraphRAG w/o reflection & 16/300 & 0.053 & [0.033, 0.085] & 0.448 & 0.410 \\
GraphRAG w/o router & 84/300 & 0.280 & [0.232, 0.333] & 0.821 & 0.723 \\
Structured-only GraphRAG & 99/300 & 0.330 & [0.279, 0.385] & 0.821 & 0.758 \\
Full Agentic GraphRAG & 22/300 & 0.073 & [0.049, 0.109] & 0.681 & 0.570 \\
\bottomrule
\end{tabularx}
\end{table*}

% Generated by scripts/phase2_generate_manuscript_tables.py from results/all_architectures_failure_pairwise_statistics.csv
\begin{table*}[t!]
\centering
\caption{Paired graph-seeded unsuccessful-outcome comparisons against Full Agentic GraphRAG.}
\label{tab:supp_failure_pairwise}
\scriptsize
\begin{tabularx}{\textwidth}{@{} >{\raggedright\arraybackslash}X c c c c c c @{}}
\toprule
\textbf{Comparison} & \textbf{N} & \textbf{Diff.} & \textbf{95\% CI} & \textbf{A fail/F pass} & \textbf{A pass/F fail} & \textbf{Holm $p$} \\
\midrule
Dense Vector-RAG minus Full & 300 & -0.073 & [-0.103, -0.047] & 0 & 22 & $<0.001$ \\
Lexical Full-Text RAG minus Full & 300 & 0.867 & [0.823, 0.903] & 262 & 2 & $<0.001$ \\
Hybrid Dense+Lexical RAG minus Full & 300 & 0.740 & [0.687, 0.790] & 223 & 1 & $<0.001$ \\
GraphRAG without reflection minus Full & 300 & -0.020 & [-0.053, 0.013] & 9 & 15 & 0.307 \\
GraphRAG without intent routing minus Full & 300 & 0.207 & [0.160, 0.253] & 64 & 2 & $<0.001$ \\
Structured-only GraphRAG minus Full & 300 & 0.257 & [0.207, 0.307] & 78 & 1 & $<0.001$ \\
\bottomrule
\end{tabularx}
\end{table*}

% Generated by scripts/phase2_generate_manuscript_tables.py from results/all_architectures_failure_summary.csv
\begin{table*}[t!]
\centering
\caption{Failure-category decomposition for the graph-seeded benchmark. Categories are non-exclusive except for the primary category recorded in the machine-readable CSV.}
\label{tab:supp_failure_categories}
\scriptsize
\begin{tabularx}{\textwidth}{@{} >{\raggedright\arraybackslash}X c c c c c @{}}
\toprule
\textbf{Architecture variant} & \textbf{Empty retrieval} & \textbf{Empty answer} & \textbf{Retrieval exec.} & \textbf{Recovery flag} & \textbf{No-answer} \\
\midrule
Dense Vector-RAG & 0 & 0 & 0 & 0 & 0 \\
Lexical Full-Text RAG & 279 & 282 & 0 & 0 & 282 \\
Hybrid Dense+Lexical RAG & 0 & 244 & 0 & 0 & 244 \\
\midrule
GraphRAG w/o reflection & 16 & 0 & 0 & 16 & 0 \\
GraphRAG w/o router & 84 & 0 & 0 & 84 & 0 \\
Structured-only GraphRAG & 98 & 0 & 1 & 99 & 3 \\
Full Agentic GraphRAG & 22 & 0 & 1 & 0 & 1 \\
\bottomrule
\end{tabularx}
\end{table*}

% Generated by scripts/phase2_generate_manuscript_tables.py from results/all_architectures_failure_faithfulness_analysis.csv
\begin{table*}[t!]
\centering
\caption{Faithfulness diagnostics on graph-seeded failed and non-failed rows.}
\label{tab:supp_failure_faithfulness}
\scriptsize
\begin{tabularx}{\textwidth}{@{} >{\raggedright\arraybackslash}X c c c c c @{}}
\toprule
\textbf{Architecture variant} & \textbf{Failed rows} & \textbf{Faith. failed} & \textbf{Faith. non-failed} & \textbf{Failed faith. $\geq$0.9} & \textbf{Failed rel./recall $\leq$0.1} \\
\midrule
Dense Vector-RAG & 0 & N/A & 0.905 & 0 & 0/0 \\
Lexical Full-Text RAG & 282 & 0.995 & 0.648 & 280 & 282/282 \\
Hybrid Dense+Lexical RAG & 244 & 0.882 & 0.863 & 200 & 243/243 \\
\midrule
GraphRAG w/o reflection & 16 & 0.131 & 0.937 & 2 & 16/16 \\
GraphRAG w/o router & 84 & 0.841 & 0.904 & 55 & 13/35 \\
Structured-only GraphRAG & 99 & 0.726 & 0.934 & 53 & 46/81 \\
Full Agentic GraphRAG & 22 & 0.940 & 0.895 & 19 & 2/5 \\
\bottomrule
\end{tabularx}
\end{table*}

% Generated by scripts/phase2_generate_manuscript_tables.py from results/all_architectures_failure_breakdown.csv
\begin{table*}[t!]
\centering
\caption{Graph-seeded unsuccessful-outcome rates by difficulty and task type.}
\label{tab:supp_failure_breakdown}
\scriptsize
\begin{tabularx}{\textwidth}{@{} >{\raggedright\arraybackslash}X L{2.3cm} L{3.1cm} c c c c @{}}
\toprule
\textbf{Architecture variant} & \textbf{Grouping} & \textbf{Value} & \textbf{N} & \textbf{Failed} & \textbf{Rate} & \textbf{Non-fail recall} \\
\midrule
Dense Vector-RAG & difficulty & Level 1 & 100 & 0 & 0.000 & 0.078 \\
Dense Vector-RAG & difficulty & Level 2 & 100 & 0 & 0.000 & 0.070 \\
Dense Vector-RAG & difficulty & Level 3 & 100 & 0 & 0.000 & 0.049 \\
Dense Vector-RAG & task type & Direct Entity Data & 100 & 0 & 0.000 & 0.078 \\
Dense Vector-RAG & task type & NameHub Entity Resolution & 100 & 0 & 0.000 & 0.070 \\
Dense Vector-RAG & task type & Temporal History & 100 & 0 & 0.000 & 0.049 \\
Full Agentic GraphRAG & difficulty & Level 1 & 100 & 1 & 0.010 & 0.522 \\
Full Agentic GraphRAG & difficulty & Level 2 & 100 & 20 & 0.200 & 0.490 \\
Full Agentic GraphRAG & difficulty & Level 3 & 100 & 1 & 0.010 & 0.683 \\
Full Agentic GraphRAG & task type & Direct Entity Data & 100 & 1 & 0.010 & 0.522 \\
Full Agentic GraphRAG & task type & NameHub Entity Resolution & 100 & 20 & 0.200 & 0.490 \\
Full Agentic GraphRAG & task type & Temporal History & 100 & 1 & 0.010 & 0.683 \\
GraphRAG w/o router & difficulty & Level 1 & 100 & 0 & 0.000 & 0.819 \\
GraphRAG w/o router & difficulty & Level 2 & 100 & 83 & 0.830 & 0.273 \\
GraphRAG w/o router & difficulty & Level 3 & 100 & 1 & 0.010 & 0.703 \\
GraphRAG w/o router & task type & Direct Entity Data & 100 & 0 & 0.000 & 0.819 \\
GraphRAG w/o router & task type & NameHub Entity Resolution & 100 & 83 & 0.830 & 0.273 \\
GraphRAG w/o router & task type & Temporal History & 100 & 1 & 0.010 & 0.703 \\
GraphRAG w/o reflection & difficulty & Level 1 & 100 & 0 & 0.000 & 0.785 \\
GraphRAG w/o reflection & difficulty & Level 2 & 100 & 15 & 0.150 & 0.036 \\
GraphRAG w/o reflection & difficulty & Level 3 & 100 & 1 & 0.010 & 0.353 \\
GraphRAG w/o reflection & task type & Direct Entity Data & 100 & 0 & 0.000 & 0.785 \\
GraphRAG w/o reflection & task type & NameHub Entity Resolution & 100 & 15 & 0.150 & 0.036 \\
GraphRAG w/o reflection & task type & Temporal History & 100 & 1 & 0.010 & 0.353 \\
Hybrid Dense+Lexical RAG & difficulty & Level 1 & 100 & 81 & 0.810 & 0.761 \\
Hybrid Dense+Lexical RAG & difficulty & Level 2 & 100 & 87 & 0.870 & 0.343 \\
Hybrid Dense+Lexical RAG & difficulty & Level 3 & 100 & 76 & 0.760 & 0.342 \\
Hybrid Dense+Lexical RAG & task type & Direct Entity Data & 100 & 81 & 0.810 & 0.761 \\
Hybrid Dense+Lexical RAG & task type & NameHub Entity Resolution & 100 & 87 & 0.870 & 0.343 \\
Hybrid Dense+Lexical RAG & task type & Temporal History & 100 & 76 & 0.760 & 0.342 \\
Lexical Full-Text RAG & difficulty & Level 1 & 100 & 98 & 0.980 & 1.000 \\
Lexical Full-Text RAG & difficulty & Level 2 & 100 & 87 & 0.870 & 0.389 \\
Lexical Full-Text RAG & difficulty & Level 3 & 100 & 97 & 0.970 & 0.610 \\
Lexical Full-Text RAG & task type & Direct Entity Data & 100 & 98 & 0.980 & 1.000 \\
Lexical Full-Text RAG & task type & NameHub Entity Resolution & 100 & 87 & 0.870 & 0.389 \\
Lexical Full-Text RAG & task type & Temporal History & 100 & 97 & 0.970 & 0.610 \\
Structured-only GraphRAG & difficulty & Level 1 & 100 & 1 & 0.010 & 0.807 \\
Structured-only GraphRAG & difficulty & Level 2 & 100 & 97 & 0.970 & 0.033 \\
Structured-only GraphRAG & difficulty & Level 3 & 100 & 1 & 0.010 & 0.731 \\
Structured-only GraphRAG & task type & Direct Entity Data & 100 & 1 & 0.010 & 0.807 \\
Structured-only GraphRAG & task type & NameHub Entity Resolution & 100 & 97 & 0.970 & 0.033 \\
Structured-only GraphRAG & task type & Temporal History & 100 & 1 & 0.010 & 0.731 \\
\bottomrule
\end{tabularx}
\end{table*}

% Generated by scripts/phase2_generate_manuscript_tables.py
% from results/all_architectures_pairwise_statistics.csv

\begin{table*}[t!]
\centering
\caption{Paired answer-quality comparisons against Full Agentic GraphRAG
across benchmark items. Benchmark abbreviations: GS, graph-seeded;
MC, manually curated; CONV, conversational. Architecture abbreviations:
FA, Full Agentic GraphRAG; DV, Dense Vector-RAG; LF, Lexical Full-Text RAG;
HY, Hybrid Dense+Lexical RAG; NR, GraphRAG without reflection;
NI, GraphRAG without intent routing; SO, Structured-only GraphRAG.}
\label{tab:supp_pairwise_quality_all_architectures}

\begingroup
\scriptsize
\setlength{\tabcolsep}{4.5pt}
\renewcommand{\arraystretch}{1.02}

\begin{tabularx}{\textwidth}{
    @{}
    >{\hsize=0.55\hsize\raggedright\arraybackslash}X
    >{\hsize=0.75\hsize\raggedright\arraybackslash}X
    >{\hsize=2.15\hsize\raggedright\arraybackslash}X
    >{\hsize=0.45\hsize\centering\arraybackslash}X
    >{\hsize=1.65\hsize\centering\arraybackslash}X
    >{\hsize=0.45\hsize\centering\arraybackslash}X
    @{}
}
\toprule
\textbf{Bench.} &
\textbf{Comp.} &
\textbf{Metric} &
\textbf{N} &
\textbf{Delta [95\% CI]} &
\textbf{$p$} \\
\midrule

GS & FA--DV & Faithfulness       & 300 & $-0.007\;[-0.043,\phantom{-}0.028]$ & 0.704 \\
GS & FA--DV & Answer Relevance   & 300 & $\phantom{-}0.609\;[\phantom{-}0.558,\phantom{-}0.658]$ & $<0.001$ \\
GS & FA--DV & Information Recall & 300 & $\phantom{-}0.508\;[\phantom{-}0.458,\phantom{-}0.561]$ & $<0.001$ \\

GS & FA--LF & Faithfulness       & 300 & $-0.077\;[-0.104,-0.047]$ & $<0.001$ \\
GS & FA--LF & Answer Relevance   & 300 & $\phantom{-}0.642\;[\phantom{-}0.592,\phantom{-}0.690]$ & $<0.001$ \\
GS & FA--LF & Information Recall & 300 & $\phantom{-}0.544\;[\phantom{-}0.497,\phantom{-}0.590]$ & $<0.001$ \\

GS & FA--HY & Faithfulness       & 300 & $\phantom{-}0.019\;[-0.020,\phantom{-}0.060]$ & 0.348 \\
GS & FA--HY & Answer Relevance   & 300 & $\phantom{-}0.563\;[\phantom{-}0.505,\phantom{-}0.616]$ & $<0.001$ \\
GS & FA--HY & Information Recall & 300 & $\phantom{-}0.483\;[\phantom{-}0.429,\phantom{-}0.534]$ & $<0.001$ \\

GS & FA--NR & Faithfulness       & 300 & $\phantom{-}0.003\;[-0.035,\phantom{-}0.042]$ & 0.859 \\
GS & FA--NR & Answer Relevance   & 300 & $\phantom{-}0.264\;[\phantom{-}0.198,\phantom{-}0.332]$ & $<0.001$ \\
GS & FA--NR & Information Recall & 300 & $\phantom{-}0.185\;[\phantom{-}0.120,\phantom{-}0.255]$ & $<0.001$ \\

GS & FA--NI & Faithfulness       & 300 & $\phantom{-}0.011\;[-0.019,\phantom{-}0.043]$ & 0.499 \\
GS & FA--NI & Answer Relevance   & 300 & $-0.053\;[-0.109,\phantom{-}0.002]$ & 0.058 \\
GS & FA--NI & Information Recall & 300 & $-0.008\;[-0.064,\phantom{-}0.052]$ & 0.798 \\

GS & FA--SO & Faithfulness       & 300 & $\phantom{-}0.033\;[-0.002,\phantom{-}0.067]$ & 0.070 \\
GS & FA--SO & Answer Relevance   & 300 & $\phantom{-}0.038\;[-0.021,\phantom{-}0.095]$ & 0.219 \\
GS & FA--SO & Information Recall & 300 & $\phantom{-}0.040\;[-0.022,\phantom{-}0.102]$ & 0.215 \\

\addlinespace[2pt]
\midrule
\addlinespace[1pt]

MC & FA--DV & Correctness        & 60 & $\phantom{-}0.685\;[\phantom{-}0.565,\phantom{-}0.790]$ & $<0.001$ \\
MC & FA--DV & Answer Relevance   & 60 & $\phantom{-}0.640\;[\phantom{-}0.514,\phantom{-}0.758]$ & $<0.001$ \\
MC & FA--DV & Information Recall & 60 & $\phantom{-}0.719\;[\phantom{-}0.618,\phantom{-}0.807]$ & $<0.001$ \\

MC & FA--LF & Correctness        & 60 & $\phantom{-}0.591\;[\phantom{-}0.473,\phantom{-}0.700]$ & $<0.001$ \\
MC & FA--LF & Answer Relevance   & 60 & $\phantom{-}0.565\;[\phantom{-}0.431,\phantom{-}0.691]$ & $<0.001$ \\
MC & FA--LF & Information Recall & 60 & $\phantom{-}0.577\;[\phantom{-}0.449,\phantom{-}0.688]$ & $<0.001$ \\

MC & FA--HY & Correctness        & 60 & $\phantom{-}0.566\;[\phantom{-}0.441,\phantom{-}0.682]$ & $<0.001$ \\
MC & FA--HY & Answer Relevance   & 60 & $\phantom{-}0.418\;[\phantom{-}0.273,\phantom{-}0.560]$ & $<0.001$ \\
MC & FA--HY & Information Recall & 60 & $\phantom{-}0.600\;[\phantom{-}0.490,\phantom{-}0.701]$ & $<0.001$ \\

MC & FA--NR & Correctness        & 60 & $\phantom{-}0.276\;[\phantom{-}0.174,\phantom{-}0.390]$ & $<0.001$ \\
MC & FA--NR & Answer Relevance   & 60 & $\phantom{-}0.324\;[\phantom{-}0.212,\phantom{-}0.441]$ & $<0.001$ \\
MC & FA--NR & Information Recall & 60 & $\phantom{-}0.278\;[\phantom{-}0.173,\phantom{-}0.393]$ & $<0.001$ \\

MC & FA--NI & Correctness        & 60 & $\phantom{-}0.121\;[\phantom{-}0.039,\phantom{-}0.213]$ & 0.004 \\
MC & FA--NI & Answer Relevance   & 60 & $\phantom{-}0.111\;[\phantom{-}0.041,\phantom{-}0.190]$ & $<0.001$ \\
MC & FA--NI & Information Recall & 60 & $\phantom{-}0.134\;[\phantom{-}0.052,\phantom{-}0.227]$ & 0.001 \\

MC & FA--SO & Correctness        & 60 & $\phantom{-}0.283\;[\phantom{-}0.177,\phantom{-}0.393]$ & $<0.001$ \\
MC & FA--SO & Answer Relevance   & 60 & $\phantom{-}0.155\;[\phantom{-}0.073,\phantom{-}0.242]$ & $<0.001$ \\
MC & FA--SO & Information Recall & 60 & $\phantom{-}0.260\;[\phantom{-}0.152,\phantom{-}0.370]$ & $<0.001$ \\

\addlinespace[2pt]
\midrule
\addlinespace[1pt]

CONV & FA--DV & Turn Success Rate          & 10 & $\phantom{-}0.375\;[\phantom{-}0.150,\phantom{-}0.575]$ & 0.001 \\
CONV & FA--DV & Context Carryover Accuracy & 10 & $\phantom{-}0.250\;[\phantom{-}0.033,\phantom{-}0.467]$ & 0.030 \\

CONV & FA--LF & Turn Success Rate          & 10 & $\phantom{-}0.433\;[\phantom{-}0.267,\phantom{-}0.592]$ & $<0.001$ \\
CONV & FA--LF & Context Carryover Accuracy & 10 & $\phantom{-}0.283\;[\phantom{-}0.133,\phantom{-}0.450]$ & $<0.001$ \\

CONV & FA--HY & Turn Success Rate          & 10 & $\phantom{-}0.433\;[\phantom{-}0.183,\phantom{-}0.642]$ & $<0.001$ \\
CONV & FA--HY & Context Carryover Accuracy & 10 & $\phantom{-}0.333\;[\phantom{-}0.083,\phantom{-}0.533]$ & 0.009 \\

CONV & FA--NR & Turn Success Rate          & 10 & $\phantom{-}0.292\;[\phantom{-}0.108,\phantom{-}0.450]$ & $<0.001$ \\
CONV & FA--NR & Context Carryover Accuracy & 10 & $\phantom{-}0.167\;[\phantom{-}0.000,\phantom{-}0.333]$ & 0.111 \\
CONV & FA--NR & Tool Transition Accuracy   & 10 & $\phantom{-}0.100\;[\phantom{-}0.000,\phantom{-}0.300]$ & 0.677 \\

CONV & FA--NI & Turn Success Rate          & 10 & $\phantom{-}0.200\;[\phantom{-}0.025,\phantom{-}0.375]$ & 0.047 \\
CONV & FA--NI & Context Carryover Accuracy & 10 & $\phantom{-}0.167\;[\phantom{-}0.000,\phantom{-}0.333]$ & 0.111 \\
CONV & FA--NI & Tool Transition Accuracy   & 10 & $\phantom{-}0.000\;[\phantom{-}0.000,\phantom{-}0.000]$ & 1.000 \\

CONV & FA--SO & Turn Success Rate          & 10 & $\phantom{-}0.133\;[-0.058,\phantom{-}0.317]$ & 0.167 \\
CONV & FA--SO & Context Carryover Accuracy & 10 & $\phantom{-}0.033\;[-0.167,\phantom{-}0.200]$ & 0.767 \\
CONV & FA--SO & Tool Transition Accuracy   & 10 & $\phantom{-}0.000\;[\phantom{-}0.000,\phantom{-}0.000]$ & 1.000 \\

\bottomrule
\end{tabularx}

\endgroup
\end{table*}

These tables should be read together with the main results. The graph-seeded benchmark shows that graph-mediated retrieval substantially improves relevance and recall relative to flat retrieval, but also reveals different reliability profiles among graph variants. The manually curated benchmark provides the clearest answer-quality comparison because each item has a verified reference answer. The conversational benchmark is smaller and should be interpreted as preliminary evidence of context carryover rather than a comprehensive user study. The trajectory diagnostics connect retrieval behavior to evidence-bearing context and final answer correctness.

\FloatBarrier
\subsection{Answer-Quality Top-$k$ Sensitivity}
\label{sec:supp_vector_topk_answer_quality}

The first sensitivity analysis reruns the complete vector-RAG answer-quality pipeline on the 60-question manually curated benchmark at $k=5$, $k=10$, and $k=20$. This is an end-to-end answer-quality evaluation: for each value of $k$, the baseline retrieves documents from the full-corpus Chroma index, generates an answer with GPT-4o mini, and scores the result with the same LLM-as-a-judge protocol using GPT-5. No embeddings are rebuilt, and no alternative embedding model is introduced. These runs regenerate answers and judge scores rather than reusing the vector-RAG outputs from the main all-architecture comparison. Therefore, the $k=5$ row should be interpreted as a new sensitivity run, not as an exact duplicate of the dense Vector-RAG row in the main manuscript. The purpose of the experiment is narrower: to determine whether the vector-RAG result in the main comparison was primarily caused by insufficient retrieval depth. Table~\ref{tab:phase2_vector_topk_answer_quality} reports the complete results.

\begin{table*}[htbp]
\centering
\caption{Vector-RAG answer-quality sensitivity by retrieval depth on the manually curated benchmark.}
\label{tab:phase2_vector_topk_answer_quality}
\small
\begin{tabular}{ccccccccc}
\toprule
$k$ & Faith. & Corr. & Rel. & Recall & Mean s & Median s & P95 s & Avg. Docs \\
\midrule
5 & 0.858 & 0.082 & 0.142 & 0.064 & 16.97 & 16.23 & 23.26 & 4.65 \\
10 & 0.875 & 0.103 & 0.197 & 0.082 & 16.61 & 16.09 & 24.43 & 9.25 \\
20 & 0.887 & 0.112 & 0.201 & 0.086 & 16.86 & 16.44 & 24.17 & 18.50 \\
\bottomrule
\end{tabular}
\end{table*}

The results show only modest gains as retrieval depth increases. Correctness rises from 0.082 at $k=5$ to 0.112 at $k=20$, Answer Relevance from 0.142 to 0.201, and Information Recall from 0.064 to 0.086. Faithfulness remains high across settings, increasing from 0.858 to 0.887, which indicates that generated answers generally remain grounded in the retrieved chunks even when those chunks do not contain the reference answer. Mean latency is similar across settings, while the average number of retrieved documents increases from 4.65 to 18.50 and the average context size increases from approximately 1,512 to 5,940 characters.

The generated answers, aggregate metrics, and evaluation metadata are available in the project's code repository.
This experiment should be interpreted conservatively. It shows that increasing dense-retrieval depth alone does not substantially change the answer-quality outcome for the evaluated vector baseline. It does not show that all vector-RAG systems are weak, and it should be read together with the lexical and hybrid baselines reported below rather than as a complete baseline analysis on its own.

\FloatBarrier
\subsection{Lexical Full-Text Baseline}
\label{sec:supp_lexical_answer_quality}

The lexical baseline uses a Neo4j Lucene full-text index over BaseNode properties \texttt{name}, \texttt{text}, and \texttt{rubric}. Returned evidence is built primarily from \texttt{Event.text}, with node identifiers, labels, Lucene scores, dates, rubrics, sub-rubrics, and names included when available. The answer-generation and judging protocol is identical to the vector baseline: answers are generated with GPT-4o mini and scored by GPT-5. The evaluation is run on the 60-question manually curated benchmark at $k=5$, $k=10$, and $k=20$. Table~\ref{tab:phase2_lexical_answer_quality} reports the results.

\begin{table*}[htbp]
\centering
\caption{Lexical full-text answer-quality sensitivity by retrieval depth on the manually curated benchmark.}
\label{tab:phase2_lexical_answer_quality}
\small
\begin{tabular}{ccccccccc}
\toprule
$k$ & Faith. & Corr. & Rel. & Recall & Mean s & Median s & P95 s & Avg. Docs \\
\midrule
5 & 0.974 & 0.134 & 0.195 & 0.117 & 13.57 & 13.24 & 19.04 & 1.67 \\
10 & 0.992 & 0.215 & 0.312 & 0.221 & 16.53 & 15.67 & 28.37 & 2.98 \\
20 & 0.977 & 0.237 & 0.322 & 0.261 & 14.29 & 13.92 & 21.05 & 4.62 \\
\bottomrule
\end{tabular}
\end{table*}

The lexical baseline improves over the vector-only baseline at all tested retrieval depths. At $k=20$, Correctness increases to 0.237, Answer Relevance to 0.322, and Information Recall to 0.261. The generated answers, aggregate metrics, and evaluation metadata are available in the project's code repository. Token usage was logged for answer generation and judging; the estimated OpenAI cost for the three lexical runs combined is approximately \$1.50.

\FloatBarrier
\subsection{Hybrid Dense+Lexical Baseline}
\label{sec:supp_hybrid_answer_quality}

The hybrid baseline combines the Chroma all-MiniLM-L6-v2 dense index with the Neo4j full-text index. For each question and final $k$, the system retrieves dense candidates and lexical candidates, fuses them using reciprocal rank fusion with constant $c=60$, keeps the top-$k$ fused evidence items, and then applies the same answer-generation and LLM-as-a-judge protocol used for the vector and lexical baselines. Table~\ref{tab:phase2_hybrid_answer_quality} reports the results.

\begin{table*}[htbp]
\centering
\caption{Hybrid dense+lexical answer-quality sensitivity by retrieval depth on the manually curated benchmark.}
\label{tab:phase2_hybrid_answer_quality}
\small
\begin{tabular}{ccccccccc}
\toprule
$k$ & Faith. & Corr. & Rel. & Recall & Mean s & Median s & P95 s & Avg. Docs \\
\midrule
5 & 0.909 & 0.195 & 0.310 & 0.171 & 20.13 & 17.73 & 34.26 & 5.00 \\
10 & 0.963 & 0.203 & 0.404 & 0.169 & 16.78 & 15.67 & 24.13 & 10.00 \\
20 & 0.958 & 0.262 & 0.470 & 0.237 & 17.50 & 16.62 & 24.78 & 20.00 \\
\bottomrule
\end{tabular}
\end{table*}

The hybrid baseline substantially improves over vector-only retrieval, especially in Answer Relevance. At $k=20$, it reaches Correctness 0.262, Answer Relevance 0.470, and Information Recall 0.237. Compared with lexical-only retrieval at $k=20$, the hybrid baseline improves Correctness and Answer Relevance, while lexical-only retrieval has slightly higher Information Recall (0.261 versus 0.237). The generated answers, aggregate metrics, and evaluation metadata are available in the project's code repository. Token usage was logged for answer generation and judging; the estimated OpenAI cost for the three hybrid runs combined is approximately \$2.04.

\FloatBarrier
\subsection{Retrieval-Only Top-$k$ Sensitivity}
\label{sec:supp_vector_topk_retrieval_only}

The second sensitivity analysis is retrieval-only. It evaluates the existing full-corpus Chroma index at $k=5$, $k=10$, and $k=20$ on the manually curated benchmark without using OpenAI, Neo4j, backend/API execution, answer generation, or LLM judging. Unlike the answer-quality sensitivity analysis above, this diagnostic only evaluates whether retrieved chunks contain lexical evidence from the reference answer. Table~\ref{tab:phase2_vector_topk_retrieval_only} reports the retrieval diagnostics.

\begin{table}[htbp]
\centering
\caption{Retrieval-only top-$k$ sensitivity for the vector baseline. No answer generation or LLM judging is used.}
\label{tab:phase2_vector_topk_retrieval_only}
\small
\begin{tabular}{lcccc}
\toprule
$k$ & Exact Hit & Token Recall & Token Hit@50\% & Nonempty \\
\midrule
5 & 0.000 & 0.259 & 0.217 & 1.000 \\
10 & 0.000 & 0.279 & 0.233 & 1.000 \\
20 & 0.000 & 0.300 & 0.233 & 1.000 \\
\bottomrule
\end{tabular}
\end{table}

The exact reference-answer retrieval hit is a strict indicator equal to 1 when the normalized full reference answer appears verbatim in the retrieved chunks and 0 otherwise. Because many reference answers are compositional summaries, this metric is intentionally conservative and should not be overinterpreted. Reference-token recall measures the average fraction of non-stopword reference-answer tokens that appear in the retrieved chunks. The 50-percent reference-token coverage rate measures the fraction of questions for which at least 50\% of the reference-answer tokens are recovered. Finally, the nonempty retrieval rate measures whether the vector index returned at least one retrieved chunk.

These metrics evaluate whether retrieved chunks contain lexical evidence from the reference answer. They do not evaluate whether the system can synthesize the final answer, perform multi-hop reasoning, or produce a correct response.

\FloatBarrier
\subsection{Evidence-linked trajectory diagnostics}
\label{sec:supp_evidence_linked_trajectory}

The evidence-linked trajectory diagnostics complement the final answer-quality evaluation by testing whether the agent's retrieval path actually contained answer-bearing evidence. These metrics are computed on the 60-question manually curated benchmark by linking recorded tool trajectories to the reference answers and existing Correctness scores. They do not introduce new answer-generation or judging runs. Table~\ref{tab:evidence_linked_trajectory_metrics} reports the resulting trajectory-level measures.

\begin{table}[t!]
\centering
\caption{Evidence-linked trajectory diagnostics for the full agentic GraphRAG on the manually curated benchmark. These diagnostics connect recorded tool traces to reference-answer evidence and existing correctness scores; they do not replace final answer-quality metrics.}
\label{tab:evidence_linked_trajectory_metrics}
\small
\begin{tabular}{lll}
\toprule
Metric & Operational definition & Value \\
\midrule
Evidence-bearing retrieval success & Evidence in retrieved context & 0.9500 \\
Correct-answer supported trajectory & Correctness $\geq 0.80$ and evidence-bearing & 0.8000 \\
Non-evidence retrieval rate & Returned data, but no answer-bearing evidence & 0.0500 \\
Tool error / empty-result rate & Empty/error tool calls over all tool calls & 0.0132 \\
Reflection recovery rate & Later evidence after weak first retrieval & 0.6000 \\
Duplicate tool-call rate & Repeated same tool and arguments & 0.0000 \\
\bottomrule
\end{tabular}
\end{table}

Evidence-bearing retrieval success measures whether the retrieved tool context contains conservative answer-specific overlap with the reference answer, such as company or person names, CHE identifiers, dates, numeric values, legal event terms, or distinctive non-generic answer tokens. Correct-answer supported trajectory combines this retrieval-side condition with the existing final-answer Correctness score: a trajectory is counted when the answer is substantially correct and the retrieved context is evidence-bearing. Non-evidence retrieval rate captures cases where a tool returned data but the returned context did not contain answer-bearing evidence. Tool error or empty-result rate is computed at the tool-call level. Reflection recovery is computed only among trajectories requiring recovery, namely multi-call trajectories in which the first retrieval step was empty, failed, or non-evidence-bearing. Duplicate tool-call rate counts exact repeated calls with the same tool and normalized arguments.

On the 60-question manually curated benchmark, $95.0\%$ of trajectories retrieved evidence overlapping with the reference answer, indicating that the tool layer usually returned substantively relevant information rather than merely non-empty results. Moreover, $80.0\%$ of questions achieved a supported correct trajectory, meaning that the final answer was substantially correct and the retrieved context contained answer-bearing evidence. The remaining gap between evidence-bearing retrieval success and supported correct trajectories suggests that some errors occur after relevant evidence has already been retrieved, likely during synthesis, disambiguation, or completeness-sensitive answer formulation.

Operationally, the tool layer was stable: the tool error or empty-result rate was only $1.32\%$ at the tool-call level, and no duplicate tool calls were detected. The reflection recovery rate of $60.0\%$ among trajectories requiring recovery further supports the value of the bounded reflection loop: when an initial retrieval step was weak or non-evidence-bearing, later retrieval steps often recovered useful evidence. Together with the no-reflection ablation, these diagnostics suggest that reflection is not merely an added cost, but contributes to evidence recovery and trajectory reliability.

These diagnostics should be interpreted conservatively. Evidence-bearing retrieval is identified from recorded contexts using answer-specific overlap rather than full human annotation of every trajectory. Therefore, the metrics connect retrieval trajectories to final answer quality, but they do not constitute exhaustive human trajectory validation.

\FloatBarrier
\section{Manual Validation of LLM Weak-Node Extraction}
\label{sec:supp_extraction_validation_protocol}

We manually reviewed a validation sample of 120 SHAB notices to evaluate the LLM-based weak-node extraction stage independently from downstream retrieval performance. The sample covers the event types used by the weak-node extraction pipeline, including rubrics HR01, KK02, KK03, KK06, LS01, and LS02, with notices drawn across German, French, and Italian where available. The evaluated fields are extracted persons, extracted companies, semantic roles, and entity-role relations.

The evaluation is restricted to the weak-node extraction task. Main registered subject companies are treated as structured entities because they are captured by deterministic ingestion from registry metadata. They are therefore excluded from weak-company evaluation on both the prediction and gold sides. Person and company extraction are scored with fuzzy name matching to account for minor orthographic differences. Role extraction is scored with a coarse-compatible criterion aligned with the extractor's intended broad role vocabulary. For example, multilingual or near-equivalent roles are treated as compatible when they describe the same functional role. Entity-role relations require both an entity match and a compatible role match. Table~\ref{tab:supp_llm_extraction_validation} reports the validation results.

\begin{table*}[t!]
\centering
\caption{Manual validation of LLM weak-node extraction on 120 manually reviewed SHAB notices. Main registered subject companies are excluded from the weak-company evaluation because they are captured by deterministic structured ingestion.}
\label{tab:supp_llm_extraction_validation}
\begin{tabularx}{\textwidth}{@{} >{\raggedright\arraybackslash}X c c c >{\raggedright\arraybackslash}X @{}}
\hline
\textbf{Extracted item} & \textbf{Precision} & \textbf{Recall} & \textbf{F1} & \textbf{Matching criterion} \\
\hline
Persons & 0.989 & 0.956 & 0.972 & Fuzzy name match \\
Companies & 0.933 & 0.651 & 0.767 & Fuzzy name match, excluding structured subject entities \\
Roles & 0.723 & 0.611 & 0.662 & Coarse-compatible role match \\
Entity-role relations & 0.701 & 0.482 & 0.571 & Coarse-compatible entity-role match \\
\hline
\end{tabularx}
\end{table*}

These scores validate the LLM weak-node extraction stage only; they do not validate complete graph correctness. The results show high precision for recovered persons and organizations, while company and relation recall remain lower. This pattern supports the interpretation of weak-node extraction as a conservative graph-enrichment mechanism rather than a complete reconstruction of every actor and legal relation in the source notices. The accompanying result file also reports strict-string and ontology-aligned diagnostics; these are conservative lower-bound checks and are not used as the primary role/relation scores because the extraction stage was designed to recover broad actor roles rather than a fully normalized legal-role ontology.

\subsection{Language-Specific Extraction Results}

Table~\ref{tab:supp_llm_extraction_by_language} provides a descriptive breakdown for the notices labeled as French, German, or Italian. Person extraction remains stable across the three languages, with F1 scores of 1.000, 0.963, and 0.966, respectively. Company and entity-role relation extraction vary more: company F1 ranges from 0.625 to 0.917, and relation F1 ranges from 0.471 to 0.651. For 15 notices, the archived validation records did not provide a reliable language label, and the text-based heuristic could not assign one confidently. These notices remain in the aggregate evaluation but are excluded from the language-specific comparison. Because the language strata are modest, the language-specific values should be interpreted as descriptive diagnostics rather than as evidence of statistically significant differences between languages.

\begin{table*}[t!]
\centering
\caption{Language-specific manual validation of LLM weak-node extraction. Each metric cell reports precision/recall/F1. Person and company names use fuzzy matching; roles and entity-role relations use coarse-compatible matching.}
\label{tab:supp_llm_extraction_by_language}
\small
\begin{tabularx}{\textwidth}{@{} >{\raggedright\arraybackslash}X c c c c c @{} }
\hline
\textbf{Language} & \textbf{$N$} & \textbf{Persons} & \textbf{Companies} & \textbf{Roles} & \textbf{Relations} \\
\hline
French & 36 & 1.000/1.000/1.000 & 1.000/0.846/0.917 & 0.800/0.696/0.744 & 0.765/0.542/0.634 \\
German & 36 & 1.000/0.929/0.963 & 0.714/0.556/0.625 & 0.829/0.723/0.773 & 0.818/0.540/0.651 \\
Italian & 33 & 0.966/0.966/0.966 & 1.000/0.538/0.700 & 0.658/0.532/0.588 & 0.571/0.400/0.471 \\
\hline
\multicolumn{6}{@{}p{\textwidth}@{}}{\footnotesize For 15 notices, the archived validation records did not provide a reliable language label, and the text-based heuristic could not assign one confidently. These notices are included in the aggregate 120-notice results but excluded from this language-specific comparison.} \\
\hline
\end{tabularx}
\end{table*}

\section{Company UID-Based Entity-Resolution Diagnostic}
\label{sec:supp_entity_resolution_recall}

The Tier 1 experiment in the main paper reports orthographic consistency for names grouped under the same NameHub. That check evaluates whether grouped names satisfy the predefined orthographic criterion, but it does not measure missed aliases. To provide a stricter external-identifier diagnostic for companies, we additionally evaluate company identity continuity using official Swiss UID values where available.

For this diagnostic, two Company nodes are treated as a predicted same-entity pair when they share the same NameHub. They are treated as a true same-entity pair when they share the same official UID. Under this definition, true positives are company pairs that share both a NameHub and an official UID; false positives are pairs that share a NameHub but have different UIDs; and false negatives are pairs that share an official UID but are assigned to different NameHub nodes. Table~\ref{tab:company_entity_resolution_uid} reports the UID-based metrics.

\begin{table*}[t!]
\centering
\caption{Company entity-resolution validation using official CHE UIDs as an external identity diagnostic. Pairwise recall is UID-continuity recall and should not be interpreted as true person/entity recall for the full NameHub layer.}
\label{tab:company_entity_resolution_uid}
\small
\begin{tabular}{lr}
\toprule
Metric & Value \\
\midrule
Company nodes evaluated & 1,039,198 \\
Unique official UIDs & 806,520 \\
NameHubs evaluated & 908,203 \\
True-positive pairs & 114,556 \\
False-positive pairs & 41,837 \\
False-negative pairs & 157,006 \\
Pairwise precision & 0.7325 \\
Pairwise recall & 0.4218 \\
Pairwise F1 & 0.5354 \\
Contaminated NameHubs & 12,796 (0.0141) \\
Fragmented UIDs & 109,318 (0.1355) \\
\bottomrule
\end{tabular}
\end{table*}

The UID-based diagnostic is stricter than the Tier 1 orthographic consistency audit. On Company nodes with usable official identifiers, the NameHub strategy achieves precision 0.733, UID-continuity recall 0.422, and F1 0.535. These values do not contradict the Tier 1 consistency rate because the two evaluations answer different questions. Tier 1 measures whether sampled grouped names satisfy the orthographic criterion; the UID diagnostic measures whether company identity remains continuous across official-identifier-linked records. The lower recall confirms that the conservative NameHub strategy misses a substantial share of UID-continuous aliases, consistent with the precision-first design of exact alphabetical tokenization.
For persons, no comparable official identifier is consistently available in the SHAB records. We therefore do not report true person-level recall or F1. This limitation is distinct from the company UID diagnostic: official UIDs provide an external continuity signal for many Company nodes, whereas Person nodes generally lack an equivalent registry-wide identifier. As a result, person-level NameHub evaluation remains limited to conservative orthographic consistency and qualitative error analysis rather than full recall estimation.

\section{Statistical Testing and Bootstrap Confidence Intervals}
\label{sec:supp_statistical_testing}

The statistical audit computes paired bootstrap confidence intervals over the per-question and per-conversation result files. Table~\ref{tab:phase2_confidence_intervals} reports the full GraphRAG mean with a 95\% bootstrap interval and the paired full-minus-baseline delta for the two-system comparison against the dense vector baseline. The all-architecture paired comparisons reported in Table~\ref{tab:supp_pairwise_quality_all_architectures} extend this logic to the complete architecture set. These intervals quantify sampling uncertainty over the benchmark items, but they do not estimate judge variance because repeated judging and second-judge annotations were not collected.

\begin{table*}[htbp]
\centering
\caption{Bootstrap confidence intervals and paired graph-minus-baseline deltas from the two-system comparison between the full agentic GraphRAG and the dense vector baseline. Dataset abbreviations: GS, graph-seeded automated benchmark; MC, manually curated benchmark; CONV, conversational benchmark.}
\label{tab:phase2_confidence_intervals}
\small
\begin{tabular}{llllll}
\toprule
Dataset & Metric & Graph Mean [95\% CI] & $\Delta$ & $\Delta$ 95\% CI & $p$ \\
\midrule
GS & Faithfulness & 0.898 [0.874, 0.920] & -0.007 & [-0.044, 0.029] & 0.693 \\
GS & Answer Relevance & 0.689 [0.646, 0.730] & 0.609 & [0.557, 0.658] & $<0.001$ \\
GS & Information Recall & 0.574 [0.528, 0.618] & 0.508 & [0.456, 0.561] & $<0.001$ \\
MC & Correctness & 0.828 [0.735, 0.913] & 0.685 & [0.565, 0.797] & $<0.001$ \\
MC & Answer Relevance & 0.887 [0.822, 0.941] & 0.641 & [0.517, 0.756] & $<0.001$ \\
MC & Information Recall & 0.838 [0.746, 0.919] & 0.719 & [0.611, 0.819] & $<0.001$ \\
CONV & Turn Success Rate & 0.742 [0.633, 0.850] & 0.375 & [0.142, 0.575] & 0.001 \\
CONV & Conversation Goal Completion Rate & 0.800 [0.500, 1.000] & 0.000 & [0.000, 0.000] & 1.000 \\
CONV & Context Carryover Accuracy & 0.683 [0.583, 0.800] & 0.250 & [0.017, 0.467] & 0.039 \\
CONV & Tool Transition Accuracy & 0.900 [0.700, 1.000] & -- & [--, --] & -- \\
\bottomrule
\end{tabular}
\end{table*}

\section{Additional Conversational Examples}
\label{sec:supp_conversational_examples}

The conversational benchmark contains follow-up turns that test entity memory, compression, counting, temporal memory, and subject changes. Table~\ref{tab:phase2_conversation_stats} summarizes the evaluated conversation set, followed by representative context-carryover examples.

\begin{table*}[htbp]
\centering
\caption{Descriptive statistics for the multi-turn conversational benchmark.}
\label{tab:phase2_conversation_stats}
\small
\begin{tabular}{ll}
\toprule
Statistic & Value \\
\midrule
Conversations & 10 \\
Total turns & 36 \\
Mean turns & 3.60 \\
Minimum turns & 3 \\
Maximum turns & 4 \\
Std. turns & 0.49 \\
\bottomrule
\end{tabular}
\end{table*}

The per-conversation results include both successful and weak context-carryover cases.

Conversation C007 is a successful carryover case: the graph agent preserves the active entity across follow-up turns and obtains a context-carryover score of 1.000.

Conversation C008 is a weaker case: a local interpretation error propagates into later turns, producing a context-carryover score of 0.500. This illustrates why turn-level success can be lower than answer fluency in multi-turn settings.

\section{Latency and Tool-Call Statistics}
\label{sec:supp_latency}

Table~\ref{tab:phase2_latency} reports latency and tool-call statistics computed from the execution traces of the two-system evaluation runs, covering the full agentic GraphRAG and the dense vector baseline. Latency depends on sequential LLM calls and tool executions, so these values should be read as implementation-dependent indicators rather than optimized serving performance.

\begin{table*}[htbp]
\centering
\caption{Latency and tool-call statistics computed from the recorded evaluation traces. Dataset abbreviations: GS, graph-seeded automated benchmark; MC, manually curated benchmark; CONV, conversational benchmark. Token usage was not logged for these runs, so API cost is not estimated.}
\label{tab:phase2_latency}
\small
\begin{tabular}{llllllll}
\toprule
Dataset & System & $N$ & Mean s & Median s & P95 s & Mean Tools & Max Tools \\
\midrule
GS & Full Agentic GraphRAG & 300 & 12.53 & 11.36 & 26.93 & 1.67 & 4 \\
GS & Dense Vector-RAG & 300 & 32.45 & 32.35 & 37.66 & -- & -- \\
MC & Full Agentic GraphRAG & 60 & 10.27 & 9.90 & 20.11 & 1.27 & 3 \\
MC & Dense Vector-RAG & 60 & 15.62 & 15.39 & 27.92 & -- & -- \\
CONV & Full Agentic GraphRAG & 10 & 30.60 & 28.37 & 61.63 & 3.70 & 7 \\
CONV & Dense Vector-RAG & 10 & 37.06 & 35.06 & 58.64 & -- & -- \\
\bottomrule
\end{tabular}
\end{table*}

\section{Implementation Identifier Mapping}
\label{sec:supp_identifier_mapping_sec}

Table~\ref{tab:supp_identifier_mapping} maps the implementation identifiers used in the released code and result files to the reader-facing terminology used in the manuscript.

\begin{table*}[htbp]
    \centering
    \footnotesize
    \caption{Mapping between implementation identifiers and reader-facing terminology. Graph-schema labels and properties are intentionally excluded because they identify the database model.}
    \label{tab:supp_identifier_mapping}
    \begin{tabularx}{\textwidth}{@{} >{\raggedright\arraybackslash}p{3.2cm} >{\raggedright\arraybackslash}p{6.5cm} >{\raggedright\arraybackslash}X @{}}
        \toprule
        \textbf{Category} & \textbf{Implementation identifier} & \textbf{Reader-facing terminology} \\
        \midrule
        Agent endpoint & \texttt{search\_companies} & Entity-disambiguation endpoint \\
        Agent endpoint & \texttt{explore\_network} & Network-exploration endpoint \\
        Agent endpoint & \texttt{get\_node\_history} & Event-history endpoint \\
        Agent endpoint & \texttt{global\_text\_search} & Lexical full-text search endpoint \\
        Agent endpoint & \texttt{get\_top\_entities} & Top-entity aggregation endpoint \\
        Agent endpoint & \texttt{count\_entities\_by\_event} & Event-conditioned counting endpoint \\
        Agent endpoint & \texttt{execute\_custom\_cypher} & Read-only custom-query endpoint \\
        Index interface & \texttt{global\_search} & Neo4j full-text index \\
        Router mode & \texttt{all\_tools} & Unrestricted multi-hop exploration \\
        Algorithm & \texttt{generate\_hub\_key} & Alphabetical hub-key generation algorithm \\
        State variable & \texttt{current\_uid} & Active entity identifier \\
        State variable & \texttt{has\_network} & Network-explored flag \\
        State variable & \texttt{has\_history} & History-explored flag \\
        Failure field & \texttt{has\_failure} & Failure-detection rule \\
        Trace field & \texttt{trace\_log} & Execution trace \\
        Result field & \texttt{error\_status} & Error-status field \\
        Result field & \texttt{empty\_answer} & Empty-answer indicator \\
        Result field & \texttt{failed\_tool\_calls} & Failed tool-call count \\
        Conversation tag & \texttt{context\_carryover} & Context carryover \\
        Conversation tag & \texttt{entity\_memory} & Entity memory \\
        Conversation tag & \texttt{temporal\_memory} & Temporal memory \\
        Conversation field & \texttt{expected\_tool\_behavior} & Expected tool behavior \\
        Retrieval metric & \texttt{exact\_expected\_answer\_hit\_at\_k} & Exact reference-answer retrieval hit \\
        Retrieval metric & \texttt{token\_recall\_at\_k} & Reference-token recall \\
        Retrieval metric & \texttt{token\_hit\_50\_at\_k} & 50-percent reference-token coverage rate \\
        Retrieval metric & \texttt{retrieved\_nonempty\_at\_k} & Nonempty retrieval rate \\
        LLM parameter & \texttt{top\_p} & top-$p$ \\
        LLM parameter & \texttt{frequency\_penalty} & Frequency penalty \\
        LLM parameter & \texttt{presence\_penalty} & Presence penalty \\
        API control & \texttt{parallel\_tool\_calls} & Parallel tool execution \\
        \bottomrule
    \end{tabularx}
\end{table*}

\end{document}